# Ionic-to-electronic current amplification in hybrid perovskite solar cells: ionically gated transistor-interface circuit model explains hysteresis and impedance of mixed conducting devices

Davide Moia[1†*], Ilario Gelmetti[2,3†*], Phil Calado[1], William Fisher[1], Michael Stringer[4], Onkar Game[4], Yinghong Hu[5], Pablo Docampo[5,6], David Lidzey[4], Emilio Palomares[2,7], Jenny Nelson[1], Piers R. F. Barnes[1*]

[1]Department of Physics, Imperial College London, London SW7 2AZ, UK
[2]Institute of Chemical Research of Catalonia (ICIQ), Barcelona Institute of Science and Technology (BIST), Avda. Països Catalans 16, 43007 Tarragona, Spain
[3]Departament d'Enginyeria Electrònica, Elèctrica i Automàtica, Universitat Rovira i Virgili, Avda. Països Catalans 26, 43007 Tarragona, Spain
[4]Department of Physics and Astronomy, University of Sheffield, Sheffield S3 7RH, UK
[5]Department of Chemistry and Center for NanoScience (CeNS), LMU München, Butenandtstrasse 5-13, 81377 München, Germany
[6]Physics Department, School of Electrical and Electronic Engineering, Newcastle University, Newcastle upon Tyne NE1 7RU, UK
[7]ICREA, Passeig Lluís Companys, 23, Barcelona, Spain

* davide.moia11@imperial.ac.uk
* igelmetti@iciq.es
* piers.barnes@imperial.ac.uk
† These authors contributed equally to this study

## Abstract
Mobile ions in hybrid perovskite semiconductors introduce a new degree of freedom to electronic devices suggesting applications beyond photovoltaics. An intuitive device model describing the interplay between ionic and electronic charge transfer is needed to unlock the full potential of the technology. We describe the perovskite-contact interfaces as transistors which couple ionic charge redistribution to energetic barriers controlling electronic injection and recombination. This reveals an amplification factor between the out of phase electronic current and the ionic current. Our findings suggest a strategy to design thin film electronic components with large, tuneable, capacitor-like and inductor-like characteristics. The resulting simple equivalent circuit model, which we verified with time-dependent drift-diffusion simulations of measured impedance spectra, allows a general description and interpretation of perovskite solar cell behaviour.

## Broader context
Highly efficient solar cells made using hybrid perovskite semiconductors may prove commercially viable. The success of these cheap materials is in part due to their ability to tolerate high concentrations of crystal defects associated with processing at low temperature while retaining excellent electronic properties. The presence of these electrically charged defects, some of which are mobile, has an interesting side-effect: the solar cell performance will vary with time following a change in the operating conditions (often referred to as hysteresis). This significantly complicates the measurement and analysis of these materials. Hysteresis means that the diode equivalent circuit model, which is successfully used as the simplest description of virtually all other photovoltaic technologies, is not applicable to most hybrid perovskite devices. We show that the interfaces of solar cells and related devices containing inert mobile ions can be very well described if the diode model is replaced by a transistor model. In this description, the 'gate' of the transistor is controlled by the accumulation of mobile charged defects. Consequently, if the time dependent variation of the ionic charge can be understood then the electrical properties of the device can be predicted. This powerful model provides a framework to allow new material/interfaces to be screened for solar cells and other devices laden with inert mobile defects, it also provides a means to unlock the potential of impedance spectroscopy for characterisation, and a method to determine ionic conductivities in hybrid perovskites.

**Table of contents graphic**

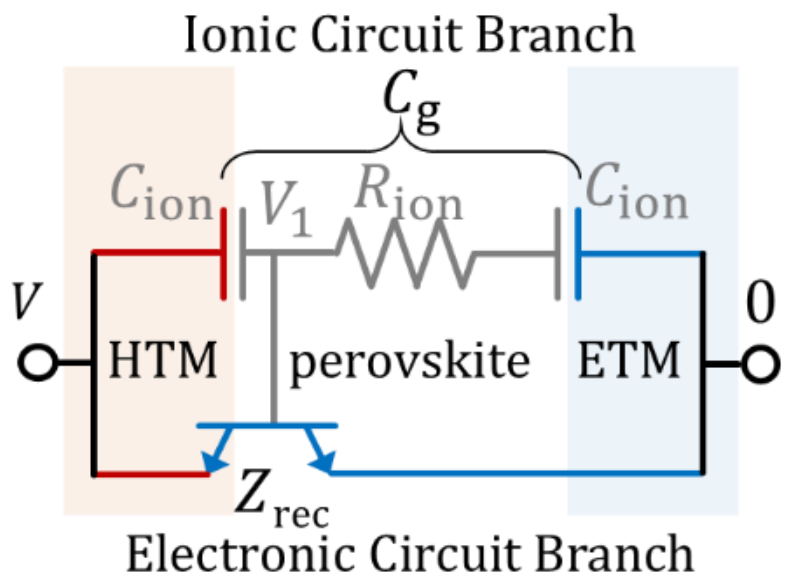

The time and frequency dependent behaviour of hybrid perovskite solar cells is described by an interfacial-transistor circuit model which couples electronic charge transfer to mobile ions.

**Main Text**

The exponential increase in current with the voltage applied across a semiconductor junction arises from the asymmetric change in the energy barrier to charge transfer in each direction across the junction (Fig. 1a). This realisation was a pivotal step in human history. It underpinned the success of the diode and led to the development of transistors and optoelectronic devices such as light emitting diodes and solar cells. Representing solar cells as diodes in equivalent circuit models neatly encapsulates their behaviour[1] and has helped facilitate the worldwide deployment of photovoltaics. However, solar cells based on the rapidly developing technology of hybrid perovskite semiconductors[2, 3] do not generally display pure diode-like characteristics. Identifying an accurate equivalent circuit model describing their behaviour is a priority, both to unravel their unique history-dependent properties, and to enable development and application of new electronic devices utilising these properties. Mobile ionic defects in the perovskite semiconductor phase are thought to underlie the hysteresis often seen in current-voltage sweeps and step-measurements characteristics[4-7] but a physically meaningful equivalent circuit explaining the very large capacitive (> $10^{-3}$ F $cm^{-2}$) and inductive (> 1 H $cm^{-2}$) behaviour reported in perovskite devices is lacking[3, 8-12]. Ferroelectric effects, a photoinduced giant dielectric constant[13], and accumulation of ionic charge[7, 14] have all been discounted as explanations[15-17]. Bisquert *et al.* have proposed that giant capacitances and inductances[17-19] could arise from phase-shifted accumulation or release of electronic charge from within a degenerate layer induced by fluctuations in the surface polarisation due to ionic charge. However, interfacial degeneracy is unlikely to exist under normal operating conditions.[20] More promisingly, Pockett *et al*. have highlighted the link between rate of recombination and varying ion distribution as an explanation for the low frequency behaviour of perovskite impedance spectra[10]. Previous attempts to model the interaction between electronic and ionic charge have used capacitive elements which cannot describe the influence of one species on the electrostatic potential barriers that control fluxes of the other species. This intrinsically limits the applicability of equivalent circuit models of mixed conductors such as perovskites.

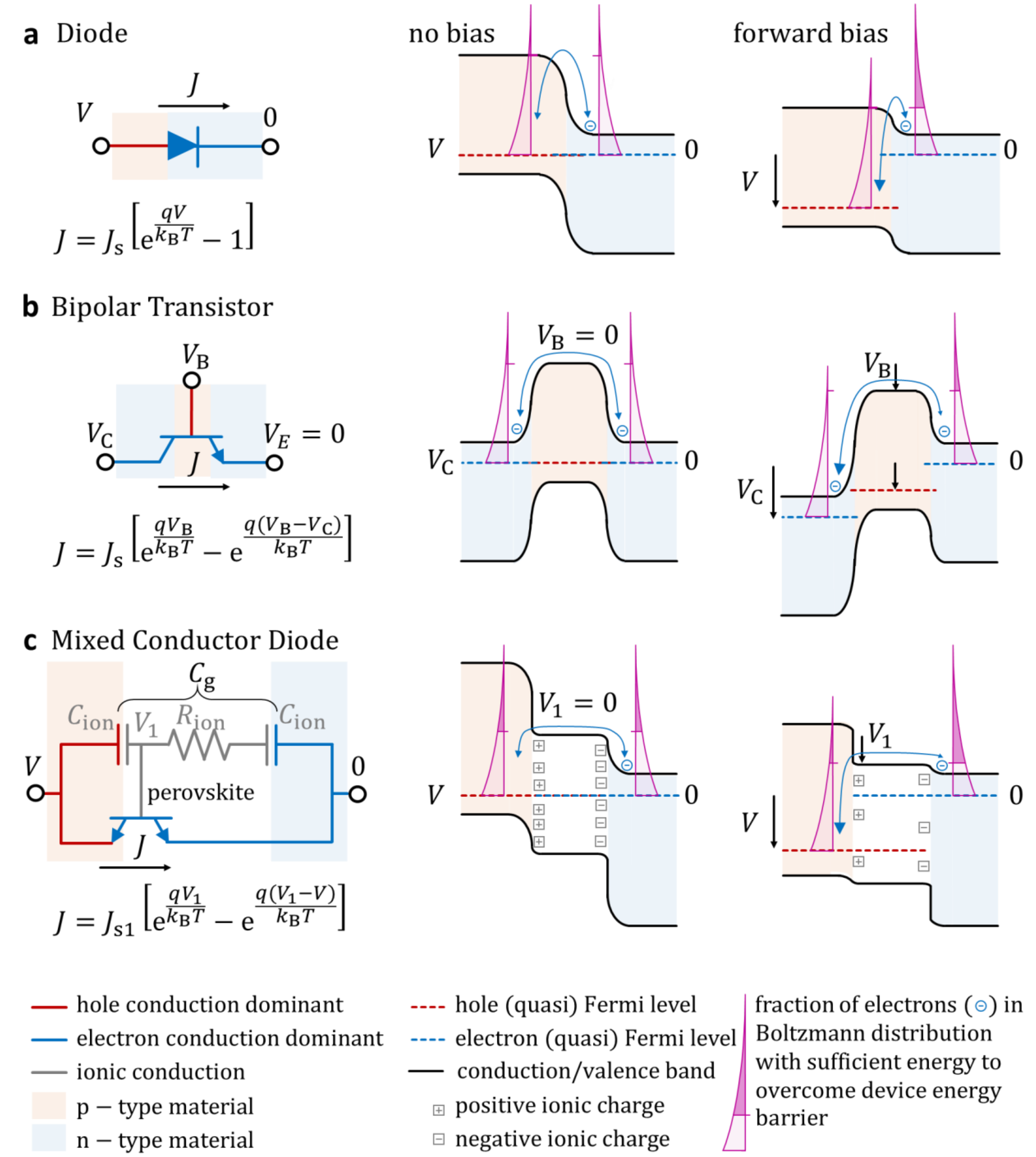

**Fig. 1 Device circuit models and schematic energy level diagrams of a diode, bipolar transistor, and perovskite solar cell in their unbiased and forward biased steady states.** In the energy level diagrams, a positive voltage difference is in the downward direction. All voltages are referenced to the Fermi level of the n-type material on the right-hand side (in contact with the cathode) defined to be zero. The difference in magnitude of the flux of electrons across the energy barrier in each direction is indicated by relative size of the arrow heads on the curved blue lines. (**a**) A p-n junction diode. A forward bias voltage applied across the diode reduces the barrier to electron transfer from the n-type region by $V$, exponentially increasing the flux in this direction, while the flux from the p-type region is unchanged ($J_s$) resulting in a total current density $J$. (**b**) A bipolar n-p-n transistor where the electrical potentials on collector, base and emitter

terminals are $V_C$, $V_B$ and $V_E$ respectively, we define $V_E = 0$ and the base to collector current gain to be infinite. In the unbiased state $V_C = V_B = V_E = 0$. The barrier height of the p-type region can be modulated by varying $V_B$ which exponentially changes the flux of electrons overcoming the barrier from each side, resulting in a total current between the collector and emitter of $J$. The recombination of electrons with holes in the base is neglected. (**c**) A perovskite solar cell forming a mixed ionic-electronic conducting diode. Changes from the dark equilibrium distribution of mobile ionic charge (which occurs with time constant $R_{ion}C_{ion}/2$) result in a *change* in electrostatic potential, $V_1$, *relative to dark equilibrium*. This gates electronic charge transfer across the perovskite HTM interface in a manner analogous to the base of a bipolar transistor (c.f. **b**). The overall device has only two external terminals, $V_1$ is voltage on the base of the transistor element in the circuit model, $J_{s1}$ is the saturation current density of interface at dark equilibrium.

Here we show that the interfaces in perovskite solar cells behave like bipolar transistors[21] (Fig. 1b) in which the electronic energy barriers to injection and recombination are modulated by the accumulation/depletion of ionic charge at the interfaces (Fig. 1c)[22]. Using drift-diffusion simulations of impedance measurements which include mobile ions, we find that: (i) an oscillating voltage applied to the solar cell naturally introduces an out of phase, capacitive ionic current; (ii) the associated changes in electrostatic potential from ion redistribution across the perovskite modulate the rates of electronic recombination and injection across the interfaces. The resulting out of phase electronic current is related to the ionic current through a trans-carrier amplification factor with either a positive sign (for recombination) or a negative sign (for injection or specific recombination cases) and causes capacitor-like or inductor-like behaviour without accumulation of electronic charge at the interfaces. Modelling this amplification effect using ionically gated transistor elements incorporated in a simple equivalent circuit (Fig. 1c) allows us to efficiently explain and physically interpret the many peculiar features observed in the small and large perturbation transient behaviour of perovskite devices (including impedance and current-voltage sweeps). In this context ionic gating is the control of the electronic charge transfer rate across an interface in response to changes in electrostatic potential from mobile ionic charge in the active layer.

The ionically gated interface-transistor model for mixed conductor devices has similar explanatory power to the diode model applied to standard semiconductor devices. It incorporates the key physics of the device to provide a coherent general description of both the time, and frequency dependent behaviour of perovskite solar cells. The resulting insights open the possibility of engineering a new class of mixed conducting electronic devices whose behaviour is controlled by the properties of mobile ions in the active layer. It also reduces the need for far more complex drift-diffusion models and enables key performance parameters of interfaces to be meaningfully assessed using techniques such as electrochemical impedance spectroscopy.

**Measured and simulated impedance spectra characteristics**

To demonstrate the application of the interface-transistor model and the ionic-to-electronic current amplification behaviour at device interfaces we measured impedance spectra of perovskite solar cells. Impedance spectroscopy involves the application of a voltage, $V$, across the external terminals of the device, which includes small periodic voltage perturbation, $v$, superimposed on a background voltage, $\bar{V}$, and measurement of the amplitude and phase shift of the induced oscillating current, $j$, superimposed on a background current $\bar{J}$. The complex impedance ($Z = Z' + iZ''$) is given by $Z = |v/j|\exp(i\theta)$ where $\theta$ is the phase shift of $v$ relative to $j$. This is evaluated at different angular frequencies ($\omega$) resulting in a spectrum $Z(\omega)$.

Fig. 2a and b show impedance data collected from a stable perovskite solar cell equilibrated at open circuit for different light intensities (see complete spectra in Fig. S1, ESI and the effects of stabilisation time which reduces loop artefacts in Fig. S2a-d, ESI). The measurements indicate that, at low frequencies, there is a significant out of phase component in the induced current ($j''$) which results in a large apparent device capacitance, as defined by $\omega^{-1}\mathrm{Im}(Z^{-1})$. This increases linearly with light intensity and thus exponentially with the bias voltage (Fig. 2b), consistent with previous observations[8, 11, 18, 23]. Similar behaviour was also seen at short circuit, or with different applied biases in the dark (Fig. S2d, e, h, i, ESI) ruling out a significant contribution from photoinduced changes in ionic conductivity[24, 25] (Fig. S3, ESI).

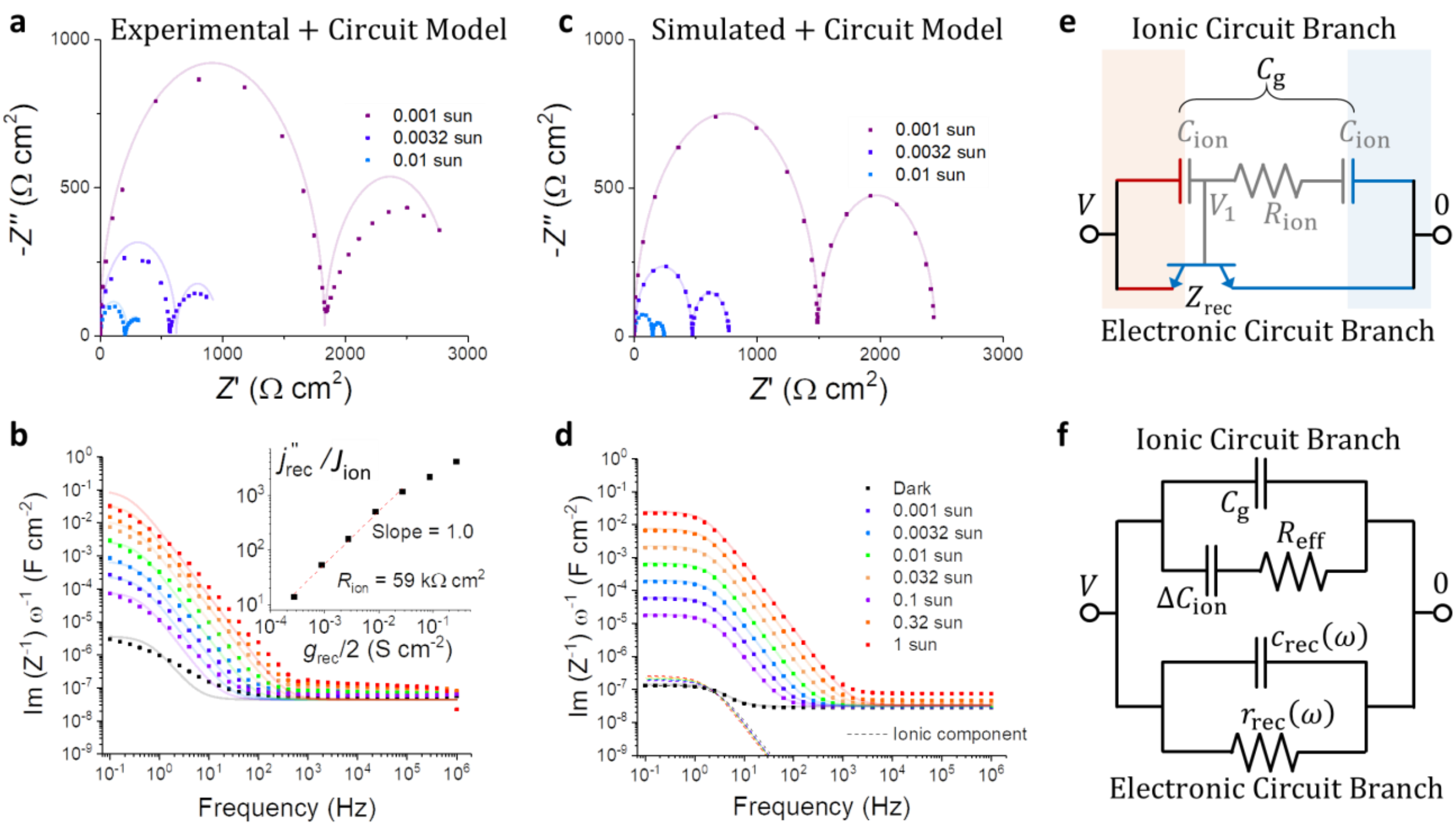

**Fig. 2 Measured and simulated impedance spectra of a perovskite solar cell, and transistor-interface recombination circuit model.** (**a**) Nyquist plot of the real (Z') *vs* imaginary (Z'') impedance components, and (**b**) apparent capacitance, $\omega^{-1}\mathrm{Im}(Z^{-1})$ *vs*

frequency of a spiro-OMeTAD/ $Cs_{0.05}FA_{0.81}MA_{0.14}PbI_{2.55}Br_{0.45}/TiO_2$ solar cell measured around the open circuit voltage with a perturbation amplitude of 20 mV, illuminated with constant bias light intensities (legends of **c** and **d** respectively, for $V_{OC}$ values see Table S1, ESI, and for device details see Methods 1.1, ESI). The devices were stabilised to avoid loops in the Nyquist plot arising as artefacts due to incomplete stabilisation of the device during data collection (see Fig. S2, and for characterisation and stabilisation details see Methods 2.2, ESI,). The inset of **b** shows the out of phase electronic to ionic current ratio, $j''_{\mathrm{rec}}/J_{\mathrm{ion}}$ plotted against half the recombination interface transconductance, $g_{\mathrm{rec}} = \mathrm{d}J_{\mathrm{rec}}/\mathrm{d}(V_1 - V_{\mathrm{n}})$, evaluated from the measurements (Methods 5, ESI). The log-log slope of 1 indicates a linear relationship. (**c**, **d**) Corresponding simulated impedance measurements (filled squares) determined from a drift-diffusion model of a p-type/intrinsic/n-type (p-i-n) device structure containing mobile ionic charge. The dashed lines indicate the simulated contribution to the capacitance from mobile ionic charge. (**e**) The mixed conductor diode circuit model containing an ionically gated transistor used for the simultaneous 5 parameter global fit (continuous lines) to the experimental data (filled squares in **a** and **b**) and to the drift-diffusion simulated data (filled squares in **c** and **d**). The fit parameters are given in Table S1 (ESI). (**f**) An alternative representation of the equivalent circuit model shown in **e**. The elements in the ionic circuit branches are related by $\Delta C_{\mathrm{ion}} = C_{\mathrm{ion}}/2 - C_{\mathrm{g}}$ and $R_{\mathrm{eff}} = R_{\mathrm{ion}}(1 + C_{\mathrm{g}}/C_{\mathrm{ion}})$. The apparent capacitance and recombination resistance elements in the electronic circuit branch, $c_{\mathrm{rec}}(\omega)$ and $r_{\mathrm{rec}}(\omega)$, have a frequency dependence controlled by the ionic circuit branch as derived from the transistor model (see equation 4 and Table 1).

To underpin these measurements with a physical interpretation we simulated impedance spectroscopy measurements using our open source time-dependent drift-diffusion semiconductor model (Driftfusion) which includes the effect of mobile ionic defects [26, 27]. The drift-diffusion simulation solves for the time-evolution of free electron, hole, and mobile ionic defect concentration profiles, as well as the electrostatic potential in a p-i-n device in response to illumination and a varying voltage between the terminals as a boundary condition (Methods 3, ESI). In these simulations, we define the dominant recombination mechanism to be via trap states located in the interfacial regions between the p-type hole transporting material (HTM) and the perovskite, and between n-type electron transporting materials (ETM) and the perovskite. The simulation parameters are listed in Table S2, ESI. We have defined the mobility of the ions to be about 11 orders of magnitude lower than the electrons and holes. As a result, the distribution of electrons will maintain a dynamic equilibrium with respect to any changes in electrostatic potential due to ion redistribution. The positive mobile ionic charge is compensated by a uniform distribution of negative static charge, so that the total ionic charge in the perovskite is zero mimicking Schottky vacancies where one defect species is mobile. We confine the mobile ionic defects to the perovskite layer. The concentration of mobile ions is defined to be about 12 orders of magnitude greater than the intrinsic electronic carrier concentration in the perovskite so that ionic conductivity is approximately 6 times greater than the intrinsic electronic conductivity of the semiconductor at room temperature in the dark. However, under illumination, or with a forward bias, the increase in electronic charge concentration will result in the electronic conductivity significantly exceeding the ionic conductivity. The qualitative behaviour of the simulations that follow is not sensitive to these numbers as long as electronic conductivity significantly exceeds ionic conductivity under operation and the mean concentration of mobile ionic charge exceeds the mean electronic charge concentration.

Fig. 3a shows an example of the simulated steady state profiles of the conduction band, valence band, and quasi Fermi levels under 1 sun equivalent illumination with an applied d.c. voltage boundary condition ($\bar{V}$) equal to the steady state open circuit voltage ($V_{\mathrm{OC}}$). There is no electric field in the bulk of the perovskite layer since the mobile ionic charge has migrated to accumulate at the interfaces screening the built-in potential (Fig. 3a insets) consistent with previous observations and simulations explaining hysteresis.[26, 28-32] Note that, even at 1 sun at open circuit conditions, the majority of the photogenerated electronic charge accumulates in the HTM and ETM at steady state. The amount of electronic charge built up in the perovskite layer is small relative to the amount of mobile ionic charge available to screen changes in potential. Consequently, the changes in electrostatic potential associated with ionic redistribution control the local concentration of electrons and holes in the perovskite. This is important because the concentration of free electrons in the perovskite at the perovskite/HTM interface and the concentration of holes in the perovskite at perovskite/ETM interface determine the rate of recombination via interfacial traps to the respective hole and electron populations in the HTM and ETM layers. Stated another way: the electrostatic potential profile due to ionic charge controls the rate of electron-hole recombination at the interfaces, and this in turn controls current-voltage characteristics of the device.

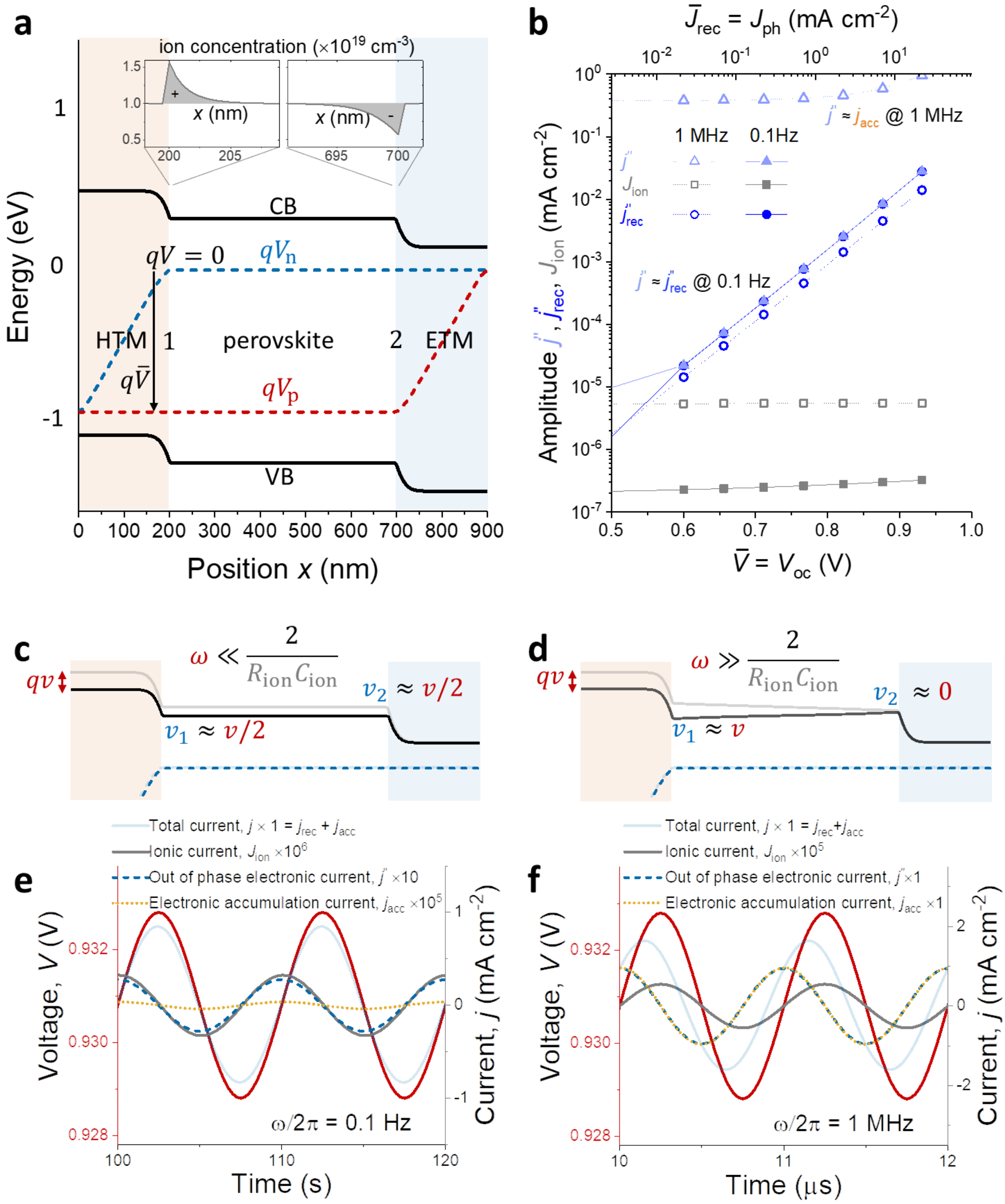

**Fig. 3. Drift-diffusion simulations of energy level diagram and ionic/electronic currents during impedance measurements.** (**a**) The steady state electrostatic energy level profile of the conduction band (CB) and valence band (VB) corresponding to the simulations in Fig. 2c and d at open circuit under 1 sun equivalent illumination, the insets show net accumulation of ionic charges at the HTM/perovskite and perovskite/ETM interfaces screening the internal electric field. (**b**) The simulated oscillation amplitudes of the out of phase component of the cell current, $j''$, the out of phase component of recombination current, $j''_{rec}$, and the ionic current, $J_{ion}$, in response to $v$ (with amplitude ± 2 mV) at 1MHz and 0.1 Hz, plotted against steady state bias voltage $\bar{V} = V_{OC}$ and recombination current $\bar{J}_{rec} = J_{ph}$. Effect of an applied voltage perturbation with amplitude $v$ superimposed on $\bar{V} = V_{OC}$ (**c**) at low frequency (0.1 Hz)

and (**d**) at high frequency (1 MHz) on the conduction band energy profile (limits indicated by the black and grey lines). The amplitude of the electrostatic potential oscillations at each interface, $v_1$ and $v_2$, in response to $v$ are indicated. The corresponding simulated electronic currents (total, $j$, out of phase, $j''$, and accumulation, $j_{acc}$) and ionic current ($J_{ion}$) in response to (**e**) the low frequency and (**f**) the high frequency applied voltage oscillation ($V_{OC} + v$, red line) vs time. At 1 MHz $j'' \approx j_{acc}$, but at 0.1 Hz $j'' \approx j_{rec}$.

We superimposed small oscillating voltages ($v$) on the applied background bias voltage ($\bar{V}$) boundary condition and simulated the resulting oscillations in current ($j$) for a range of angular frequencies ($\omega$) and bias $\bar{V}$. The impedance, $Z(\omega)$, evaluated from these simulations (Fig. 2c and d) shows remarkably similar behaviour to the impedance measurements in Fig. 2a and b. Analysis of the simulations shows that in the dark, with no bias voltage or light, the capacitance of the device, evaluated as $\omega^{-1}\mathrm{Im}(Z^{-1})$, is dominated by contributions from the movement of ionic charge accumulating at the interfaces at low frequencies in response to $v$ (dashed lines in Fig. 2d). However, the exponential increase of $\omega^{-1}\mathrm{Im}(Z^{-1})$ at low frequencies when the steady state voltage $\bar{V}$ across the device was increased by light (or applied voltage in the dark, Fig. S2f and g, ESI) does not arise directly from the ions, and is also not due to an accumulation of electronic charge (see Fig. S2j-l, ESI and the magnitude of electronic accumulation current in Fig. 3e). Instead, this apparent capacitance arises from current due to the out of phase modulation of electronic recombination at the interfaces.

The explanation for this is seen in Figs 3c and d which show that ionic redistribution influences the electrostatic potential profile dropping across the perovskite layer when the applied voltage perturbation ($v$) oscillates at sufficiently low frequencies for the ions to move. The electronic carrier concentration profile responds to form a dynamic equilibrium with the changing electrostatic potential due to mobile ions. At low frequency, the voltage screening effect of ionic redistribution (with associated capacitive ionic current $J_{ion}$) occurs out of phase with $v$ resulting in out of phase modulation of the interfacial recombination of electronic charge ($j''_{\mathrm{rec}}$), and thus current through the device ($j'' \approx j''_{\mathrm{rec}}$, Fig. 3e and b). At high frequencies the ionic redistribution is too slow for ions to compensate the rapid changes in applied potential, so recombination only varies in phase with $v$; in this case the out of phase current component arises primarily from capacitive accumulation of electronic charge in the HTM and ETM contacts ($j'' \approx j_{\mathrm{acc}}$, Fig. 3f and b).

The changes in electrostatic potential due to the oscillation of ionic charge at low frequencies can be viewed as varying the magnitude of the barrier to charge transfer through the device from each direction. This interfacial charge transfer is mediated by the processes of interfacial recombination and thermal generation similar to a standard diode. However, it is as if the built-in potential barrier of the diode junction is being modulated in addition to the voltage being applied across it (c.f. Fig. 1a and 1c). 'Barrier' in this context refers to the energy that would be required to promote an electron (or hole) from the quasi Fermi energy on either side of the interface to the conduction (or valence) band of the perovskite at the interface (Fig. 1c). The local electrostatic potential, which arises from the solution of Poisson's equation accounting for the profile of ionic charge (in addition to the electrons and holes), thus varies this 'barrier' and determines the local change in concentration of electrons and holes by the corresponding Boltzmann factor ($\exp[qv_1/k_BT]$ and $\exp[-qv_1/k_BT]$ respectively at interface 1, Fig. 3c) when the system is in a dynamic equilibrium. This in turn controls the charge transfer rate of recombination and generation.

**Ionically gated interfacial transistor**

We now develop simple expressions to describe the characteristics and impedance of the interfaces in a semiconductor with mixed ionic and electronic conduction by considering how the current across each interface will vary with the externally applied voltage in the presence of inert mobile ions. We will show that these expressions, represented by the circuit model shown in Fig. 2e (or Fig. 4g or h for more complete descriptions), give an excellent approximation to the results of the ionically coupled drift-diffusion simulations described above. This allows the time or frequency dependent behaviour of hybrid perovskite solar cells in response to changing biases to be easily evaluated.

In these devices, the interfacial electronic currents can be related to the processes of charge injection, collection, thermal generation and recombination between the active semiconductor layer and the hole transporting material (HTM) or electron transporting material (ETM) layers. The currents related to these processes are indicated in Fig. 4a. Under most circumstances one of these processes will dominate the impedance of the device, either for the free electron or free hole species (c.f. Note S6, ESI). We assume resistance to free electron and hole transport in the perovskite is low relative to the recombination/generation and the injection/collection impedances, consistent with measurements showing long diffusion lengths observed in these materials[33, 34]. We also assume that the influence of ionic defect accumulation on the recombination rate constant is of secondary importance relative to the electrostatic effect of the ions, although it could have an influence in some cases[35]. In cases where interfacial recombination centres are passivated, photogenerated charge can accumulate in the perovskite layer and 'screen' hysteresis[26, 29]. This passivation could be modulated by a varying concentration of ionic defects, but we neglect any such effects here.

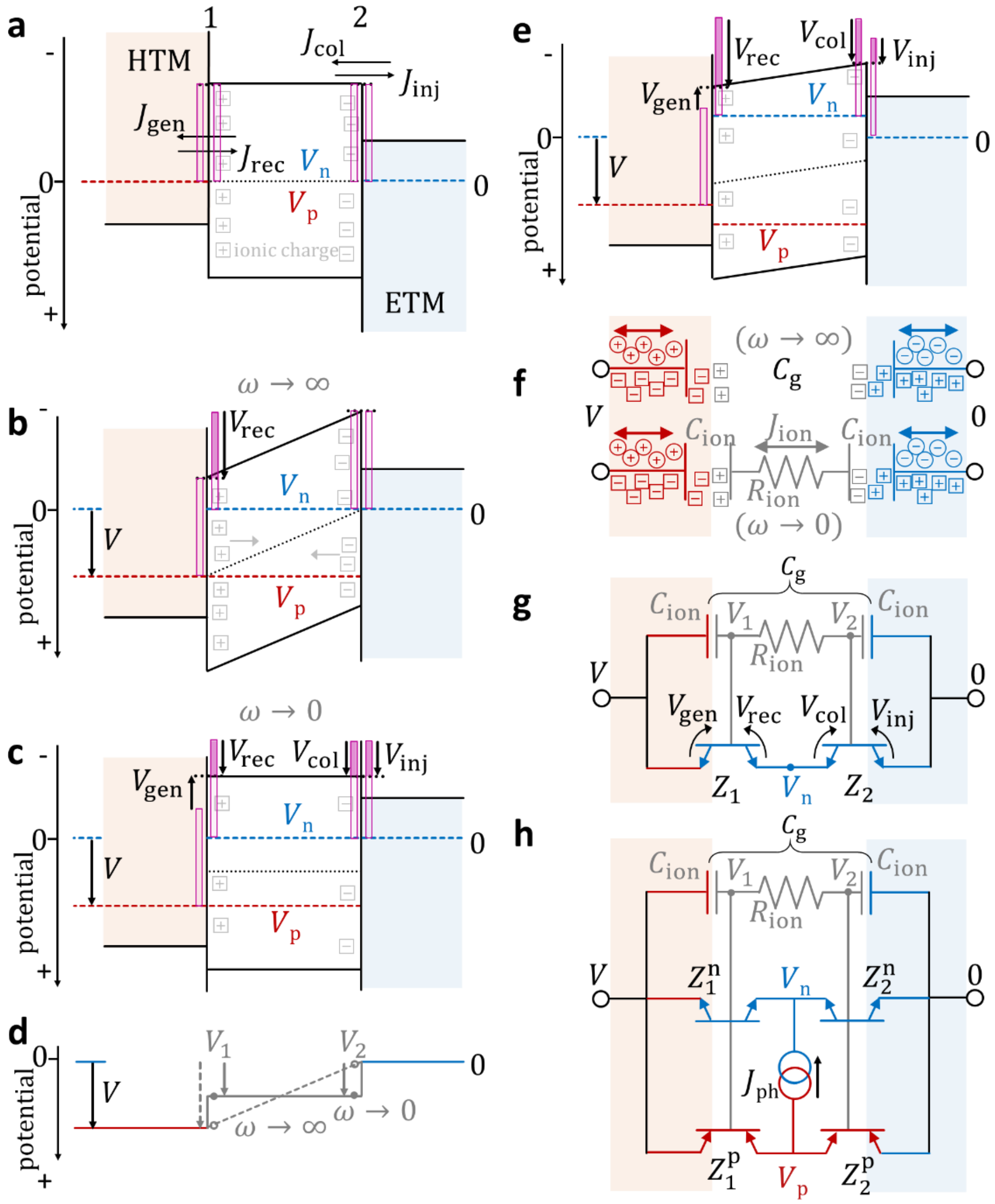

1 HTM/perovskite interface
$V$ cell potential
$V_n$ electron quasi Fermi level
$V_p$ hole quasi Fermi level
▯ equilibrium barrier potential
$V_1$ change in electrostatic potential at 1
$V_2$ change in electrostatic potential at 2
$V_{gen}$ change in generation potential
$V_{rec}$ change in recombination potential
$V_{col}$ change in hole collection potential
$V_{inj}$ change in injection potential
$Z_1^n$ ($Z_1^p$) interface 1 electron (hole) impedance

2 perovskite/ETM interface
$J_{gen}$ thermal generation current density
$J_{rec}$ recombination current density
$J_{col}$ collection current density
$J_{inj}$ injection current density
$J_{ion}$ ionic current density
$J_{ph}$ photogeneration current density
$R_{ion}$ ionic resistance
$C_{ion}$ interfacial space charge capacitance
$C_g$ geometric capacitance of cell
$\omega$ angular frequency of perturbation
$Z_2^n$ ($Z_2^p$) interface 2 electron (hole) impedance

**Fig. 4 Simplified energy level diagrams and circuit models using transistors to describe the ionic gating of electron processes at different interfaces.** The dark equilibrium barrier height is indicated by the unfilled purple rectangles. In non-equilibrium situations, a reduction in barrier height is indicated by the filled section. On application of a cell voltage the pink filled section of the rectangle represents the reduction in this energy barrier. (**a**) The energy levels of the conduction and valence bands in the dark

after equilibration of ionic charge. Due to detailed balance the interfacial currents are equal and opposite ($J_{\mathrm{gen}} = J_{\mathrm{rec}} = J_{s1}$ at interface 1 and $J_{\mathrm{col}} = J_{\mathrm{inj}} = J_{s2}$ at interface 2) at dark equilibrium. The corresponding energy level profiles after applying a voltage $V$, in the dark ($V_n$ = 0), for a device whose impedance is limited by electron recombination (**b**) immediately after the voltage is applied ($\omega\rightarrow\infty$) and (**c**) after the redistribution of ionic charge has reached steady state ($\omega\rightarrow0$). The changes in barrier heights ($V_{\mathrm{gen}}$, $V_{\mathrm{rec}}$, $V_{\mathrm{col}}$ and $V_{\mathrm{inj}}$) for the various interfacial electron transfer processes in response to an applied potential $V$ and the electron quasi-Fermi potential ($V_n$) are indicated. (**d**) The corresponding change in the electrostatic potential profile (dashed line – instantaneous, solid line – steady state). The changes in electrostatic potential at interfaces 1 and 2 are indicated by $V_1$ and $V_2$. The relationship between these changes is given in Table 1. (**e**) A general example for a device in the light (where the electron quasi Fermi level $V_n \neq 0$). In this case the device impedance has contributions from both interfaces and the ions have not reached a steady state distribution. (**f**) The equivalent circuit model for the impedance of the ionic circuit branch in response to high frequency voltage perturbation, $v(\omega \rightarrow \infty)$, where perovskite ions are effectively frozen, and at lower frequencies, $v(\omega < \infty)$ where perovskite ionic motion is described by $C_{\mathrm{ion}}$-$R_{\mathrm{ion}}$-$C_{\mathrm{ion}}$ series elements. Here we assume the dopant ions in the HTM and ETM (red and blue squares) are static. (**g**) An equivalent circuit model for the device in which the impedance to electron transfer for both interfaces are modelled as bipolar transistors with impedance $Z_1$ and $Z_2$, the base terminals are gated by the ionic potentials $V_1$ and $V_2$. The curved arrows indicate the potential differences between the 'terminals' on the transistor elements. (**h**) General circuit model considering both electrons (n) and holes (p) with a (negative) photogeneration current ($J_{\mathrm{ph}}$), where the potential of the electrons ($V_n$) and holes ($V_p$) in the perovskite layer correspond to the electron and hole quasi Fermi levels.

Initially we consider the impedance related to the recombination (and thermal generation) of electrons at the interface with p-type HTM (interface 1) assuming electron injection and collection is not limiting. Close to the interface, where most recombination is thought to occur[36, 37], electrons in the perovskite phase with concentration $n_1$ may be considered a minority species relative to the holes in the neighbouring HTM. Here, for simplicity we assume the electron recombination current density from the perovskite to HTM can be approximated by the first order process, $J_{\mathrm{rec}} \propto n_1$. For the fits to data described later we explicitly account for the ideality factor of the interfaces, allowing for non-linear recombination, see Methods 4.2 (ESI).

The recombination current density $J_{\mathrm{rec}}$ varies exponentially with the potential 'barrier' given by the difference between the conduction band edge of the perovskite at interface 1 and electron quasi Fermi level in the perovskite ($V_n$) which we reference to the ETM Fermi level (0 V). At dynamic equilibrium, this barrier height controls the population of free electrons in the perovskite available to recombine at the interface by the corresponding Boltzmann factor (see Fig. 1c). In addition to $J_{\mathrm{rec}}$, there will also be a thermal generation current, $-J_{\mathrm{gen}}$, of electrons from the HTM to the perovskite. This current density, in the opposite direction to $J_{rec}$ across the interface, varies exponentially with the potential barrier given by the difference between the conduction band of the perovskite at interface 1 and the Fermi level in the HTM ($V$, the cell bias voltage – since the Fermi level of the ETM is defined to be zero). Similarly, under dynamic equilibrium conditions, this barrier height determines the population of electrons in the HTM at the perovskite interface with sufficient energy to be promoted to the perovskite conduction band from the HTM. At dark equilibrium ($V = 0$) the barrier for the two processes is the same (see the open pink bars on either side of the interface in Fig. 4a and no bias case of Fig. 1c). Since the system must obey the principle of detailed balance at equilibrium, there will be equal and opposite current densities across the barrier with magnitude $J_{\mathrm{rec}} = -J_{gen} = J_{s1}$. Here, $J_{s1}$ is the saturation current density of recombination for interface 1.

We refer to the *changes* in the potential barrier relative to the dark equilibrium case for the recombination and the generation current as $V_{\mathrm{rec}}$ and $V_{\mathrm{gen}}$ respectively (at dark equilibrium $V_{\mathrm{rec}} = V_{\mathrm{gen}} = 0$). The net electron recombination at this interface is then given by:

$$J_1 = J_{\mathrm{rec}} - J_{\mathrm{gen}} = J_{\mathrm{s1}} e^{\frac{qV_{\mathrm{rec}}}{k_{\mathrm{B}}T}} - J_{\mathrm{s1}} e^{\frac{qV_{\mathrm{gen}}}{k_{\mathrm{B}}T}} \qquad 1$$

where $q$ is the electronic charge, $k_B$ is Boltzmann's constant, and $T$ is temperature (see Fig. 4 and Fig. S4, ESI). Without mobile ions in the system, a potential, $V$, applied across the cell would be fully experienced by the electrons in the perovskite at interface 1 so that $V_{\mathrm{rec}} = V$ with no change in the barrier to thermal generation ($V_{\mathrm{gen}} = 0$) so equation 1 would become the standard diode equation: $J_1 = J_{\mathrm{s1}}(\exp[qV/k_{\mathrm{B}}T] - 1)$.

However, as observed in the simulations, the electrostatic potential at the interfaces in hybrid perovskites devices depend both on the applied potential $V$ and also on the effect of the redistribution of mobile ions. Ionic redistribution modifies the electrostatic potential and thus the barrier height at the HTM perovskite interface. This influences the values of both $V_{rec}$ and $V_{gen}$ as illustrated in Figs 4b-e. Here, we refer to the *changes* in the electrostatic potential at the interfaces 1 and 2 relative to the values at dark equilibrium as $V_1$ and $V_2$ (as indicated schematically in Fig. 4d). The relationships between these various changes in potential is expressed in Table 1 and will be discussed below.

**Table 1 Expressions for potentials driving electron transfer processes, and circuit branch impedances.** The terms in the equations are illustrated in Fig. 4 and defined in the text. *The impedance for the electronic branch of the circuit is given for the specific case where impedance due to recombination of a single carrier dominates (more general cases are discussed in the Methods, Notes S2 and S6, and Tables S3 and S4, ESI). The impedance of the electronic circuit branch, $Z_{rec}$, is given in terms of the apparent capacitance and resistance of the interface $c_{rec}$ and $r_{rec}$ which are represented in Fig. 2f.

| Change in barrier potential for electron transfer relative to equilibrium (V) | | | |
|---|---|---|---|
| Electron generation | $V_{\mathrm{gen}}$ | $=$ | $V_1 - V$ |
| Electron recombination | $V_{\mathrm{rec}}$ | $=$ | $V_1 - V_{\mathrm{n}}$ |
| Electron collection | $V_{\mathrm{col}}$ | $=$ | $V_2 - V_{\mathrm{n}}$ |
| Electron injection | $V_{\mathrm{inj}}$ | $=$ | $V_2$ |

| Electrostatic potential from ionic circuit (V) | | | |
|---|---|---|---|
| Interface 1 | $V_1$ | $=$ | $\frac{\bar{V}}{2} + \frac{v}{2}\left(2 - \frac{1}{1 + i\omega R_{\mathrm{ion}}\,C_{\mathrm{ion}}/2}\right)$ |
| Interface 2 | $V_2$ | $=$ | $\frac{\bar{V}}{2} + \frac{v}{2}\left(\frac{1}{1 + i\omega R_{\mathrm{ion}}\,C_{\mathrm{ion}}/2}\right)$ |

| Impedance of ionic circuit branch (Ω cm²) | | | |
|---|---|---|---|
| | $Z_{\mathrm{ion}}$ | $=$ | $\left[i\omega C_{\mathrm{g}} + \frac{i\omega\left(C_{\mathrm{ion}}/2 - C_{\mathrm{g}}\right)}{1 + i\omega R_{\mathrm{ion}}\,C_{\mathrm{ion}}/2}\right]^{-1}$ |

| Impedance of electronic circuit branch* (Ω cm²) | | | |
|---|---|---|---|
| Interface 1 | $Z_{\mathrm{rec}}$ | $=$ | $\left[\frac{1}{r_{\mathrm{rec}}} + i\omega c_{\mathrm{rec}}\right]^{-1}$ |
| | $r_{\mathrm{rec}}$ | $=$ | $\frac{2 + \omega^2 R_{\mathrm{ion}}^2 C_{\mathrm{ion}}^2}{\left(1 + \omega^2 R_{\mathrm{ion}}^2 C_{\mathrm{ion}}^2\right)} \frac{k_{\mathrm{B}}T}{q\bar{J}_{\mathrm{rec}}(\bar{V})}$ |
| (F cm⁻²) | $c_{\mathrm{rec}}$ | $=$ | $\frac{R_{\mathrm{ion}} C_{\mathrm{ion}}}{4 + \omega^2 R_{\mathrm{ion}}^2 C_{\mathrm{ion}}^2} \frac{q\bar{J}_{\mathrm{rec}}(\bar{V})}{k_{\mathrm{B}}T}$ |

In the simple case of a p-i-n device with ion blocking contacts and symmetric capacitances at each contact, ion redistribution occurs with a time constant approximated by $R_{\mathrm{ion}}C_{\mathrm{ion}}/2$. $R_{\mathrm{ion}}$ is the specific resistance (Ω cm$^2$) to ionic motion across the perovskite layer. $C_{\mathrm{ion}}$ is the specific capacitance (F cm$^{-2}$) of the interfacial space charge layer corresponding to the accumulation of mobile ionic defects in the perovskite and uncompensated static dopant ions in the HTM or ETM (Fig. 4f). If the concentration of mobile ionic defects is large relative to the concentration of free electrons and holes in the active layer then the ionic distribution will determine the electrostatic potential profile in the perovskite layer. The change in electrostatic potential at each interface, $V_1$ and $V_2$, can be found by analysing the voltage drop on either side of the resistor in the $C_{\mathrm{ion}}$-$R_{\mathrm{ion}}$-$C_{\mathrm{ion}}$ series when a voltage $V = \bar{V} + v(\omega)$ is applied across the whole series as shown in Table 1. This description assumes that changes in electrostatic potential across the interfaces due to ionic accumulation predominantly drop within each contact material (as sketched in Fig. 1c and simulated in Fig. 3a). This will be the case when the mobile ionic defect concentration is significantly greater than the doping concentration of the contact materials. We discuss the case where there is a significant drop in electrostatic potential in the perovskite (depicted in Figure

S1a, ESI) as well as including the dependence of $C_{ion}$ on $\bar{V}$ due to variation in the space charge layer widths in the contacts in the Methods section S4 (ESI). We treat cases with different capacitances at each interface in Note S6 and Table S4 (ESI).

Based on these assumptions, equation 1 gives a general expression for the net electron recombination current across interface 1 in terms of the applied potential $V$, the electron quasi Fermi level $V_n$, and the change in electrostatic potential of the interface $V_1$ by substituting them into the expressions for $V_{gen}$ and $V_{rec}$ (Table 1):

$$J_1 = J_{s1}\left[e^{\frac{q(V_1 - V_n)}{k_B T}} - e^{\frac{q(V_1 - V)}{k_B T}}\right]. \quad (2)$$

This is analogous to the expression used to describe a bipolar n-p-n transistor (c.f. Figs 1b and c where $V_E$ and $V_n$ are zero) where the electrostatic potential of interface 1 behaves like the transistor base (so the device only has two 'external' terminals). The voltage of this conceptual base, $V_1$, relative to the Fermi level of the ETM/cathode (0 V), arises from any change in ionic accumulation at interface 1. Under dark forward bias conditions ($V > 0$) there is net flux of electrons from the perovskite (which acts as the emitter with potential $V_n$) to the HTM (which acts as the collector with potential $V$). The potential differences of the base-emitter ($V_{BE} = V_B - V_E$) and base-collector ($V_{BC} = V_B - V_C$) junctions are equivalent to $V_{rec}$ and $V_{gen}$ respectively (c.f. Fig. 1b). We have modified the conventional bipolar transistor symbol to emphasise that the net electronic current through the transistor may be in either direction according to the electrical and light bias conditions. If $V_{rec} < V_{gen}$ (e.g. under reverse bias), then the conventional assignment of the terms 'emitter' and 'collector' to the two sides of the interface would be reversed. If there is no chemical reaction between ionic and electronic charge at the interface and no ionic penetration into the HTM, then the ionic-to-electronic current gain of the transistor, $\beta_{ion\text{-}electron}$, is infinite. In bipolar transistors $\beta$ is defined by the ratio of the collector current to the base current. In this basic case, only electronic charge may be transferred across the interface (collector current = $J_1$) and ionic charge is confined to the perovskite and cannot cross the interface (base current = 0) despite the possibility of an ionic current, $J_{ion}$, in the rest of the perovskite.

These observations naturally result in the simple equivalent circuit illustrated in Fig. 1c and 2e where an 'ionic circuit' branch is connected in parallel to an 'electronic circuit' branch. The complex impedance of the ionic branch of the circuit ($Z_{ion}$) behaves analogously to an insulating material which shows dielectric loss through Debye relaxation of charge (in this case ionic polarisation) with an equivalent series resistance corresponding to the $R_{ion}$. At high frequencies, when the ionic charge is effectively static, $Z_{ion}$ is dominated by the device's geometric capacitance ($C_g$) but at lower frequencies the ionic motion in the $C_{ion}$-$R_{ion}$-$C_{ion}$ series dominates (Fig. 4f). There is a continuous transition between these two regimes centred around the frequency $2/(R_{ion}C_{ion})$. This results in the expression for the ionic impedance presented in Table 1 which is derived in Methods 4.1 (ESI). We have represented the ionic branch with the capacitor resistor series components curly bracketed by $C_g$ to denote the transition between the frequency regimes using physically meaningful circuit elements. As discussed, the $C_{ion}$-$R_{ion}$-$C_{ion}$ series components enable the straightforward evaluation of $V_1$ (and $V_2$) in terms of $V$ (these are given in Table 1). The change in electrostatic potential due to ion redistribution, $V_1$, controls the base of transistor element and consequently the impedance of the electronic branch of the equivalent circuit. This will be discussed in detail further below. Strikingly, virtually all the features related to the electronic behaviour of a perovskite solar cell, under the conditions described above, can be summarised through the use of this single circuit element coupling the electrostatic potential due to ions to electronic charge transfer, i.e. a transistor.

An alternative, more conventional, representation of this same circuit is shown in Fig. 2f, however, the physical meaning of the elements is less intuitive. In the ionic circuit branch $\Delta C_{\mathrm{ion}} = C_{\mathrm{ion}}/2 - C_{\mathrm{g}}$ and $R_{\mathrm{eff}} = R_{\mathrm{ion}}(1 + C_{\mathrm{g}}/C_{\mathrm{ion}})$ as discussed in Methods 4.1 (ESI). The apparent capacitance and recombination resistance elements in the electronic circuit branch, $c_{rec}(\omega)$ and $r_{rec}(\omega)$, have a frequency dependence controlled by the ionic circuit branch as derived from the transistor model discussed in the following sections (the expressions for them are given in Table 1). We now consider the implications of a transistor-like interface for the behaviour of the device.

**Ionic-to-electronic current amplification**

Amplification is a key property shown by bipolar transistors[21], where changes in electronic energy barriers induced by the gating terminal (base) amplify the flux of electrons or holes between the emitter and collector terminals. The simulated impedance spectroscopy results show that, at sufficiently low frequency voltage oscillations, the out of phase component of the electronic current oscillations is directly proportional to the ionic current in the device (c.f. solid grey and dashed blue curves in Fig. 3e). The amplitude of this out of phase electronic current scales in direct proportion to the steady state current $\bar{J}_{\mathrm{rec}}$ across the interface (see Fig. 3b where $\bar{J}_{\mathrm{rec}} = J_{\mathrm{ph}}$, the photogenerated current, since each simulation is around $V_{OC}$). This implies that there is an ionic-to-electronic current amplification process that can occur in mixed conducting devices such as perovskite solar cells and the effect varies exponentially in magnitude with the steady state bias voltage dropping across the interface. We now examine the mechanism underlying this effect.

At low frequencies when $\omega \ll (R_{\mathrm{ion}}C_{\mathrm{ion}}/2)^{-1}$ the impedance due to $R_{ion}$ becomes negligible relative to that of the $C_{ion}$ elements in series so that the ionic current can be approximated by $J_{\mathrm{ion}} = v/Z_{\mathrm{ion}} \approx i\omega C_{\mathrm{ion}}v/2$. This ionic current induces an out of phase change in potential at the interface of $v_1'' = J_{\mathrm{ion}}R_{\mathrm{ion}}/2$ due to the potential dropped across $R_{ion}$ with the small ionic current $J_{ion}$ flowing through the perovskite. Since oscillations in the potential controlling recombination rate are equal to the changes in potential at the interface , $v_{rec} = v_1$ (since the potential of the electrons in the perovskite is pinned to the potential of the ETM, i.e. $v_n = V_n = 0$), there will be an out of phase component to the electronic current given by $j''_{\mathrm{rec}} = v''_{\mathrm{rec}} g_{\mathrm{rec}} = J_{\mathrm{ion}} R_{\mathrm{ion}} g_{\mathrm{rec}}/2$. Here $g_{\mathrm{rec}}$ is the recombination transconductance which describes the change in interfacial current in response to changes in $V_{\mathrm{rec}}$ given by $g_{\mathrm{rec}}(\bar{V}) = \mathrm{d}J_{\mathrm{rec}}/\mathrm{d}(V_1 - V_{\mathrm{n}}) = q\bar{J}_{\mathrm{rec}}/k_{\mathrm{B}}T$, where $V_{\mathrm{n}}$ = 0 V in this example. Taking the ratio between the out of phase electronic and ionic currents gives an ionic-to-electronic transcarrier amplification factor:

$$\frac{j''_{\mathrm{rec}}}{J_{\mathrm{ion}}} = \frac{R_{ion}}{2} g_{\mathrm{rec}}(\bar{V}) = \frac{R_{\mathrm{ion}}}{2} \frac{q\bar{J}_{\mathrm{rec}}(\bar{V})}{k_{\mathrm{B}}T} \quad (3)$$

analogous to the classic result for an amplification circuit using a bipolar transistor. The magnitude of $j''_{\mathrm{rec}}$ across the interface is proportional to $R_{\mathrm{ion}}$, independent of the value of $C_{\mathrm{ion}}$, and will also increase exponentially with background bias voltage, $\bar{V}$.

Rearranging equation 3 gives $R_{\mathrm{ion}} = 2j''_{\mathrm{rec}}/(J_{\mathrm{ion}} g_{\mathrm{rec}})$. Interestingly, this result implies that $R_{\mathrm{ion}}$ (and thus ionic conductivity) can be inferred from measurements of the device's apparent capacitance ($c_{\mathrm{rec}}$) due to modulated electron recombination. This is because both the out of phase electronic recombination current, $j''_{\mathrm{rec}}$, and the ionic current, $J_{\mathrm{ion}}$ (as $\omega \rightarrow 0$), are directly proportional to the measured capacitance of the device so that $j''_{\mathrm{rec}}/J_{\mathrm{ion}} = 2c_{\mathrm{rec}}(\bar{V})/C_{\mathrm{ion}}$. The meaning of $c_{\mathrm{rec}}$ which results in this relationship is discussed further in the next section. Experimentally, $g_{\mathrm{rec}}(\bar{V})$ can be found if $\bar{J}_{\mathrm{rec}}$ can be estimated from the measured data (see Methods 5, ESI). $C_{\mathrm{ion}}$ can easily be determined from the measurements of the low frequency device capacitance in dark conditions with $V$ = 0 V and $c_{\mathrm{rec}}$ determined from the apparent capacitance with a bias voltage $\bar{V}$ (in the light or dark). The inset of Fig. 2b shows that this method predicts a value of $R_{\mathrm{ion}} \approx 60$ kΩ cm$^2$ (ionic conductivity of about $10^{-9}$ S cm$^{-1}$) for the cell under consideration.

**Capacitor-like and inductor-like behaviour**

The ionic gating effect at the interfaces results in out of phase electronic currents causing the device to display very large apparent capacitances or inductances at low frequencies. We now explore the implications of this. Under forward bias ($V > 0$) conditions $J_{\mathrm{rec}} \gg J_{\mathrm{gen}}$ so the second term of equation 2 can be neglected such that $J_1 \approx J_{\mathrm{rec}} = J_{\mathrm{s1}} \exp[qV_1/k_{\mathrm{B}}T]$ when $V_{\mathrm{n}}$ = 0 and the expression for $V_1(V, R_{\mathrm{ion}}, C_{\mathrm{ion}}, \omega)$ is given in Table 1. Substituting this in, and differentiating $J_{\mathrm{rec}}$ with respect to the applied voltage $V$ gives an expression for the reciprocal of the recombination impedance, which in the small voltage perturbation ($v$) limit can be written:

$$\frac{1}{Z_{\mathrm{rec}}(\bar{V})} = \frac{\mathrm{d}j_{\mathrm{rec}}}{\mathrm{d}v} = \frac{1}{2}\left(2 - \frac{1}{1 + i\omega R_{\mathrm{ion}} C_{\mathrm{ion}}/2}\right) \frac{q}{k_{\mathrm{B}}T} \bar{J}_{\mathrm{rec}}(\bar{V}) = \frac{1}{r_{\mathrm{rec}}} + i\omega c_{\mathrm{rec}} \quad (4)$$

where the background recombination current across the interface with a potential difference $\bar{V}$ at steady state ($\omega$=0) is $\bar{J}_{\mathrm{rec}}(\bar{V}) = J_{\mathrm{s1}} \exp[q\bar{V}/(2k_{\mathrm{B}}T)]$. Separating $1/Z_{\mathrm{rec}}$ into its real and imaginary parts enables expressions for the small perturbation recombination resistance of the interface, $r_{\mathrm{rec}}$, and the apparent electronic capacitance of the interface, $c_{\mathrm{rec}}$ to be determined in terms of $R_{ion}$, $C_{ion}$ and $\omega$ (these expressions are written out in Table 1). Several features of $r_{rec}$ and $c_{rec}$ are noteworthy. First, $r_{rec}$ shows a dependence on frequency since the amplitude of the interfacial barrier ($v_1 - v_n$) oscillations is frequency dependent so that $r_{rec}$ $(\omega \rightarrow 0) = 2r_{rec}$ $(\omega \rightarrow \infty)$. Second, the interface behaves like a frequency dependent capacitor despite no accumulation of electronic charge being required; using the expression for $c_{rec}$ in Table 1 and $g_{rec}(\bar{V}) = q\bar{J}_{rec}(\bar{V})/k_B T$ from equation 3 we see that at low frequency $c_{rec}(\omega \rightarrow 0) = R_{ion}C_{ion}g_{rec}/4$ but at high frequency $c_{rec}(\omega \rightarrow \infty) = 0$. Third, the magnitude of $c_{rec}$ is proportional to $\bar{J}_{rec}(\bar{V})$ and so increases exponentially with the voltage (which may be photoinduced) across the interface allowing variable control of the apparent capacitance. Although this capacitive behaviour could not be used for energy storage, the effect offers a route to achieve at least $10^3$ times greater effective capacitance per unit area than the capacitance achieved by state-of-the-art hafnium oxide capacitors used in electronic circuitry (~$2\times10^{-6}$ F $cm^{-2}$)[38].

Global fits to both experimental measurements and drift-diffusion simulated measurements are shown in Fig. 2 using the expression for $Z_{rec}$ based on equation 4 incorporated in the circuit model shown in Fig. 1c. Only five free parameters are required to simultaneously fit all measurement conditions. The complete equation for the circuit model fit to the data is given in the Methods 4.3 (ESI). The inputs to the fitting model are: the measured impedance spectra, $Z(\omega)$; the bias voltages ($\bar{V} = V_{OC}$, for open circuit measurements) at which these were collected; and the steady state ideality factor, $m_{ss}$, determined from the $V_{OC}$ *vs* light intensity relationship of the device. The free fitting parameters in the model are: $C_g$, $R_{ion}$, $C_{ion}$, $J_{s1}$, and $f_c$, the fraction of interfacial electrostatic potential dropping within the contacts (Table S1, ESI, presents the values of the fitting parameters). Since we define the current gain of the transistor ($\beta_{ion\text{-}electron}$) to be infinite, the transistor element is described by only two parameters, $J_{s1}$, and its ideality factor, $m_1$, which is related to $m_{ss}$ and $f_c$ (a more detailed explanation of $f_c$ and $m_1$ is given in Methods 4.2, ESI).

Agreement is seen between the values of $C_{ion}$ and $R_{ion}$ determined from the equivalent circuit fit and the values derived from the inputs to the Driftfusion model, helping to validate our interpretation of the system. The frequency dependence of $Z_{rec}$ displayed in a Nyquist plot gives rise to a low frequency semicircle in agreement with the observations of Pockett *et al.*[10]. The details are illustrated in Figs S4 and S5. In addition to yielding the ionic conductivity from $R_{ion}$, fitting of our model to the impedance measurement enables quantification of the $J_{s1}$, $C_g$, $C_{ion}$ and $f_c$ parameters. $C_g$, $C_{ion}$ and $f_c$ are related to both the concentration of mobile ionic charge in the perovskite, and the concentration of dopants in the contacts as well as the dielectric constants of the materials (excluding any contributions from surface polarisation by mobile ions). These control where electrostatic potential drops and, in combination with $R_{ion}$, the magnitude, and timescale of hysteresis effects that a given cell will produce. The saturation current density, $J_{s1}$, parameterises non-radiative recombination at the interface and is likely to be related to the density and depth of interfacial traps, a factor critical for assessing the relative performance of different interface combinations.

The expression we have derived for $Z_{rec}$ (equation 4) explains the majority of unusual features observed in the impedance spectroscopy measurements of hybrid perovskite solar cells. Similar arguments can be used to derive expressions for the impedance to recombination of holes at the perovskite/ETM interface which also yield capacitive behaviour (see Methods section S4 and the general case in Note S6, ESI). However, in some perovskite devices, inductor-like behaviour is seen in their impedance spectra[11, 19] and is also apparent in the slow evolution of current towards a new steady state in response to step changes of voltage or light[4]. The capacitor-like form of $Z_{rec}$ in equation 3 is unable to explain this inductive behaviour.

The description of the electronic impedance so far assumed that the rate of injection and collection is sufficiently fast (also shown by $V_{\mathrm{n}} \approx 0$) such that the electronic impedance is dominated by the recombination process (Fig. 4b and c). If this were not the case, the electron injection ($J_{inj}$) and collection ($J_{col}$) currents at interface 2 follow an analogous dependence on the injection and collection voltages $V_{inj}$ and $V_{col}$ which are controlled by the electrostatic potential $V_2$ (Table 1, see Notes S1 and S2, ESI, and Fig. 4e).

In the limiting case where charge injection dominates the impedance of the circuit, at low frequencies, the out of phase injection current is negatively amplified by the ionic current (hypothetical examples are shown in Fig. S4c and d, ESI). The trans-carrier amplification factor is $-R_{\mathrm{ion}}/2\,[qJ_{\mathrm{inj}}(V,\omega=0)/k_{\mathrm{B}}T]$ (c.f. equation 3) resulting in inductive behaviour (see Note S1, ESI). The effect opens the possibility to design thin film devices with huge tuneable effective inductances per unit volume (> $10^4$ H cm$^{-3}$) without relying on the elements coupling to a magnetic flux.

Given the influence of the ionic circuit on the electronic impedance described here, we note that more complex interactions of ionic charge with electronic charge or contact materials would also modulate interfacial electronic processes (see Note S2, Figs S1 and S6, ESI for effects of both interfaces). For example, the phase of $j_{rec}$ can lag $v$ if ionic charge penetrates, or undergoes a reversible chemical reaction, at a dominant recombination interface. Fits from an equivalent circuit allowing ion penetration into an interface to experimental data are shown in Fig. 5a. The ion penetration/reversible reaction is approximated by extending the ionic circuit branch into one of the contacts with an additional interfacial ionic transfer (or reaction) resistance ($R_{\mathrm{ion}}$) and an ionic capacitance within the contact ($C_{\mathrm{con}}$). Under these circumstances our transistor interface model implies that the ionic gating of the electronic recombination process can result in both apparent capacitive and inductive behaviour.

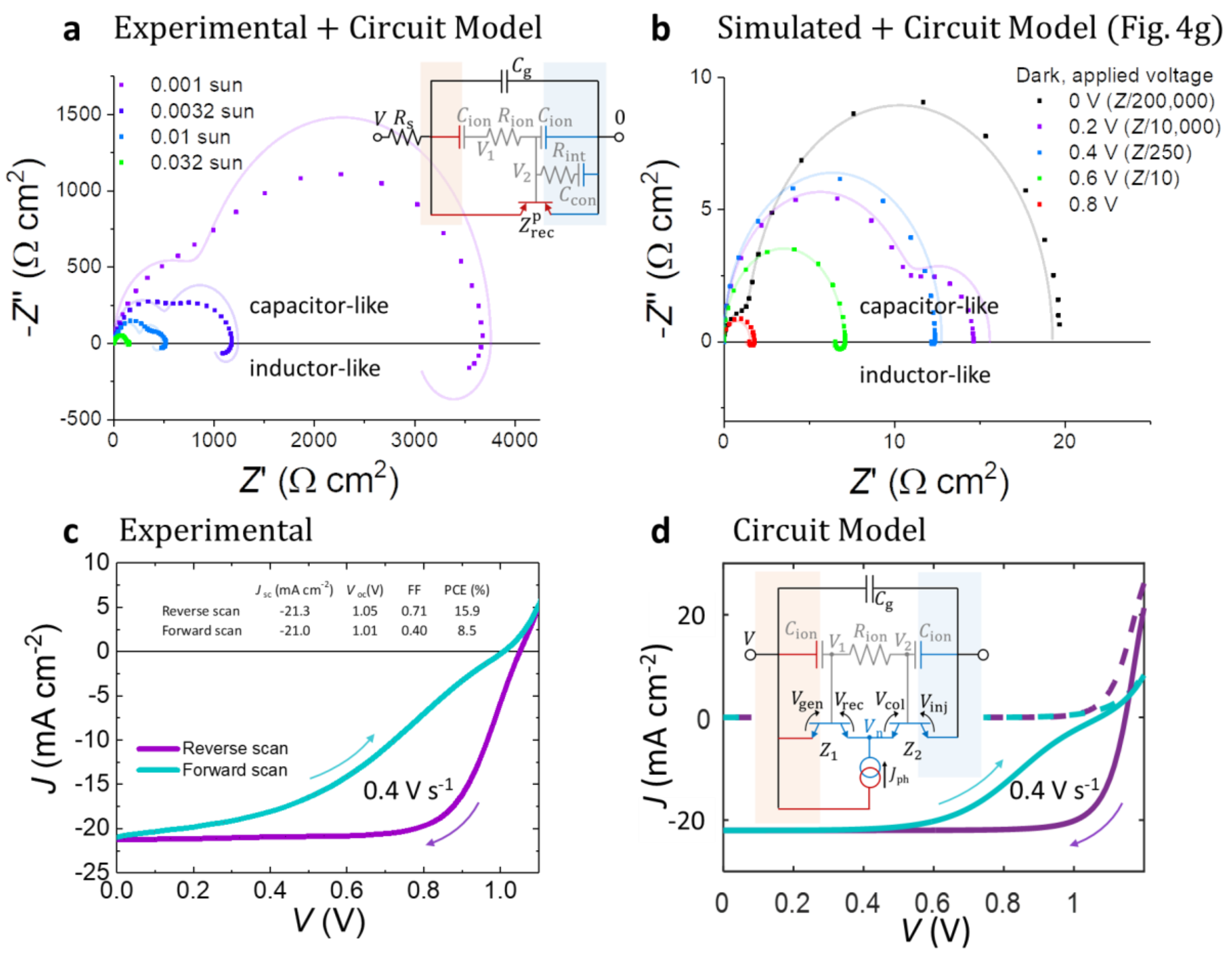

**Fig. 5 Measurements, simulations, and models of different devices showing inductive behaviour and current-voltage behaviour. a**, **b**, Nyquist plot of the real (Z') *vs* imaginary (Z'') impedance components (filled squares) for (**a**) a spiro-OMeTAD/ $FA_{0.85}MA_{0.15}PbI_3/SnO_2$ solar cell (Methods 1.2, ESI) measured around the open circuit voltage, illuminated at different constant light intensities and (**b**) a drift-diffusion simulated (different) device with low majority carrier mobility in contacts and high interfacial recombination in the dark and light. The inset in **a** shows the equivalent circuit model used for the global fit to the data, note that the ionic circuit branch is a crude approximation allowing penetration and/or reversible reaction of ions at interface 2. The equivalent circuit model used to fit the simulated data in **b** is shown in Fig. 4g. The solid lines are global fits to the data using 8 and 6 free parameters respectively (see Table S1, ESI) and the models with all data are shown in Fig. S7, ESI. (**c**) Measured current-voltage characteristics for the device characterised in Fig. 2a and b with forward and reverse voltage scans at 0.4 V $s^{-1}$ under an AM1.5 solar spectrum. (**d**) Modelled light (solid lines where $J_{ph}$ = 22 mA $cm^{-2}$) and dark (dashed lines) current-voltage characteristics using the inset equivalent circuit (similar to Fig. 4g but with photocurrent generation included explicitly). The input parameters were determined from the fit of this model to the impedance data shown in Fig. 2a and b. The global fits, using 6 free parameters are shown in Fig. S1 (ESI), with the parameters given in Table S1 (ESI). Further circuit modelled *J*-*V* curves for different scan rates are shown in Fig. S6 (ESI). These current sweeps are calculated using the approach described in Note S4.

We emphasise that the transistor element was used in the circuit model but not the simulations. The circuit model described encapsulates the key physical processes observed in simulations based on the standard current continuity equations, charge transfer processes (generation, recombination, collection and injection), and Poisson's equation with mobile ions having a higher concentration and lower conductivity than electronic charge under operation. The gating of interfacial electronic charge transfer (and thus electronic current *through* the device) by ionic redistribution (and consequent surface polarisation) explains very high low frequency apparent capacitances and inductances without accumulation of electronic charge at the interfaces. In contrast, the surface polarisation model introduced by Bisquert et al.[17-19] requires that large concentrations of electronic charge accumulate at the interfaces to explain observed cell behaviour. If this were the case, significantly lower $V_{OC}$ values than typically observed in these devices might be expected.

Our model provides a basis to include additional factors that may influence device behaviour such as: the fraction of ionic screening potential dropping within the HTM and ETM contacts, asymmetric interfacial ionic capacitances, non-ideal recombination and injection (Methods 4.2, ESI), treatment of both electrons and holes (Fig. 4h), recombination in the perovskite bulk, and the effect of interface screening by electronic charge (see Fig. S8d-f, Notes S6 and S3, ESI). The latter factor is expected to be relevant in record efficiency solar cells and at large light or electrical bias conditions. In its simplest version, our ionically coupled transistor circuit model is already able to interpret the most important features of impedance spectra observed in the literature. Additionally, it also allows simple calculation of large perturbation measurements such as *J-V* sweeps at any scan speed (see Fig. 5c and d, Fig. S6 and Note S4, ESI) and voltage step measurements (Note S5, ESI) as well as the d.c. (photo)current of the device. Such transient outputs of the circuit model could be used to parameterise measurements of device current or voltage response to voltage or light intensity steps which have been used previously to assess the influence of mobile ions.[4, 26, 32]

In cases where the impedance from each interface in the circuit model is comparable, an analytical solution is no longer accessible due to the need to numerically evaluate $V_n$ (and/or $V_p$). However, the procedure to determine the device behaviour is qualitatively similar and straightforward (see Notes S2 and S6, ESI); an example of a fit using numerical evaluation of $V_n$ to a drift-diffusion simulated device with mixed capacitor and inductor-like behaviour is shown in Fig. 5b.

To conclude, our description of the interfaces of perovskite devices as ionically gated transistors provides an intuitive framework to interpret the complicated current-voltage behaviour of these devices as well as unlocking the potential of impedance spectroscopy as a means to identify the key bottlenecks of their performance. The interfacial transistor model also has a number of interesting broader implications. The trans-carrier amplification phenomenon described suggests a strategy to design devices displaying huge, tuneable, effective capacitances or inductances without the volume required for similar physical capacitances or inductances and with the option to be powered by light. Furthermore, the model will be generally applicable to other electrochemical redox processes supported by a high concentration of low mobility inert ions as well as to ionic motion signal sensing and amplification in biological systems requiring neural interfacing in a manner related to electrochemical transistors.[39]

**References**

1. J. Nelson, *The Physics of Solar Cells*, Imperial College Press, London, 2003.
2. M. M. Lee, J. Teuscher, T. Miyasaka, T. N. Murakami and H. J. Snaith, *Science*, 2012, **338**, 643-647.
3. H.-S. Kim, C.-R. Lee, J.-H. Im, K.-B. Lee, T. Moehl, A. Marchioro, S.-J. Moon, R. Humphry-Baker, J.-H. Yum, J. E. Moser, M. Grätzel and N.-G. Park, *Scientific Reports*, 2012, **2**, 591.
4. C. Eames, J. M. Frost, P. R. F. Barnes, B. C. O'Regan, A. Walsh and M. S. Islam, *Nat Commun*, 2015, **6**, 7497.
5. Z. Xiao, Y. Yuan, Y. Shao, Q. Wang, Q. Dong, C. Bi, P. Sharma, A. Gruverman and J. Huang, *Nat Mater*, 2015, **14**, 193-198.
6. Y. Yuan and J. Huang, *Accounts of Chemical Research*, 2016, **49**, 286-293.
7. G. Gregori, T.-Y. Yang, A. Senocrate, M. Grätzel and J. Maier, in *Organic-Inorganic Halide Perovskite Photovoltaics: From Fundamentals to Device Architectures*, eds. N.-G. Park, M. Grätzel and T. Miyasaka, Springer International Publishing, Cham, 2016, pp. 107-135.
8. A. Guerrero, G. Garcia-Belmonte, I. Mora-Sero, J. Bisquert, Y. S. Kang, T. J. Jacobsson, J.-P. Correa-Baena and A. Hagfeldt, *The Journal of Physical Chemistry C*, 2016, **120**, 8023-8032.
9. T. Kazuya, *Applied Physics Express*, 2017, **10**, 059101.
10. A. Pockett, G. E. Eperon, N. Sakai, H. J. Snaith, L. M. Peter and P. J. Cameron, *Phys. Chem. Chem. Phys.*, 2017, **19**, 5959-5970.
11. O. Almora, K. T. Cho, S. Aghazada, I. Zimmermann, G. J. Matt, C. J. Brabec, M. K. Nazeeruddin and G. Garcia-Belmonte, *Nano Energy*, 2018, **48**, 63-72.
12. C. Ludmila, U. Satoshi, V. V. J. Piyankarage, K. Shoji, T. Yasutake, N. Jotaro, K. Takaya and S. Hiroshi, *Applied Physics Express*, 2017, **10**, 025701.
13. E. J. Juarez-Perez, R. S. Sanchez, L. Badia, G. Garcia-Belmonte, Y. S. Kang, I. Mora-Sero and J. Bisquert, *The Journal of Physical Chemistry Letters*, 2014, **5**, 2390-2394.
14. A. Dualeh, T. Moehl, N. Tétreault, J. Teuscher, P. Gao, M. K. Nazeeruddin and M. Grätzel, *Acs Nano*, 2014, **8**, 362-373.
15. M. N. F. Hoque, M. Yang, Z. Li, N. Islam, X. Pan, K. Zhu and Z. Fan, *ACS Energy Letters*, 2016, **1**, 142-149.
16. O. Almora, I. Zarazua, E. Mas-Marza, I. Mora-Sero, J. Bisquert and G. Garcia-Belmonte, *The Journal of Physical Chemistry Letters*, 2015, **6**, 1645-1652.
17. I. Zarazua, J. Bisquert and G. Garcia-Belmonte, *The Journal of Physical Chemistry Letters*, 2016, **7**, 525-528.

18. I. Zarazua, G. Han, P. P. Boix, S. Mhaisalkar, F. Fabregat-Santiago, I. Mora-Seró, J. Bisquert and G. Garcia-Belmonte, *The Journal of Physical Chemistry Letters*, 2016, **7**, 5105-5113.
19. E. Ghahremanirad, A. Bou, S. Olyaee and J. Bisquert, *The Journal of Physical Chemistry Letters*, 2017, **8**, 1402-1406.
20. P. Schulz, E. Edri, S. Kirmayer, G. Hodes, D. Cahen and A. Kahn, *Energy Environ. Sci.*, 2014, **7**, 1377-1381.
21. W. Shockley, *The Bell System Technical Journal*, 1949, **28**, 435-489.
22. D. Moia, I. Gelmetti, P. Calado, W. Fisher, M. Stringer, O. Game, Y. Hu, P. Docampo, D. Lidzey, E. Palomares, J. Nelson and P. R. F. Barnes, *ArXiv*, 2018, 1805.06446
23. D. Prochowicz, P. Yadav, M. Saliba, M. Saski, S. M. Zakeeruddin, J. Lewiński and M. Grätzel, *ACS Applied Materials & Interfaces*, 2017, **9**, 28418-28425.
24. Y.-C. Zhao, W.-K. Zhou, X. Zhou, K.-H. Liu, D.-P. Yu and Q. Zhao, *Light: Science & Applications*, 2017, **6**, e16243.
25. G. Y. Kim, A. Senocrate, T.-Y. Yang, G. Gregori, M. Grätzel and J. Maier, *Nature Materials*, 2018, **17**, 445-449.
26. P. Calado, A. M. Telford, D. Bryant, X. Li, J. Nelson, B. C. O'Regan and P. R. F. Barnes, *Nat Commun*, 2016, **7**, 13831.
27. P. Calado, I. Gelmetti, M. Azzouzi, B. Hilton and P. R. F. Barnes, *https://github.com/barnesgroupICL/Driftfusion*, 2018.
28. G. Richardson, S. E. J. O'Kane, R. G. Niemann, T. A. Peltola, J. M. Foster, P. J. Cameron and A. B. Walker, *Energy Environ. Sci.*, 2016, **9**, 1476-1485.
29. S. van Reenen, M. Kemerink and H. J. Snaith, *The Journal of Physical Chemistry Letters*, 2015, **6**, 3808-3814.
30. W. Tress, N. Marinova, T. Moehl, S. M. Zakeeruddin, M. K. Nazeeruddin and M. Gratzel, *Energy Environ. Sci.*, 2015, **8**, 995-1004.
31. Y. Wu, H. Shen, D. Walter, D. Jacobs, T. Duong, J. Peng, L. Jiang, Y.-B. Cheng and K. Weber, *Adv. Funct. Mater.*, 2016, **26**, 6807-6813.
32. D. Walter, A. Fell, Y. Wu, T. Duong, C. Barugkin, N. Wu, T. White and K. Weber, *The Journal of Physical Chemistry C*, 2018, **122**, 11270-11281.
33. G. Xing, N. Mathews, S. Sun, S. S. Lim, Y. M. Lam, M. Grätzel, S. Mhaisalkar and T. C. Sum, *Science*, 2013, **342**, 344-347.
34. S. D. Stranks, G. E. Eperon, G. Grancini, C. Menelaou, M. J. P. Alcocer, T. Leijtens, L. M. Herz, A. Petrozza and H. J. Snaith, *Science*, 2013, **342**, 341-344.
35. D. W. deQuilettes, W. Zhang, V. M. Burlakov, D. J. Graham, T. Leijtens, A. Osherov, V. Bulović, H. J. Snaith, D. S. Ginger and S. D. Stranks, *Nature Communications*, 2016, **7**, 11683.
36. L. Contreras-Bernal, M. Salado, A. Todinova, L. Calio, S. Ahmad, J. Idígoras and J. A. Anta, *The Journal of Physical Chemistry C*, 2017, **121**, 9705-9713.
37. P. Calado, D. Burkitt, J. Yao, J. Troughton, T. M. Watson, M. J. Carnie, A. M. Telford, B. C. O' Regan, J. Nelson and P. R. F. Barnes, *ArXiv*, 2018, DOI: arXiv:1804.09049, 1804.09049.

38. O. Mangla and V. Gupta, *Journal of Materials Science: Materials in Electronics*, 2016, **27**, 12527-12532.
39. J. Rivnay, S. Inal, A. Salleo, R. M. Owens, M. Berggren and G. G. Malliaras, *Nature Reviews Materials*, 2018, **3**, 17086.

**Acknowledgements**

We thank the EPSRC for funding this work (EP/J002305/1, EP/M025020/1, EP/M014797/1, EP/L016702/1, EP/R020590/1). I.G. and E.P. would like to thank the MINECO for the CTQ2016-80042-R project. E.P. also acknowledges AGAUR for the SGR project 2014 SGR 763 and ICREA for financial support.

**Author contributions**

I.G., D.M. and P.B. initiated the project led by P.B. D.M. measured devices fabricated and developed by M.S., O.G., H.H., D.L. and P.D.; I.G. performed the simulations on software developed with P.C.; D.M. and P.B. developed the transistor description and circuit models; W.F. and D.M. performed the equivalent circuit fitting using circuit models coded by P.B. All authors discussed the results and participated in preparation of the manuscript drafted by P.B, D.M. I.G. and P.C.

**Competing interests**

I.G., D.M., P.C. and P.B. have filed a patent application based on aspects of this work.

**Materials & Correspondence**

Correspondence and requests for materials should be addressed to P.B. (piers.barnes@imperial.ac.uk), D.M. (davide.moia11@imperial.ac.uk), and I.G. (igelmetti@iciq.es). The simulation code is available at https://github.com/barnesgroupICL/Driftfusion

**Electronic Supplementary Information**

**Ionic-to-electronic current amplification in hybrid perovskite solar cells: ionically gated transistor-interface circuit model explains hysteresis and impedance of mixed conducting devices**

Davide Moia[1†*], Ilario Gelmetti[2,3†*], Phil Calado[1], William Fisher[1], Michael Stringer[4], Onkar Game[4], Yinghong Hu[5], Pablo Docampo[5,6], David Lidzey[4], Emilio Palomares[2,7], Jenny Nelson[1], Piers R. F. Barnes[1*]

[1]Department of Physics, Imperial College London, London SW7 2AZ, UK
[2]Institute of Chemical Research of Catalonia (ICIQ), Barcelona Institute of Science and Technology (BIST), Avda. Països Catalans 16, 43007 Tarragona, Spain
[3]Departament d'Enginyeria Electrònica, Elèctrica i Automàtica, Universitat Rovira i Virgili, Avda. Països Catalans 26, 43007 Tarragona, Spain
[4]Department of Physics and Astronomy, University of Sheffield, Sheffield S3 7RH, UK
[5]Department of Chemistry and Center for NanoScience (CeNS), LMU München, Butenandtstrasse 5-13, 81377 München, Germany
[6]Physics Department, School of Electrical and Electronic Engineering, Newcastle University, Newcastle upon Tyne NE1 7RU, UK
[7]ICREA, Passeig Lluís Companys, 23, Barcelona, Spain

* davide.moia11@imperial.ac.uk
* igelmetti@iciq.es
* piers.barnes@imperial.ac.uk
† These authors contributed equally to this study

**Table of Contents**

## Methods 1 Device Fabrication

### 1.1 Spiro-OMeTAD/ $Cs_{0.05}FA_{0.81}MA_{0.14}PbI_{2.55}Br_{0.45}/TiO_2$ (Fig. 2)

Chemicals: Lead (II) Iodide ($PbI_2$, 99.99%), Lead Bromide ($PbBr_2$) were purchased from TCI UK Ltd. Formamidinium Iodide (FAI), Methylammonium Bromide (MABr), FK209 Co(III) TFSI and 30NTD $TiO_2$ paste were purchased from Greatcell Solar. Dimethylformamide (DMF anhydrous), Dimethyl sulfoxide (DMSO, anhydrous), Chlorobenzene (anhydrous), Acetonitrile (anhydrous), Titanium di-isopropoxide bis-acetylacetonate (TiPAcAc, 75 wt% in IPA), Butyl Alcohol (anhydrous), Bis(trifluoromethane)sulfonimide lithium salt (Li-TFSI), 4-tert-butyl pyridine (96%), Cesium Iodide (99.9%) were purchased from Sigma Aldrich. Spiro-MeOTAD (Sublimed grade 98%) and Fluorine doped Tin Oxide (FTO, 8Ω/□) substrates were purchased from Ossila Ltd. UK. All chemicals were used without further purification.

FTO substrates were patterned to desired geometry using chemical etching with Zinc metal powder and Hydrochloric Acid (4M, Sigma Aldrich). Substrates were cleaned by sequential ultra-sonication in diluted Hellmanex (Sigma Aldrich), Deionised water and Isopropyl-Alcohol. Compact-$TiO_2$ layer (~30 nm) was deposited on patterned FTOs using spray pyrolysis of TiPAcAc (0.5 M in butyl alcohol) at 450 °C and post-heated at 450 °C for 30 min. Mesoporous $TiO_2$ layer (~150 nm) was then deposited by spin coating 30NRD solution (1:6 wt:wt in butyl alcohol) at 5000 RPM for 30 s and heated at 150 °C for 10 min. Substrates were then heat-treated at 480 °C for 30 min to remove organic contents in the 30-NRD paste.

Triple cation ($Cs_{0.05}FA_{0.81}MA_{0.14}PbI_{2.55}Br_{0.45}$) perovskite solution was prepared using a reported protocol[1]. CsI, FAI, MABr, $PbI_2$ and $PbBr_2$ were mixed in appropriate ratio in mixed solvents DMF:DMSO (4:1 v:v) to get 1.2 M concentration of $Pb^{2+}$ ions. This solution was filtered using 0.4 µm PTFE syringe filter before use. Perovskite films were deposited by anti-solvent quenching method in which 70 µL solution was spin coated initially at 2000 RPM for 10 s (ramped 200 RPM $s^{-1}$) and then at 6000 RPM for 20 s (ramp 2000 RPM $s^{-1}$) with 100 µL chlorobenzene dripped at 10 s before the end of second spin cycle. Spin coated perovskite films were crystalised by heating at 100 °C for 30 min. After cooling, hole-transport layer (HTL) of spiro-OMeTAD was spin coated at 4000 RPM for 30 s. HTL solution was prepared by dissolving 86 mg spiro-OMeTAD (Ossila Ltd. sublime grade) in 1 mL chlorobenzene, Li-TFSI (20 µL from 500 mg $mL^{-1}$ stock solution in Acetonitrile), FK209 Co-TFSI (11 µL from 300 mg $mL^{-1}$ stock solution in acetonitrile) and tert-butyl pyridine (34 µL). HTL coated perovskite cells were aged in dry air (RH < 20 %) for 12 hours before depositing Au (80 nm) top electrodes using thermal evaporation. Fabricated devices were then encapsulated first using 250 nm $Al_2O_3$ deposited by e-beam process and then using UV-Vis curable epoxy (Ossila Ltd.) with glass cover-slip. The thickness of the perovskite layer was 550 ±20 nm. The active area of the device was 0.12 $cm^2$.

## 1.2 Spiro-OMeTAD/ $FA_{0.85}MA_{0.15}PbI_3$ /$SnO_x$ (Fig. 5a)

For the fabrication of perovskite solar cell on an $SnO_x$ compact layer, patterned and cleaned FTO-glass (7Ω/sq, Hartfordglass Inc.) was covered with a 10 nm $SnO_x$ layer using an atomic layer deposition (ALD) process. Tetrakis(dimethylamino)tin(IV) (TDMSn, Strem, 99.99%) was used as a tin precursor and held at 75 °C during depositions. The deposition was conducted at 118 °C with a base pressure of 5 mbar in a Picosun R-200 Advanced ALD reactor. Ozone gas was produced by an ozone generator (INUSA AC2025). Nitrogen (99.999%, Air Liquide) was used as the carrier and purge gas with a flow rate of 50 sccm per precursor line. The growth rate was 0.69 Å per cycle. Double cation ($FA_{0.85}MA_{0.15}PbI_3$) perovskite solution was prepared by dissolving FAI (182.7 mg, 1.06 mmol), MAI (29.8 mg, 0.19 mmol) and $PbI_2$ (576.2 mg, 1.25 mmol) in a mixture of 800 µL DMF and 200 µL DMSO. The solution was filtered using a 0.45 µm PTFE syringe filter before use. $FA_{0.85}MA_{0.15}PbI_3$ perovskite films were prepared on the compact $SnO_x$ layer by spin-coating 75 µL solution at first 1000 rpm, then 5000 rpm for 10 s and 30 s, respectively. 500 µL chlorobenzene was dripped as an anti-solvent 15 s before the end of the second spin cycle. Spin-coated perovskite films were annealed at 100 °C for 10 min. For the hole transporter layer, 1 mL of a solution of spiro-OMeTAD (Borun Chemicals, 99.8%) in anhydrous chlorobenzene (75 mg $mL^{-1}$) was doped with 10 µL 4-*tert*-butylpyridine and 30 µL of a Li-TFSI solution in acetonitrile (170 mg $mL^{-1}$) and deposited by spin-coating at 1500 rpm for 40 s and then 2000 rpm for 5 s. After storing the samples overnight in air at 25% relative humidity, 40 nm Au was deposited through a patterned shadow mask by thermal evaporation. The devices were encapsulated using epoxy (Liqui Moly GmbH) and glass cover-slips. The active area was 0.158 $cm^2$ for the impedance measurements.

# Methods 2 Device characterisation

## 2.1 Photovoltaic measurements

The current-voltage characteristics of the spiro-OMeTAD/$Cs_{0.05}FA_{0.81}MA_{0.14}PbI_{2.55}Br_{0.45}$/$TiO_2$ device was measured with forward and backward scans between -0.1 V to 1.2 V with scan rate of 400 mV $s^{-1}$ under a Newport 92251A–1000 AM 1.5 solar simulator calibrated against an NREL certified silicon reference cell. An aperture mask of 0.0261 $cm^2$ was used to define the active area, see Fig. S8a. The performance of the spiro-OMeTAD/ $FA_{0.85}MA_{0.15}PbI_3$/$SnO_x$ device is shown in Fig. S8b. An identical spiro-OMeTAD/$Cs_{0.05}FA_{0.81}MA_{0.14}PbI_{2.55}Br_{0.45}$/$TiO_2$ device showed good stability when aged using an ATLAS Suntest CPS+ solar simulator with a 1500 W xenon lamp and internal reflector assembly to provide continuous illumination (~100 mW $cm^{-2}$) to the unmasked device for 40 hours. Current-voltage measurements were made every 10 minutes (reverse sweep 1.15 V to 0V) in lifetime tester, see Fig. S8c.

## 2.2 Impedance measurements

Impedance measurements were performed using an Ivium CompactStat potentiostat. The perovskite solar cell devices were masked using an aperture slightly bigger than the total active area defined by the overlap between the FTO layer and the top metal contact. All impedance measurements were run by applying a 20 mV sinusoidal voltage perturbation to the cell superimposed on a DC voltage. The potentiostat measures the resulting current, this is used to calculate the impedance spectrum as described in the main text. The frequency of the perturbation was varied between 1 MHz to 0.1 Hz. The measurement was performed after a stabilisation time of at least 100 seconds at the (light and voltage) bias condition used in the measurement, unless stated otherwise. When different stabilisation protocols were used to investigate the effect of preconditioning on the impedance measurements, these are specified in the figure legends. Different bias light conditions were obtained using white LEDs and the sun equivalent light intensity was using a filtered silicon photodiode calibrated by an AM1.5 solar simulator. Stabilisation of the cell was performed as follows. Chronopotentiometry (for impedance measurements under light at open circuit) or chronoamperometry (for impedance measurements under light at short circuit or in the dark with an applied potential bias) measurements were collected before the stabilisation stage to monitor the cell behaviour while settling to the set measurement condition. For each measurement at open circuit under light, we ran a chronopotentiometry measurement and we used the open circuit voltage measured after at least 100 seconds as the DC voltage bias condition during the impedance measurement. This voltage was applied for an additional 100 seconds before the beginning of the impedance measurement. For measurements at short circuit under light or at an applied potential in the dark, a chronoamperometry measurement was run for 100 seconds to monitor the evolution of the current in the device at the applied voltage. The same voltage was then applied for additional 100 seconds before the start of the impedance measurement. In some cases we noticed that changes in cell potential or current still occurred after 100 second stabilisation time. One could expect that these slow variations would not significantly vary the features probed at frequencies that range down to about 10 times the inverse of the stabilisation time (in our case about 0.1 Hz). However, we found that this is not the case. In particular, some peculiar features (loops in the Nyquist plots) disappeared after sufficiently long stabilisation (see Fig. S2a-d). While these features might still be indicative of the state of the device at the time of the measurement, they represented a transient state rather than the equilibrated state. For measurements at quasi-equilibrium the influence of different stabilisation times should be recorded to assess the influence on a feature of interest in an impedance spectrum to identify the minimum time needed for the spectra to reach acceptable convergence.

## Methods 3 Drift-diffusion simulation of impedance measurements

Driftfusion is a one-dimension drift-diffusion simulation for modelling perovskite solar cells which solves for the time-dependent profiles of free electron, free hole, mobile ion and electrostatic potential. The device physics of the model are based on established semi-classical transport and continuity equations, which are described in reference [1]. The code uses MATLAB's built-in Partial Differential Equation solver for Parabolic and Elliptic equations (PDEPE) to solve the continuity equations and Poisson's equation for electron density $n$, hole density $p$, a positively charged mobile ionic charge density $a$, and the electrostatic potential $V$ as a function of position $x$ and time $t$. Positively charged mobile ions and a negatively charged static counter ions (simulating Shottky defects[2]) are confined to the intrinsic region in order to simulate the high density of mobile defects in the perovskites. High rates of recombination in the contact regions are used to simulate surface/interfacial recombination.

In order to deal with the high charge density and electrostatic potential gradients at the interfaces a piece-wise linear spatial mesh was used with a spacing of 2.54 nm outside of, and 0.55 nm within the approximate depletion regions of the device. The time mesh was evaluated with either linearly or logarithmically spaced points dependent on predicted gradients in the time dimension. A complete description of the model is given in the supporting information of reference 25. Interfacial recombination (SRH) was defined to occur within a region ± 2nm from the perovskite interfaces. The code used for simulation can be downloaded from: https://github.com/barnesgroupICL/Driftfusion where usage examples specific to impedance spectroscopy are reported in the included documentation.

For simplicity we used electron and hole transporting contacts with the same band-gap, but work functions that differ from the intrinsic perovskite, to create a built-in potential in the simulated perovskite layer. Illumination was described by a uniform rate of charge generation throughout the active layer also for simplicity.

The solution of the charge and electrostatic concentration profiles of the device under steady state operating conditions was determined to provide initial conditions for the simulated impedance spectroscopy. The impedance spectroscopy simulations were performed by applying an oscillating voltage, $v$, with amplitude, $v_{max}$ = 2 mV superimposed on a bias voltage $\bar{V}$ boundary condition:

$$\bar{V} + v = \bar{V} + v_{\text{max}} \cdot \sin(\omega t)$$

where $\omega$ = 2π×frequency. For measurement of the device around its open circuit potential, $\bar{V}$ was set to the equilibrated value of $V_{OC}$ at steady state.

The electronic current was then estimated from the solution via the continuity equations. Usually a simulation of 20 voltage periods (evaluated with 40 time points per period) was enough for extracting the impedance information from the current profile.

The amplitude and phase of the oscillating electronic current density was obtained via demodulation, mimicking the working principle of a two-phase lock-in amplifier. The current density profile was point-by-point multiplied by the voltage profile or the π/2 rad shifted voltage profile normalised by $v_{max}$ and integrated over time (typically 10 periods):

$$X = \frac{\omega}{m\pi}\int_{t_0}^{t_0+\frac{2m\pi}{\omega}} j(t)\ \cdot \sin(\omega t)dt$$
$$Y = \frac{\omega}{m\pi}\int_{t_0}^{t_0+\frac{2m\pi}{\omega}} j(t)\ \cdot \cos(\omega t)dt$$

where $m$ is the number of periods, and $t_0$ is the start of the integration time. The amplitude and phase are then given via:

$$j_{\max} = \sqrt{X^2 + Y^2}$$
$$\theta = \arctan\left(\frac{Y}{X}\right)$$

allowing the impedance to be determined by $Z = v_{\max}/j_{\max}\exp(-i\theta)$. The amplitude and phase obtained this way were confirmed by fitting $j(t)$ with a sinusoidal function.

To analyse of the output of the simulation, both the electronic accumulation current and the ionic displacement current were evaluated from the solutions for the time dependent concentration profiles of electrons, holes, and ions (see Fig. 2). The ionic displacement current, $J_{\text{ion}}$, in the device was evaluated by determining the electric field profile due only to ions $E_{\text{ion}}$ as a function of time:

$$E_{\text{ion}}(x,t) = \frac{q}{\varepsilon_0\varepsilon_r}\int_{x_1}^{x_1+x} a(x',t)dx'$$

then finding its average value as a function of time:

$$\langle E_{\text{ion}}(t)\rangle = \frac{1}{d_{prv}}\int_{x_1}^{x_2} E_{\text{ion}}(x,t)dx$$

to calculate the corresponding displacement current:

$$J_{\text{ion}} = -\varepsilon_0\varepsilon_r\frac{\partial\langle E_{\text{ion}}(t)\rangle}{\partial t}.$$

Where $a(x,t)$ is the ionic concentration profile, $x$ is the position in the device, $x_1$ is the position of the HTL/perovskite interface, $x_2$ the position of the perovskite/ETM interface, $q$ is the elementary charge, $\varepsilon_0\varepsilon_r$ is the perovskite permittivity.

The total accumulation current, $j_{\text{acc}}$, (which includes the electronic charge in the contacts compensating ionic charge in the perovskite) was determined by subtracting the net recombination current (recombination minus generation) from the total cell current:

$$j_{\text{acc}}(t) = j(t) - j_{\text{rec}}(t) + j_{\text{gen}}(t)$$

where $j_{rec}(t)$ and $j_{gen}(t)$ were evaluated by integrating the recombination/generation terms in the current continuity equations over the device thickness using electron and hole concentration profiles. The parameters used in the simulation are listed in Table S2, unless stated otherwise.

## Methods 4 Equivalent circuit model

We now describe the expressions underlying the equivalent circuit model, and the approach to fitting the data. We will initially focus on the fit to data in Fig. 2. If a single interfacial electron or hole transfer process is assumed to dominate the observed impedance of the device (see discussion in Note S6) then the quasi Fermi potentials of the electrons or holes, $V_n$ or $V_p$, may be set to 0 or $V$ and an equivalent circuit of the following form can be used to fit to the experimental data (this example is for electron recombination so we can set $V_n$ = 0 V). The appropriate equivalent circuit, arbitrarily only considering electrons, is given in Fig. 2e. The impedance of the circuit is given by:

$$Z = \left(\frac{1}{Z_{\text{ion}}} + \frac{1}{Z_{\text{rec}}}\right)^{-1}$$

where $Z_{ion}$ is the impedance of the ionic circuit branch and $Z_{rec}$ is the impedance of the electronic circuit branch, in this case specifically for the limiting process of recombination. The expressions for this simple case of these terms are presented in Table 1 of the main text, but their origins are described in more detail below.

### 4.1 Impedance of the ionic circuit branch

$Z_{ion}$ is determined by drawing an analogy with the Debye relaxation of a lossy dielectric material (the perovskite) between two conducting plates (representing the undepleted regions of the device contacts) where the dependence of the effective complex permittivity of the medium between the plates varies with the angular frequency ($\omega$) as:

$$\varepsilon(\omega) = \varepsilon_\infty + \frac{\varepsilon_s - \varepsilon_\infty}{1 - i\omega\tau_{\text{ion}}}.$$

In this expression $\varepsilon_\infty$ represents the effective permittivity of the material between the conducting plates if no mobile ions were present or when $\omega$ is too high for the ions move. The capacitance per unit area of this device at high frequency (excluding any interfacial charge transfer effects discussed elsewhere in this study) would then be given by $C_g = \varepsilon_\infty / d_g$ where $d_g$ is the combined thickness of the perovskite layer and the space charge layers in the contacts (see upper panel of Fig. 4f where $\omega \to \infty$). The term $\varepsilon_s$ represents the effective permittivity of the medium sandwiched between the plates at sufficiently low angular frequencies that the perovskite layer is fully polarised by the accumulation of mobile ions to screen the applied potential (see lower panel of Fig. 4f where $\omega \to 0$). In this case the measured capacitance of the device (again excluding any interfacial charge transfer effects) will be related to $\varepsilon_s$ by: $C_{\text{ion}}/2 = \varepsilon_s/d_g$ where $C_{\text{ion}}$ is the capacitance of the space charge region surrounding each of the two contact/perovskite interfaces so $C_{\text{ion}} = \varepsilon_{\text{sc}}/d_{\text{sc}}$ where $\varepsilon_{\text{sc}}$ is permittivity of the interfacial space charge region and $d_{\text{sc}}$ is its thickness (for simplicity here we assume the capacitance of each interface is similar so the device capacitance is given by the capacitors in series $\left[C_{\text{ion}}^{-1} + C_{\text{ion}}^{-1}\right]^{-1}$). The time constant for ionic redistribution is given by $\tau_{\text{ion}} = R_{\text{ion}}\, C_{\text{ion}}/2$ where $R_{\text{ion}}$ is the specific resistance to ion motion across the perovskite layer (related to the perovskite ionic conductivity by $\sigma_{\text{ion}} \approx d_g/R_{\text{ion}}$ (if $d_{\text{sc}} << d_g$) so that $\tau_{\text{ion}} \approx \varepsilon_s/\sigma_{\text{ion}}$ as recently highlighted by Jacobs *et al.*[3]

The complex capacitance of the device due to the ionic branch of the circuit as a function of frequency is then given by $\varepsilon(\omega)/d_g$ from which we can derive an expression for the impedance of the ionic branch of the circuit:

$$Z_{\text{ion}} = \left[i\omega C_g + \frac{i\omega(C_{\text{ion}}/2 - C_g)}{1 + i\omega R_{\text{ion}} C_{\text{ion}}/2}\right]^{-1}.$$

We have depicted this frequency dependent ionic branch of the device's circuit in Fig. 4f as a $C_{\text{ion}}$–$R_{\text{ion}}$–$C_{\text{ion}}$ series curly bracketed by $C_g$ in other figures both for compactness and to maintain the physical meaning of the circuit elements. The parameters $R_{\text{ion}}$ and $C_{\text{ion}}$ can easily be determined by fitting with an $(R_{\text{eff}}–\Delta C_{\text{ion}})||C_g$ equivalent circuit represented by the ionic circuit branch in Fig. 2f to the impedance data of a device at zero bias in the dark (assuming the impedance of the electronic circuit branch is large) and determining the parameters $\Delta C_{\text{ion}}$, $R_{\text{eff}}$ and $C_g$ to give:

$$C_{\text{ion}} = 2(\Delta C_{\text{ion}} + C_g)$$

and

$$R_{\text{ion}} = \frac{R_{\text{eff}} \Delta C_{\text{ion}}}{\Delta C_{\text{ion}} + C_g}.$$

In cases where $C_{\text{ion}} \gg C_g$ then $\Delta C_{\text{ion}} \approx C_{\text{ion}}/2$ and $R_{\text{eff}} \approx R_{\text{ion}}$, however this will not hold when the space charge layers in either the perovskite or contacts are not much smaller than the perovskite thickness.

Note that we have not included a series resistance for the contacts in this model since its magnitude was negligible relative to the other elements under consideration under most measurement conditions, however we note that it is trivial to include (for example when fitting the data in Fig. 5a). As stated above, $C_g$ is the geometric capacitance of the device at high frequency, and $C_{ion}$ is the capacitance of the space charge regions of the interfaces (assumed here to be symmetric for both interfaces see Note S6, and Tables S3 and S4 for asymmetric cases) which results from the capacitance of the electronic and ionic space charge layers on either side of the interface in series. Both $C_{ion}$ and $C_g$ will show a dependence on the d.c. voltage $\bar{V}$ across the device which will change the width of the space charge layers according to the approximations:

$$C_{\mathrm{ion}}(\bar{V}) \approx C_{\mathrm{ion}}(\bar{V}=0)\sqrt{\frac{V_{\mathrm{bi}}}{V_{\mathrm{bi}}-\bar{V}}}$$

and

$$C_{\mathrm{g}}(\bar{V}) \approx \left[\frac{2}{C_{\mathrm{ion}}(\bar{V}=0)}\left(\sqrt{\frac{V_{\mathrm{bi}}-\bar{V}}{V_{\mathrm{bi}}}}-1\right)+\frac{1}{C_{\mathrm{g}}(\bar{V}=0)}\right]^{-1}$$

where $V_{bi}$ is the built-in potential of the device corresponding to the difference in work functions between the ETM and HTM contacts (or more generally between the perovskite and each contact material if calculating $C_{ion}$ for each interface). If $V_{bi}$ is known, or can be roughly estimated, it can be used as a constant input in the model, otherwise it can be used as an optional free fitting parameter. The value of $V_{bi}$ has only a weak influence on the overall quality of the fit, and similar results will be achieved if $C_{ion}$ and $C_g$ are considered constant.

A more accurate description of the ionic branch of the circuit could be expanded to describe dispersive ionic transport, effects of a mesoporous layer, and diffusion of more than one mobile ionic species.

### 4.2 Impedance of the electronic circuit branch (dominated by recombination of one carrier type)

To determine the impedance of the electronic circuit branch it is necessary to find the effect of the electrostatic potential of the ions on the concentration of electronic charge in the perovskite. The expression for $Z_{rec}$ can be derived following the arguments in the main text based on the interfacial transistor model, in this section, we confine to considering electron transfer across interface 1 which has an electrostatic potential of $V_1$ due to the ionic distribution. As discussed, the current across the interface, $J_1$, is approximated by the recombination current, $J_{rec}$:

$$J_1 \approx J_{\mathrm{rec}} = J_{\mathrm{s1}} \exp[qV_1/m_1 k_{\mathrm{B}}T]$$

where $J_{s1}$ is the saturation current density for the interface at equilibrium in the dark. To allow a more general description of the interfacial processes in real devices, we have included an ideality factor, $m_1$, describing the non-ideal variation of recombination current across interface 1 as a function of $V_1$. To find how $J_{rec}$ varies with respect to the voltage $V$ applied across the cell we must understand the relationship between $V$ and $V_1$, the electrostatic potential due to the ions at the interface.

The ionic branch of the circuit discussed above contains the series of elements $C_{ion}$–$R_{ion}$–$C_{ion}$. When a voltage is applied across the circuit, the electrostatic potential at the HTM/perovskite interface relative to dark equilibrium, $V_1$, can be calculated from the potential drop across the remaining components in the series ($R_{ion}$–$C_{ion}$). To account for the interface being located within the space charge layer that spans the interface (the interface is located between the depletion layer in the contact and the ion accumulation layer in the perovskite, Fig. S1a) we introduce the term $f_c$. This parameterises the fraction of the electrostatic potential dropping across the interface which occurs within the contact layer to control the interfacial transfer process (in this case recombination). If recombination is localised only at the interface then $f_\mathrm{c} \approx 1 - C_\mathrm{ion}/C_\mathrm{per}$ where $C_{per}$ is the capacitance due to the accumulation or depletion of ionic charge at the perovskite interface neglecting the space charge layer in the contact. The capacitance across the space charge layers in both the contact, $C_{con}$, and perovskite, $C_{per}$, contribute to the overall low frequency capacitance of the interface as $C_\mathrm{ion} = \left[{C_\mathrm{con}}^{-1} + {C_\mathrm{per}}^{-1}\right]^{-1}$, see Fig 1a. Consequently, $f_c$ will be related to the relative permittivities and doping or ionic densities on either side of the interface as well as being weakly dependence on the spatial distribution of interfacial trap states. Here, for simplicity, we assume it is constant. For the $C_{ion}$–$R_{ion}$–$C_{ion}$ series, the steady state d.c. voltage driving recombination across interface 1 will be given by:

$$\bar{V}_\mathrm{rec} = V_1 - V_\mathrm{n} = \frac{\bar{V}}{2}(2 - f_\mathrm{c})$$

assuming no drop in the electron quasi Fermi level at the opposite interface ($V_n$ = 0). The electrostatic potential of the interface 1 in response to an applied d.c. voltage with a superimposed oscillation, $V = \bar{V} + v$, is given by considering the complex impedance of the $C_{ion}$–$R_{ion}$–$C_{ion}$ series:

$$V_\mathrm{rec} = V_1 = \frac{\bar{V}}{2}(2 - f_\mathrm{c}) + \frac{v}{2}\left(2 - \frac{f_\mathrm{c}}{1 - i\omega R_\mathrm{ion} C_\mathrm{ion}/2}\right).$$

Substituting this into the expression for $J_{rec}$ above and differentiating with respect to $V$ gives an expression for the electronic impedance of the recombination process, since when $V_n$ = 0, $\mathrm{d}J_{rec}/\mathrm{d}V = 1/Z_{rec}$:

$$Z_\mathrm{rec} = \frac{2}{\left(2 - \frac{f_\mathrm{c}}{1 + i\omega R_\mathrm{ion} C_\mathrm{ion}/2}\right)} \frac{m_1 k_\mathrm{B} T}{q J_\mathrm{rec}(\bar{V})}$$

The corresponding expressions for the other interface and cases where $V_\mathrm{n} \neq 0$ (i.e. when the electron quasi Fermi level in the perovskite is not equal to the electron quasi Fermi level in the ETM) are given in Table S3. More general cases where the ionic capacitance is not equal at interface 1 and 2 are given in Table S4. Table 1 of the main text shows the expressions for the potential at each interface, $V_1$ and $V_2$, and the electronic impedance for this circuit branch for the simple case where $f_c = 1$.

The ideality factor for the recombination current at interface 1 as a function of $V_{rec}$, $m_1$, can be estimated from the steady state ideality factor, $m_{ss}$, determined from the slope of $V_{OC}$ *vs* log(light intensity) measurements[35] using the following expression:

$$m_1 \approx m_{ss}\left(1 - \frac{f_c}{2}\right)$$

We can then evaluate the recombination current density across the interface at steady state with the expression:

$$J_{rec}(\bar{V}) = J_{s1}\exp\left(\frac{q\bar{V}_{rec}}{m_1 k_B T}\right) = J_{s1}\exp\left(\frac{q\bar{V}}{m_{ss} k_B T}\right)$$

Combining these concepts, the impedance of the recombination process in terms of the bias across the device, $\bar{V}$, and its steady state ideality factor, $m_{ss}$, becomes:

$$Z_{rec} = \frac{2}{\left(2 - \frac{f_c}{1 + i\omega R_{ion} C_{ion}/2}\right)} \frac{m_{ss}\left(1 - \frac{f_c}{2}\right) k_B T}{q J_{s1}\exp\left(\frac{q\bar{V}}{m_{ss} k_B T}\right)}$$

### 4.3 Impedance of the whole device

The complete expression for the impedance of the device can be calculated by considering the impedance of the ionic ($Z_{ion}$) and electronic ($Z_{rec}$) branches of the circuit model in parallel and including series resistance, $R_s$:

$$Z = R_s + \left\{ i\omega C_g(\bar{V}) + \frac{i\omega\left[C_{ion}(\bar{V})/2 - C_g(\bar{V})\right]}{1 + i\omega R_{ion} C_{ion}(\bar{V})/2} + \frac{1}{2}\left[2 - \frac{f_c}{1 + i\omega R_{ion} C_{ion}(\bar{V})/2}\right] \frac{q J_{s1}\exp\left(\frac{q\bar{V}}{m_{ss} k_B T}\right)}{m_{ss}\left(1 - \frac{f_c}{2}\right) k_B T} \right\}^{-1}$$

The cell bias voltage, $\bar{V}$ and the steady state ideality factor, $m_{ss}$, are known or determined independently from measurements. $C_{ion}(\bar{V})$ and $C_g(\bar{V})$ will approximately depend on $\bar{V}$ as described above using an estimation of $V_{bi}$. The unknown device parameters in this expression for $Z$ can be determined from a fit are: $R_s$, $R_{ion}$, $C_{ion}(\bar{V} = 0)$, $C_g(\bar{V} = 0)$, $J_{s1}$ and $f_c$. If $V_{bi}$ cannot be estimated, it can also be used as a fitting parameter. Since $R_s$ is typically trivial to determine from the impedance spectra this leaves only five significant parameters to describe key device physics.

A similar approach can be used to express the impedance of the device for the more general circuit for example if both recombination and injection of electrons limit impedance as described the section above:

$$Z = R_{\mathrm{s}} + \left(\frac{1}{Z_{\mathrm{ion}}} + \frac{1}{Z_{\mathrm{n}}}\right)^{-1}$$

where $Z_{\mathrm{n}}$ is the impedance of electronic current transfer through the device (given in Table S3). More generally for transfer of both electrons and holes with impedance $Z_{\mathrm{np}}$ (given in Table S4) the device impedance becomes:

$$Z = R_{\mathrm{s}} + \left(\frac{1}{Z_{\mathrm{ion}}} + \frac{1}{Z_{\mathrm{np}}}\right)^{-1}$$

We emphasise again that under most circumstances only one electronic process is likely to dominate the electronic branches of the device impedance so such a generalisation will not normally be required to describe a device around particylar operating conditions. We also emphasise that the impedance of the ionic branch of the circuit, $Z_{\mathrm{ion}}$, might differ from the expression presented above in some devices, for example if ions penetrate or react at interfaces, or if ion transport is dispersive, or if more than one mobile ionic species is present (Fig. 5a and Methods 6). Additionally, diffusive transport of ions might occur within mesoporous regions of a device which could potentially be described by a Warburg element in series with $R_{\mathrm{ion}}$.

### Methods 5. Fitting the impedance spectra to an equivalent circuit model

Global fits of the impedance circuit model for $Z$ (the electron recombination only model) to the experimental and simulated impedance spectra at all measured conditions presented in Fig. 2 and Fig. 5 were performed using a non-linear least squares fitting routine. We aimed to use the fewest parameters possible to give a reasonable representation of the data. For Fig. 2 the free parameters were $R_{\mathrm{ion}}$, $C_{\mathrm{ion}}(\bar{V} = 0)$, $C_{\mathrm{g}}(\bar{V} = 0)$, $J_{\mathrm{s1}}$ and $f_{\mathrm{c}}$. The bias voltage, $\bar{V}$, and measured ideality factor for each measurement were used as inputs. Relatively little co-variance was observed between the parameters for the overall shape of the resulting device impedance spectra, so the fits were performed in a stepwise fashion in which the range of frequencies over which each parameter was fit was limited to the regions of the spectra which responded to that particular parameter. $C_{\mathrm{g}}(\bar{V} = 0)$ was determined from the fit to the high frequency region of the dark, 0 V bias, spectrum. $C_{\mathrm{ion}}(\bar{V} = 0)$ was initially determined from the fit to the low frequency region of the dark, 0 V bias, spectrum. $R_{\mathrm{ion}}$, $J_{\mathrm{s1}}$, and $f_{\mathrm{c}}$ (the fraction of screening potential dropping within the contacts) were determined from the fit to all the spectra from low frequency to medium frequency. The fit parameters the Fig. 2 data are given in Table S1.

To estimate $R_{ion}$ directly from the measured impedance data we can use the relationship outlined in equation 3:

$$\frac{j''_{\text{rec}}}{j_{\text{ion}}} = \frac{R_{ion}}{2} f_{\text{c}} g_{\text{rec}} = \frac{R_{ion}}{2} f_{\text{c}} \frac{qJ_{\text{rec}}(\bar{V})}{m_1 k_{\text{B}} T}$$

where $j''_{\text{rec}}/j_{\text{ion}}$ at low frequency $(\omega \to 0)$ is given by:

$$\frac{j''_{\text{rec}}}{j_{\text{ion}}} = \frac{2c_{\text{rec}}}{C_{\text{ion}}} = \frac{c(\bar{V},\omega\to 0) - c(\bar{V}=0,\omega\to 0)\sqrt{\frac{V_{\text{bi}}}{V_{\text{bi}}-\bar{V}}}}{c(\bar{V}=0,\omega\to 0)\sqrt{\frac{V_{\text{bi}}}{V_{\text{bi}}-\bar{V}}}}$$

and the recombination transconductance can be evaluated from:

$$g_{\text{rec}} = \frac{q\bar{J}_{\text{rec}}(\bar{V})}{m_1 k_{\text{B}} T} = \frac{q\bar{J}_{\text{rec}}(\bar{V})}{m_{\text{ss}}\left(1 - \frac{f_{\text{c}}}{2}\right) k_{\text{B}} T}$$

The $c$ terms are given by the measured apparent capacitance, $c = \omega^{-1}\text{Im}(Z^{-1})$ at the high and low frequency limits and bias voltages indicated. If the measurement is made in the dark and recombination is assumed to dominate the electronic impedance then the cell current, $\bar{J} \approx \bar{J}_{\text{rec}}$. If the measurement is made at open circuit then $\bar{J}_{\text{rec}} \approx J_{\text{ph}}$ which may be estimated from the short circuit current or the absorbed photon flux.

## Methods 6. Circuit model resulting in inductive behaviour due to recombination at an interface where ions may penetrate, or undergo a reversible chemical reaction

If ionic defects penetrate or chemically react reversibly with an interface, this will result in an additional perturbation of the ionic distribution which may have a different time constant to $R_{\text{ion}}C_{\text{ion}}/2$ which could lead to inductive behaviour. For example, iodide ions might reversibly react with oxygen vacancies in an $SnO_x$ contact. An equivalent circuit giving an approximate description of ion penetration or a reversible reaction is shown in Fig. 5a and Fig. S7:

$R_{\text{int}}$ is the effective interfacial resistance to ion penetration or reaction, and $C_{\text{con}}$ is the effective chemical capacitance of the contact for the ions. Depending on the frequency range and values of the circuit elements, changes in $V_2$ may lead or lag changes in the applied potential $V$ resulting in apparently capacitive or inductive behaviour. Note that for simplicity we approximated the geometric capacitance by including a separate $C_{\text{g}}$ branch in this model. To determine the behaviour of the current flowing through this circuit the frequency dependence of $V_2$ must be determined by examining the ionic branch of the circuit which has an impedance:

$$Z_{\text{ion}} = \frac{1}{i\omega C_{\text{ion}}} + R_{\text{ion}} + \left(i\omega C_{\text{ion}} + \frac{1}{R_{\text{ion}} + \frac{1}{i\omega C_{\text{con}}}}\right)^{-1}$$

Ignoring $R_S$, at steady state the potentials at $V_1$ and $V_2$ where $\omega \to 0$ will be given by:

$$\bar{V}_1 = \bar{V}_2 = \frac{C_{\text{ion}}}{2C_{\text{ion}} + C_{\text{con}}}\bar{V}$$

This allows the transconductance for hole recombination to be calculated given the voltage driving recombination is $\bar{V} - \bar{V}_2$ (see Table S4):

$$g_{\text{rec}}^{\text{p}} = \frac{q}{k_{\text{B}}T} J_{\text{s2}} e^{\frac{q(\bar{V} - \bar{V}_2)}{k_{\text{B}}T}} = \frac{q}{k_{\text{B}}T} J_{\text{s2}} e^{\frac{q}{k_{\text{B}}T}\left(1 - \frac{C_{\text{ion}}}{2C_{\text{ion}} + C_{\text{con}}}\right)\bar{V}} = \frac{qJ_{\text{rec}}^{\text{p}}}{k_{\text{B}}T}$$

The small perturbation potentials $v_1$ and $v_2$ in response to $v$ are then given by:

$$v_1 = \left(1 - \frac{1}{i\omega C_{\text{ion}} Z_{\text{ion}}}\right) v$$
$$v_2 = \left(1 - \frac{1}{i\omega C_{\text{ion}} Z_{\text{ion}}} - \frac{R_{\text{ion}}}{Z_{\text{ion}}}\right) v.$$

When a small perturbation $v$ is applied across the interface the voltage driving recombination $v$ - $v_2$ can be found using the above expression. This enables the impedance to hole current recombining across the interface to be found by dividing $j_{\text{rec}}^{\text{p}} = (v - v_2) g_{\text{rec}}^{\text{p}}$ by $v$:

$$\frac{1}{Z_{\text{rec}}^{\text{p}}} = \frac{j}{v} = \left(\frac{1}{i\omega C_{\text{ion}} Z_{\text{ion}}} + \frac{R_{\text{ion}}}{Z_{\text{ion}}}\right) \frac{qJ_{\text{rec}}^{\text{p}}}{k_{\text{B}}T}$$

This can then be incorporated within the complete equivalent circuit to give the impedance of the device including series resistance $R_s$:

$$Z = R_{\text{s}} + \left(i\omega C_{\text{g}} + \frac{1}{Z_{\text{ion}}} + \frac{1}{Z_{\text{rec}}^{\text{p}}}\right)^{-1}$$

$$Z = R_{\text{s}} + \left(i\omega C_{\text{g}} + \left[\frac{1}{i\omega C_{\text{ion}}} + R_{\text{ion}} + \left(i\omega C_{\text{ion}} + \frac{1}{R_{\text{ion}} + \frac{1}{i\omega C_{\text{con}}}}\right)^{-1}\right]^{-1} + \left(\frac{1}{i\omega C_{\text{ion}} Z_{\text{ion}}} + \frac{R_{\text{ion}}}{Z_{\text{ion}}}\right) \frac{qJ_{\text{rec}}^{\text{p}}}{k_{\text{B}}T}\right)^{-1}$$

This expression can then be used in a global fit to the data.

## Supplementary Notes

### Note S1. Evaluation of inductive behaviour due to injection and negative ionic-to-electronic current transcarrier amplification

We now demonstrate that the circuit model can result in inductive behaviour due to interfacial charge injection processes coupled to ionic redistribution. Charge injection of a carrier (free electron or hole) will occur in series with the corresponding recombination process described above. Considering the electronic current across the ETM interface 2, the net current density is given by the difference between the injection and collection currents, $J_{\mathrm{inj}}$ and $J_{\mathrm{col}}$:

$$J_2 = J_{\mathrm{inj}} - J_{\mathrm{col}} = J_{\mathrm{s2}} e^{\frac{qV_{\mathrm{inj}}}{k_{\mathrm{B}}T}} - J_{\mathrm{s2}} e^{\frac{qV_{\mathrm{col}}}{k_{\mathrm{B}}T}}$$

where $J_{\mathrm{s2}}$ is the electron saturation current density of the interface at equilibrium in the dark and the changes in barrier potentials $V_{\mathrm{inj}}$ and $V_{\mathrm{col}}$ in relation to ionic redistribution are given in Table 1, Fig. 4 and Fig. S4c.

If $V_{\mathrm{n}} \approx V$ (which, given our assumptions, would hypothetically occur under forward bias in the dark where $J_{\mathrm{s2}} >> J_{\mathrm{s1}}$) then the electron collection current is negligible and the impedance of interface 2 is controlled by injection (Table 1):

$$\frac{1}{Z_{\mathrm{inj}}} = \frac{\mathrm{d}j_{\mathrm{inj}}}{\mathrm{d}v} = \frac{1}{2}\left(\frac{1}{1+i\omega R_{\mathrm{ion}}C_{\mathrm{ion}}/2}\right)\frac{qJ_{\mathrm{inj}}(\bar{V})}{k_{\mathrm{B}}T}$$

Comparing this with equation 3 shows that ionic motion causes $Z_{\mathrm{inj}}$ to vary with an imaginary component π rad out of phase with $Z_{\mathrm{rec}}$ so that the interface will behave like an inductor despite no release of accumulated electronic charge. The real part of this $Z_{\mathrm{inj}}$ is given by:

$$r_{\mathrm{inj}} = Z'_{\mathrm{inj}} = \frac{2k_{\mathrm{B}}T}{qJ_{\mathrm{inj}}(\bar{V})}$$

The corresponding negative value of the imaginary part of $Z_{\mathrm{inj}}$ divided by the angular frequency gives an expression which is analogous to an apparent inductance to injection $l_{\mathrm{inj}}$ of charge carriers across the interface:

$$l_{\mathrm{inj}} = -\frac{Z''_{\mathrm{inj}}}{\omega} = \frac{k_{\mathrm{B}}TR_{\mathrm{ion}}C_{\mathrm{ion}}}{qJ_{\mathrm{inj}}(\bar{V})}$$

This has the potential to lead to loops in Nyquist plots (Fig. S4c). As discussed in the main text, this result also implies the presence of a transcarrier amplification factor based on the following argument. At low frequency when $\omega \ll (R_{\mathrm{ion}}C_{\mathrm{ion}}/2)^{-1}$ the ionic current will be out of phase with $v$ is given by $J_{\mathrm{ion}} \approx i\omega C_{\mathrm{ion}}v/2$ so that the out of phase component of the voltage perturbation at interface 2 is $v''_2 = -J_{\mathrm{ion}}R_{\mathrm{ion}}/2$ due to the electrostatic drop in potential across the perovskite. This results in an out of phase electronic current of $j''_{\text{inj}} = -J_{\text{ion}}R_{\text{ion}}g_{\text{inj}}$ where $g_{\text{inj}}$ is the injection transconductance of the interface given by $\mathrm{d}J_{\text{inj}}/\mathrm{d}V_2 = qJ_{\text{inj}}(\bar{V})/(k_{\text{B}}T)$. Taking the ratio of these currents gives the ionic-to-electronic transcarrier amplification of the ionic current as mentioned in the main text:

$$\frac{j''_{\text{inj}}}{J_{\text{ion}}} = -R_{\text{ion}}g_{\text{inj}} = -R_{\text{ion}}\frac{qJ_{\text{inj}}(V,\omega=0)}{k_{\text{B}}T}$$

**Note S2. Calculating the impedance of both interfaces considering only electrons**

In cases where the impedance of both interface 1 and interface 2 are comparable, the value of $V_{\text{n}}$ will no longer be $V_{\text{n}} \approx 0$ (for a recombination dominated impedance) or $V_{\text{n}} \approx V$ (for injection dominated impedance) so it must be determined in order to quantify $Z_1$ and $Z_2$. The inclusion of both $Z_{\text{rec}}(V,J_{\text{ph}},\omega)$ (capacitor-like) and $Z_{\text{inj}}(V,J_{\text{ph}},\omega)$ (inductor like) elements within an equivalent circuit model can result in loops within Nyquist plots under some circumstances (see Fig. S4d). Table S3 (which is a more complete extension of Table 1) summarises the changes in potential barriers, electrostatic interface potentials, and small perturbation impedances considering electrons only. The value of $V_{\text{n}}$ is evaluated by substituting the expressions for the interfacial currents at steady state (i.e. $\omega = 0$) in Table S3 into the following current continuity equation for the electronic interfacial currents using the steady state values of $\bar{V}_1$ and $\bar{V}_2$ where $\bar{V}_1 = \bar{V}_2 = \bar{V}/2$ (if $f_{\text{c}} = 1$):

$$J_{\text{n}} = J_{\text{rec}} - J_{\text{gen}} + J_{\text{ph}} = J_{\text{inj}} - J_{\text{col}}$$

and solving numerically for $V_{\text{n}}$. In the small perturbation regime current continuity must also be obeyed so that:

$$j_{\text{n}} = j_{\text{rec}} - j_{\text{gen}} = j_{\text{inj}} - j_{\text{col}}$$

where the photogeneration current need not be considered as it is not perturbed. The above expression can be rewritten in terms of the product of the voltage perturbation driving each process (Table S3) with the transconductance for each process as:

$$j_{\text{n}} = v(1 - A - B_{\text{n}})\frac{J_{\text{rec}}}{k_{\text{B}}T} + vA\frac{J_{\text{gen}}}{k_{\text{B}}T} = vA\frac{J_{\text{inj}}}{k_{\text{B}}T} - v(B_{\text{n}} - A)\frac{J_{\text{col}}}{k_{\text{B}}T}$$

where $B_{\text{n}} = v_{\text{n}}/v$. Since $A$ is known (as defined in Table S3), this can be solved for $B_{\text{n}}$ to give:

$$B_{\text{n}} = \frac{J_{\text{rec}} + A(J_{\text{gen}} - J_{\text{rec}} + J_{\text{col}} - J_{\text{inj}})}{J_{\text{rec}} + J_{\text{col}}}$$

The small perturbation impedance (for electrons) of the two interfaces in series can then be found by dividing $v$ by $j_n$ to give:

$$Z_n = Z_1 + Z_2 = \left((1 - A - B_n)\frac{J_{rec}}{k_B T} + A\frac{J_{gen}}{k_B T}\right)^{-1}$$

The impedances of each interface and individual process are separately listed in Table S3 (which also includes the process ideality factors) should they need to be evaluated separately. Almost identical arguments can be used if only hole processes dominate the impedance of the device. Bulk recombination can also be easily included by adding the appropriate expression to the current continuity equation as described for the general case in the Note S6, Table S4, and Fig. S8.

**Note S3. Accounting for accumulating electronic charge in the perovskite layer**

The model we have proposed assumes that the concentration of electronic charge in the active layer is negligible relative to the background concentration of mobile ionic defects. Particularly at higher bias voltages the concentration of electronic charge may become comparable to the ionic charge. Since the electronic charge is highly mobile relative to the ionic defects it will rapidly move to screen changes in the ionic charge distribution. This will have the consequence of screening any modulation in the values of $V_1$ and $V_2$ and thus modulation out of phase components of interfacial charge transfer. To approximately describe this screening behaviour for a simplified model considering just electrons and ions we can modify the equivalent circuit as shown in Fig. S8d.

As the value of the screening capacitance, $C_n$, the amplitude of the modulation electrostatic potential by the ions at $V_1$ and $V_2$ is reduced, removing the amplification behaviour from the out of phase currents across the interfaces resulting so that the Nyquist plot returns to a single semicircle (see Fig. S8e). Additionally, this electronic screening capacitance, $C_n$, also contributes the increase in overall device capacitance at high frequencies as the bias voltage increases.

**Note S4. Calculating large perturbation current-voltage sweep behaviour**

The time varying potential in the perovskite layer close to each interface can be evaluated for large perturbations. For example, the current response of the device in response to a linear voltage sweep can be found by considering the ionic branch of the circuit and its coupling to the electron branch for the circuit shown in Fig. 4g.

A linear voltage sweep with scan rate $s$ is applied across the device terminals results in charge $Q_{ion}$ that accumulates at the interfacial capacitances $C_{ion}$ with time, this can be found by solving the differential equation:

$$\frac{\mathrm{d}Q_{\mathrm{ion}}}{\mathrm{d}t} = \frac{1}{R_{\mathrm{ion}}}\left(V_{\mathrm{initial}} + st - \frac{2Q_{\mathrm{ion}}}{C_{\mathrm{ion}}}\right)$$

with the initial condition that $Q_{\mathrm{ion}}(t = 0) = Q_0$ and $V(t = 0) = V_{\mathrm{initial}}$ is the initial potential. When the scan starts $Q_0$ need not be in equilibrium with $V_{\mathrm{initial}}$, this is particularly relevant to cases where the cell is preconditioned with a forward bias prior to measurement.

$$Q_{\mathrm{ion}}(t) = s\frac{C_{\mathrm{ion}}}{2}t - sR_{\mathrm{ion}}\left(\frac{C_{\mathrm{ion}}}{2}\right)^2 + \left(sR_{\mathrm{ion}}\left(\frac{C_{\mathrm{ion}}}{2}\right)^2 + Q_0 - V_{\mathrm{initial}}\frac{C_{\mathrm{ion}}}{2}\right)e^{-\frac{2t}{R_{\mathrm{ion}}C_{\mathrm{ion}}}}$$

$Q_0$ is the initial charge on $C_{\mathrm{ion}}$ relative to equilibrium in the dark (in which case we define $Q_0 = 0$). The electrostatic potentials at $V_1$ and $V_2$ are given by:

$$V_1(t) = V(t) - \frac{Q_{\mathrm{ion}}(t)}{C_{\mathrm{ion}}}$$
$$V_2(t) = \frac{Q_{\mathrm{ion}}(t)}{C_{\mathrm{ion}}}$$

This allows the current through the interfaces to be calculated by numerically solving the following expression to give $V_{\mathrm{n}}$ and thus $J_{\mathrm{n}}$ by substituting in the expressions for interfacial currents and potentials given in Table S3 (assuming $V_n$ it is not set to 0 for cases where injection is not limiting):

$$J_{\mathrm{n}}(t) = J_{\mathrm{rec}}(t) - J_{\mathrm{gen}}(t) + J_{\mathrm{ph}} = J_{\mathrm{inj}}(t) - J_{\mathrm{col}}(t)$$

If only one process limits the interfacial currents then the interfacial electron current, $J_{\mathrm{n}}$, can be found more simply, for example if electron recombination limits the current through the interfaces ($V_{\mathrm{n}}$ = 0 V) and:

$$J_{\mathrm{n}}(t) = \frac{J_{\mathrm{S1}}}{k_{\mathrm{B}}T}e^{\frac{Q_{\mathrm{ion}}(t)}{C_{\mathrm{ion}}}\frac{q}{k_{\mathrm{B}}T}}$$

The device current, $J$, can then be found from the sum of the ionic current, $J_{\mathrm{ion}}$, the geometric charging current $J_{\mathrm{g}}$ and the interfacial electronic current $J_{\mathrm{n}}$:

$$J(t) = J_{\mathrm{n}}(t) + J_{\mathrm{ion}}(t) + J_{\mathrm{g}}(t)$$

For a linear voltage sweep with rate $s$ these currents are:

$$J_{\mathrm{g}}(t) \approx sC_{\mathrm{g}}$$
$$J_{\mathrm{ion}}(t) = \frac{sC_{\mathrm{ion}}}{2} - \frac{2\left(sR\left(\frac{C_{\mathrm{ion}}}{2}\right)^2 + Q_0\right)}{R_{\mathrm{ion}}C_{\mathrm{ion}}}e^{-\frac{2t}{R_{\mathrm{ion}}C_{\mathrm{ion}}}}$$

Examples of the modelled $J$ using this approach are shown in Fig. 5d and Fig. S6 for cyclic voltammograms, they shows the resulting hysteresis in the current-voltage behaviour. These simulated current voltage sweeps based on the parameters determined from fitting the experimental impedance spectrum show very good agreement with the experimentally measured current voltage sweeps in Fig. S8a for the same scan rate.

**Note S5. Calculating large perturbation current-voltage step behaviour**

The response of the circuit to a voltage step may also be calculated by considering the response of the ions to a step change in cell potential from $V_{\mathrm{initial}}$ to $V_{\mathrm{final}}$. The differential equation for the evolution of ionic charge is given by:

$$V_{\mathrm{final}} - V_{\mathrm{initial}} = R_{\mathrm{ion}}\frac{\mathrm{d}Q}{\mathrm{d}t} + \frac{(2Q - C_{\mathrm{ion}}V_{\mathrm{initial}})}{C_{\mathrm{ion}}}$$

With the initial condition $Q(t=0) = C_{\mathrm{ion}}V_{\mathrm{initial}}/2$, which has the solution:

$$Q_{\mathrm{ion}}(t) = \frac{C_{\mathrm{ion}}}{2}\left[V_{\mathrm{final}} - (V_{\mathrm{final}} - V_{\mathrm{initial}})e^{\frac{-2t}{R_{\mathrm{ion}}C_{\mathrm{ion}}}}\right]$$

The electrostatic potentials at $V_1$ and $V_2$ are given by:

$$V_1(t) = V_{\mathrm{final}} - \frac{Q_{\mathrm{ion}}(t)}{C_{\mathrm{ion}}}$$

$$V_2(t) = \frac{Q_{\mathrm{ion}}(t)}{C_{\mathrm{ion}}}$$

Again, this allows the current through the interfaces to be calculated by numerically solving the following expression to give $V_{\mathrm{n}}$ and thus $J_{\mathrm{n}}$ (as described above for linear sweep voltammetry) by substituting in the expressions for interfacial currents and potentials given in Table S3:

$$J_{\mathrm{n}}(t) = J_{\mathrm{rec}}(t) - J_{\mathrm{gen}}(t) + J_{\mathrm{ph}} = J_{\mathrm{inj}}(t) - J_{\mathrm{col}}(t)$$

The currents in the other branches of the device circuit, $J_{\mathrm{ion}}$ and $J_{\mathrm{g}}$ are given by:

$$J_{\mathrm{ion}}(t) = \frac{2(V_{\mathrm{final}} - V_{\mathrm{initial}})}{R_{\mathrm{ion}}}e^{\frac{-2t}{R_{\mathrm{ion}}C_{\mathrm{ion}}}}$$

$$J_{\mathrm{g}} = \frac{2(V_{\mathrm{final}} - V_{\mathrm{initial}})}{R_{\mathrm{s}}}e^{\frac{-2t}{R_{\mathrm{s}}C_{\mathrm{g}}}}$$

Assuming that $R_{\mathrm{s}} << R_{\mathrm{ion}}$, giving $J(t) = J_{\mathrm{n}}(t) + J_{\mathrm{ion}}(t) + J_{\mathrm{g}}(t)$. The resulting current (or photocurrent transients) may display apparently capacitive or inductive behaviour.

**Note S6. General description of interfaces considering electrons, holes, bulk recombination, interface idealities, asymmetric ionic capacitance, partial ionic screening within the perovskite layer.**

In the main text we assumed that under most circumstances a single electron or hole interfacial transfer process would dominate the observed impedance behaviour under a given operating condition. If the contributions to the impedance from the processes at all interfaces are considered then the total impedance of the combined interfaces will be given by:

$$Z_{\mathrm{np}} = \left(\frac{1}{Z_1^{\mathrm{n}}+Z_2^{\mathrm{n}}} + \frac{1}{Z_1^{\mathrm{p}}+Z_2^{\mathrm{p}}}\right)^{-1}$$

where $Z_1^{\mathrm{n}}$ and $Z_2^{\mathrm{n}}$ are the electron transfer impedances of interfaces 1 and 2, and $Z_1^{\mathrm{p}}$ and $Z_2^{\mathrm{p}}$ are the corresponding hole transfer impedances (see Table S4). Note that in these expressions and those that follow the superscripts ‘n’ and ‘p’ are used to distinguish processes related electrons or holes, *they do not refer to exponents*. The value of $Z_{\mathrm{np}}$ will be dominated by the process with the highest impedance within the branch showing the lowest impedance, interface dominating impedance may vary for different operating conditions.

Under some circumstances more than one process may contribute to the observed impedance in which case a complete expression for $Z_{\mathrm{np}}$ may be evaluated. In the main text, and in the expression for $Z_{\mathrm{np}}$ above we also assumed that recombination only occurred at interfaces. We now describe the method to evaluate a more general version of the interface model, containing electrons, holes and bulk recombination (represented by a diode which describes recombination processes that depend only on the quasi Fermi level splitting such as band-to-band bimolecular recombination), see circuit diagram in Fig. S8f.

To find the impedance, the background steady state currents of each interfacial process must be established, this requires the values of $V_{\mathrm{n}}$, $V_{\mathrm{p}}$ and $J_{\mathrm{np}}$ to be determined where $J_{\mathrm{np}}$ is the steady state electronic current due to both electrons and holes. We define the photogeneration current, $J_{\mathrm{ph}}$ to be negative. These quantities can be found by numerically solving a system of three simultaneous equations arising from Kirchhoff’s laws:

$$J_{\mathrm{np}} = J_{\mathrm{rec}}^{\mathrm{n}} - J_{\mathrm{gen}}^{\mathrm{n}} + J_{\mathrm{inj}}^{\mathrm{p}} - J_{\mathrm{col}}^{\mathrm{p}}$$
$$J_{\mathrm{np}} = J_{\mathrm{rec}}^{\mathrm{p}} - J_{\mathrm{gen}}^{\mathrm{p}} + J_{\mathrm{inj}}^{\mathrm{n}} - J_{\mathrm{col}}^{\mathrm{n}}$$
$$J_{\mathrm{rec}}^{\mathrm{n}} - J_{\mathrm{gen}}^{\mathrm{n}} + J_{\mathrm{bulk}} + J_{\mathrm{ph}} = J_{\mathrm{inj}}^{\mathrm{n}} - J_{\mathrm{col}}^{\mathrm{n}}$$

with the appropriate expressions substituted into the terms which are given in Table S4. $V_{\mathrm{n}}$, $V_{\mathrm{p}}$ and $J_{\mathrm{np}}$ allow the steady state interfacial currents to be evaluated and used to calculate the evaluate transconductances described below. Similar equations govern the current continuity in the small perturbation regime, without the need to include photocurrent (we note that the model could also be applied to describe intensity modulate photocurrent and photovoltage measurements (IMPS and IMVS) by including a small perturbation photocurrent):

$$j_{\mathrm{np}} = j_{\mathrm{rec}}^{\mathrm{n}} - j_{\mathrm{gen}}^{\mathrm{n}} + j_{\mathrm{inj}}^{\mathrm{p}} - j_{\mathrm{col}}^{\mathrm{p}}$$
$$j_{\mathrm{np}} = j_{\mathrm{rec}}^{\mathrm{p}} - j_{\mathrm{gen}}^{\mathrm{p}} + j_{\mathrm{inj}}^{\mathrm{n}} - j_{\mathrm{col}}^{\mathrm{n}}$$
$$j_{\mathrm{rec}}^{\mathrm{n}} - j_{\mathrm{gen}}^{\mathrm{n}} + j_{\mathrm{bulk}} = j_{\mathrm{inj}}^{\mathrm{n}} - j_{\mathrm{col}}^{\mathrm{n}}$$

These can be rewritten in terms of the voltage perturbation driving each process and the corresponding transconductances:

$$\frac{1}{Z_{\mathrm{np}}} = \frac{j_{\mathrm{np}}}{v} = (1 - A_1 - B_{\mathrm{n}})\frac{qJ_{\mathrm{rec}}^{\mathrm{n}}}{m_1 k_{\mathrm{B}} T} + A_1 \frac{qJ_{\mathrm{gen}}^{\mathrm{n}}}{m_1 k_{\mathrm{B}} T} + A_1 \frac{qJ_{\mathrm{inj}}^{\mathrm{p}}}{m_1 k_{\mathrm{B}} T} - \left(B_{\mathrm{p}} + A_1 - 1\right)\frac{qJ_{\mathrm{col}}^{\mathrm{p}}}{m_1 k_{\mathrm{B}} T}$$
$$\frac{1}{Z_{\mathrm{np}}} = \frac{j_{\mathrm{np}}}{v} = \left(B_{\mathrm{p}} - A_2\right)\frac{qJ_{\mathrm{rec}}^{\mathrm{p}}}{m_2 k_{\mathrm{B}} T} + A_2 \frac{qJ_{\mathrm{gen}}^{\mathrm{p}}}{m_2 k_{\mathrm{B}} T} + A_2 \frac{qJ_{\mathrm{inj}}^{\mathrm{n}}}{m_2 k_{\mathrm{B}} T} - (A_2 - B_{\mathrm{n}})\frac{qJ_{\mathrm{col}}^{\mathrm{n}}}{m_2 k_{\mathrm{B}} T}$$
$$(1 - A_1 - B_{\mathrm{n}})\frac{qJ_{\mathrm{rec}}^{\mathrm{n}}}{m_1 k_{\mathrm{B}} T} + A_1 \frac{qJ_{\mathrm{gen}}^{\mathrm{n}}}{m_1 k_{\mathrm{B}} T} + (B_{\mathrm{p}} - B_{\mathrm{n}})\frac{qJ_{\mathrm{bulk}}}{k_{\mathrm{B}} T} = A_2 \frac{qJ_{\mathrm{inj}}^{\mathrm{n}}}{m_2 k_{\mathrm{B}} T} - (A_2 - B_{\mathrm{n}})\frac{qJ_{\mathrm{col}}^{\mathrm{n}}}{m_2 k_{\mathrm{B}} T}$$

given that $A_1$ and $A_2$ are known (see Table S4) this system of equations can be solved analytically to give $Z$, $B_{\mathrm{n}}$ and $B_{\mathrm{p}}$ where $B_{\mathrm{n}} = v_{\mathrm{n}}/v$ and $B_{\mathrm{p}} = v_{\mathrm{p}}/v$. Here, $Z_{\mathrm{np}}$ is the impedance of the two interfaces in series for electrons and holes. The resulting analytical solutions are rather long and thus not reproduced here, however they are straightforward to evaluate using analytical mathematics software. The impedances of the individual processes and interfaces are listed in Table S4.

**Supplementary references**

1. H. Tan, A. Jain, O. Voznyy, X. Lan, F. P. García de Arquer, J. Z. Fan, R. Quintero-Bermudez, M. Yuan, B. Zhang, Y. Zhao, F. Fan, P. Li, L. N. Quan, Y. Zhao, Z.-H. Lu, Z. Yang, S. Hoogland and E. H. Sargent, *Science*, 2017, **355**, 722-726.
2. A. Walsh, D. O. Scanlon, S. Chen, X. G. Gong and S.-H. Wei, *Angewandte Chemie (International Ed. in English)*, 2015, **54**, 1791-1794.
3. D. A. Jacobs, H. Shen, F. Pfeffer, J. Peng, T. P. White, F. J. Beck and K. R. Catchpole, *arXiv*, 2018, 1807.00954.
4. J. J. Ebers and J. L. Moll, *Proceedings of the IRE*, 1954, **42**, 1761-1772.

## Supplementary Figures

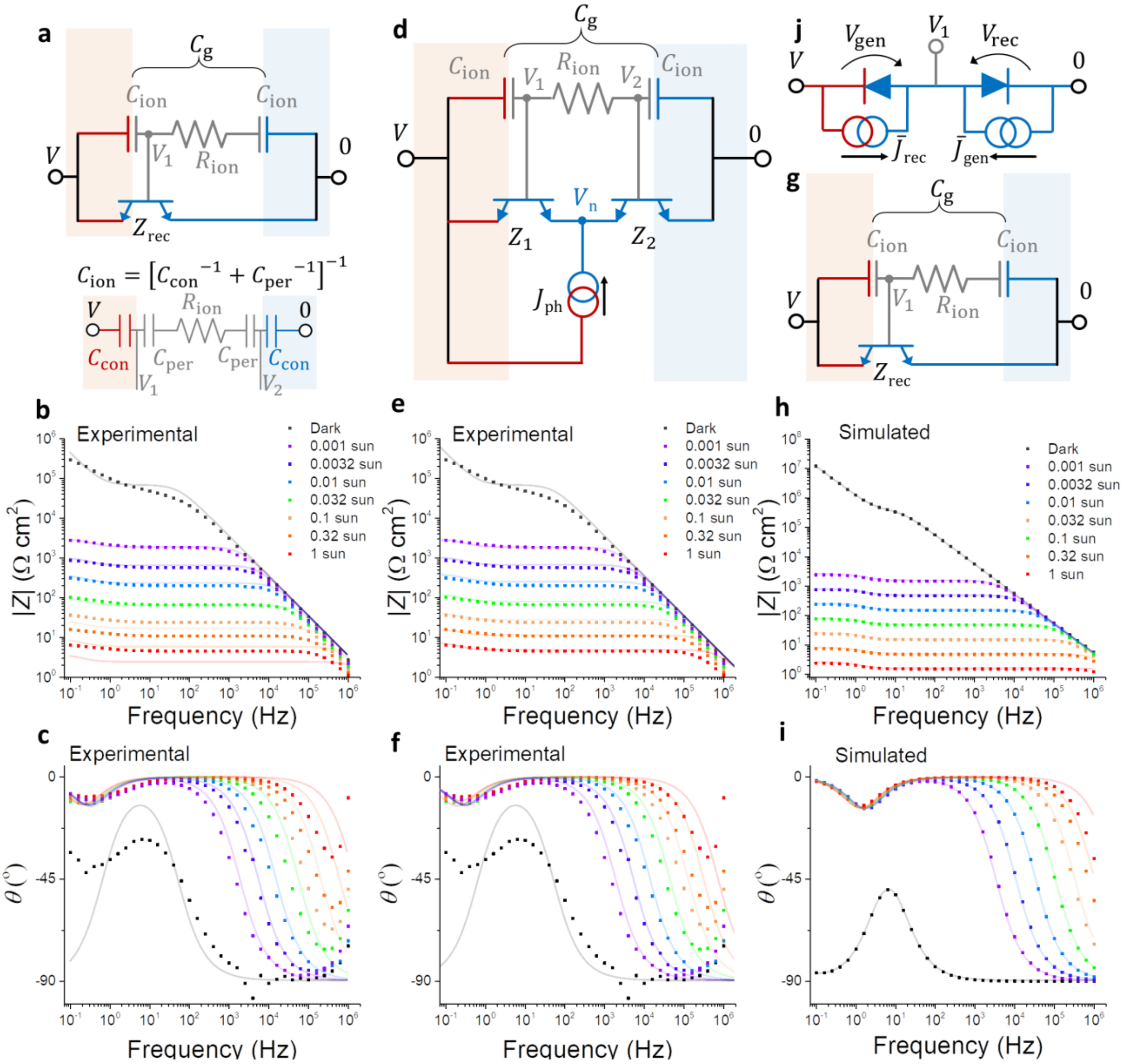

**Fig. S1 Complete measured and simulated impedance spectra corresponding to Fig 2 with equivalent circuit model fits.** The solid lines show the global fit to the measured and simulated data sets using the parameters listed in Table S1. **a-c**, The circuit model and measured impedance for the spiro-OMeTAD/ $Cs_{0.05}FA_{0.81}MA_{0.14}PbI_{2.55}Br_{0.45}/TiO_2$ solar cell in Fig. 2a and b and 5 free parameter global fit. At low frequencies it is apparent that the contribution from the transport of ionic defects is somewhat dispersive (ion movement with a range of time constants) whereas the circuit model and simulations assume non-dispersive transport. Some of the dispersive behaviour may be related to the presence of a thin (150 nm) mesoporous $TiO_2$ layer in this device which is not accounted for in the simulation or circuit model. Fine tuning the details of the ionic conduction model in the device and simulation would enable more precise characterisation of measured devices. The deviation of the fits at higher light intensities is likely to be related to either electronic screening of the interfaces by photogenerated charge (Note S3, ESI) and/or an increasing contribution from injection/collection

impedance to the measured impedance. The lower panel of **a** shows the detailed capacitive contributions to $C_{ion}$ from the space charge layer of the contact, $C_{con}$, and the ionic accumulation layer in the perovskite, $C_{per}$, as well as the consequences for determining the electrostatic potential at the interfaces, $V_1$ and $V_2$, if $C_{con}$ and $C_{per}$ are of comparable magnitudes instead of when $C_{ion} \approx C_{con}$ as is implicitly assumed in the upper part of the panel. **d-f,** Circuit model and 6 free parameter global fit to the experimental data in **b** and **c** including photogeneration (where $J_{ph}$ is defined to be negative) and an injection/collection transistor element. **g-i**, Circuit model and 5 free parameter global fit to the simulated measurements in Fig. 2c and d. The global fit parameters for each case are given in Table S1. **j**, Ebbers-Moll representation[4] of the transistor model of interface 1 assuming infinite ionic-to-electronic current gain.

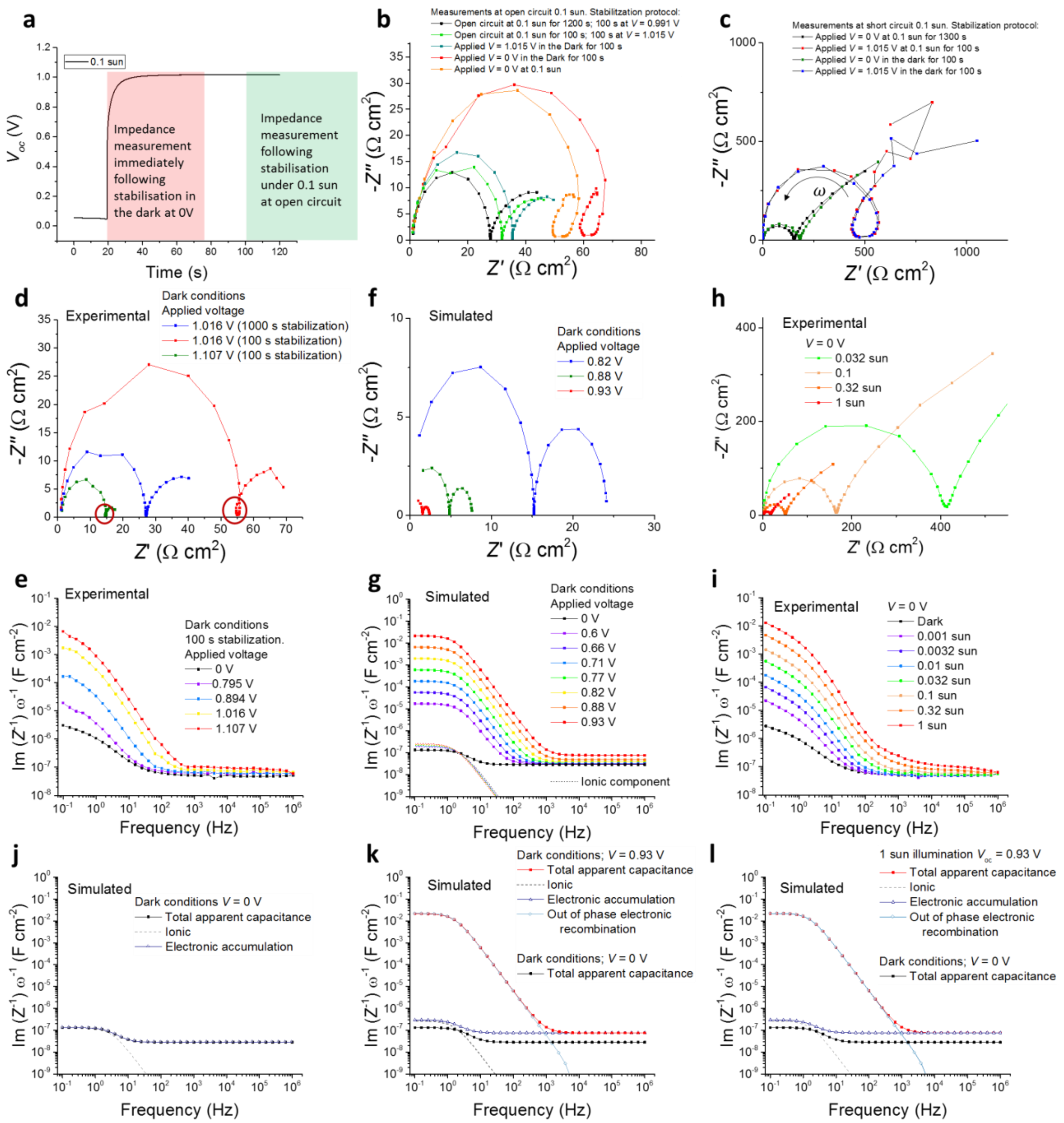

**Fig. S2 The effect of stabilisation time, light, and bias voltage in dark on impedance measurements, and the contributions to the apparent capacitance.** Measurements performed on the spiro-OMeTAD/ $Cs_{0.05}FA_{0.81}MA_{0.14}PbI_{2.55}Br_{0.45}/TiO_2$ solar cell and the simulated device in Fig. 2. **a**, Measured $V_{OC}$ *vs* time for 0.1 sun illumination following preconditioning at 0 V in the dark. **b**, Measured Nyquist plot of the imaginary *vs* real parts of the impedance over a frequency range 0.1 Hz to 1 MHz, showing effects of different stabilisation protocols prior to measurement at open circuit. **c**, Measured Nyquist plots showing effects of stabilisation protocol for measurements at short circuit. The individual impedance measurements were collected in order of decreasing frequency (opposite direction to arrow). **d - i**, Measured and simulated Nyquist plots and apparent capacitances, $\omega^{-1}\mathrm{Im}(Z^{-1})$, against frequency. (**d**, **e**) The effects of bias voltage in the dark for the measured device, and (**f**, **g**) the simulated device. Loops are

seen in the measured Nyquist plot (highlighted by the red circles) if the cell was only left to stabilise for 100 s prior to measurement at each voltage, but this loop disappeared if a longer stabilisation period of 1000 s was used prior to measurement. (**h**, **i**) The impedance spectra of the device measured at short circuit with the light intensities indicated in Fig. 2 show qualitatively similar behaviour as at open circuit, though with higher impedances. **j - l**, The different contributions to the apparent capacitance for the device simulated in Fig. 2 and Fig. 3. (**j**) Simulated under dark conditions with zero bias voltage, (**k**) dark with an applied voltage, and (**l**) with 1 sun equivalent conditions at open circuit conditions. Comparison between the electronic accumulation capacitance with an applied voltage or under light at open circuit and the total capacitance evaluated at 0 V in the dark illustrates the effect of the electronic charge in the perovskite on the 'geometric' capacitance (visible experimentally at high frequency in Fig. 2b).

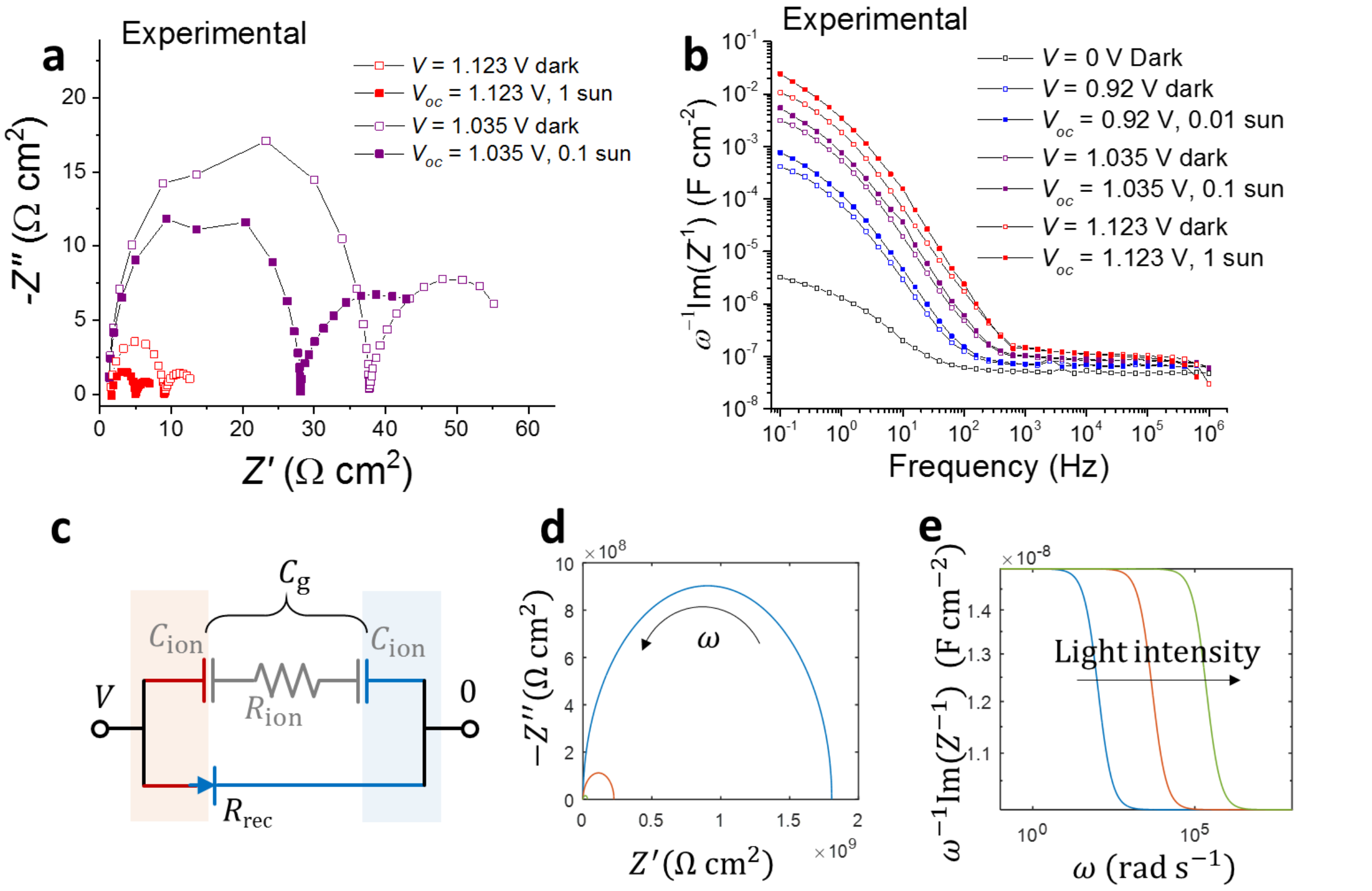

**Fig. S3 Possible consequences of photoinduced changes in ionic resistance for impedance spectra of a simplified hybrid perovskite solar cell calculated using an equivalent circuit model assuming $C_{\text{ion}}$ is constant. a**, **b**, Measured impedance in the light and the dark at the same bias voltage for the device shown in Fig. 2. The results indicate there is modest difference between the (**a**) magnitude of the impedance of the two states which might partly be explained by the consequences of optical heating or drift in cell behaviour (see Fig. S1), although could also be related to an photoinduced change in ionic conductivity[23, 24]. There is also a small change in the apparent capacitance (**b**). **c**, In this equivalent circuit model, the interfacial transistor element seen in Fig. 2e has been replaced with a diode element representing a conventional recombination process. Three light intensities are shown corresponding to potentials $V$ across the device of 0.1 V (blue), 0.2 V (red), and 0.3 V (green) and respective ionic resistances of $R_{\text{ion}} = 2 \times 10^{6}$, $4 \times 10^{4}$, $1 \times 10^{3}$ Ω cm$^{2}$. The other elements are $C_{\text{ion}} = 1 \times 10^{-8}$ F cm$^{-2}$, $C_{\text{g}} = 1 \times 10^{-8}$ F cm$^{-2}$ and $J_{s1} = 1 \times 10^{-11}$ A cm$^{-2}$. (**d**) and (**e**) show the resulting modelled impedance and capacitance. It is apparent that although the capacitance of the device shows a shift in its frequency dependence, there is no change in the magnitude of the device capacitance at low frequencies This is in contrast to observation where the apparent capacitance increases at low frequency but there is no significant shift in the frequency of this feature (Fig. 2b and Fig. S1). We note that if there were also photoinduced changes in $C_{\text{ion}}$ then it is possible that $C_{\text{ion}}$ and $R_{\text{ion}}$ could co-vary such that the time constant of the ionic response remained unchanged. However, since $C_{\text{ion}}$ will be predominantly controlled by the width of the interfacial space charge regions, which have contributions from both the accumulation/depletion of mobile ions in the perovskite as well as a contribution

from depletion of electrons or holes in the contacts. Any change in $C_{ion}$ is likely to be dominated by changes in the electronic depletion layer which to a first approximation scales with $(V_{bi}/(V_{bi} - V))^{1/2}$. Thus perfect co-variance of $C_{ion}$ and $R_{ion}$ is unlikely.

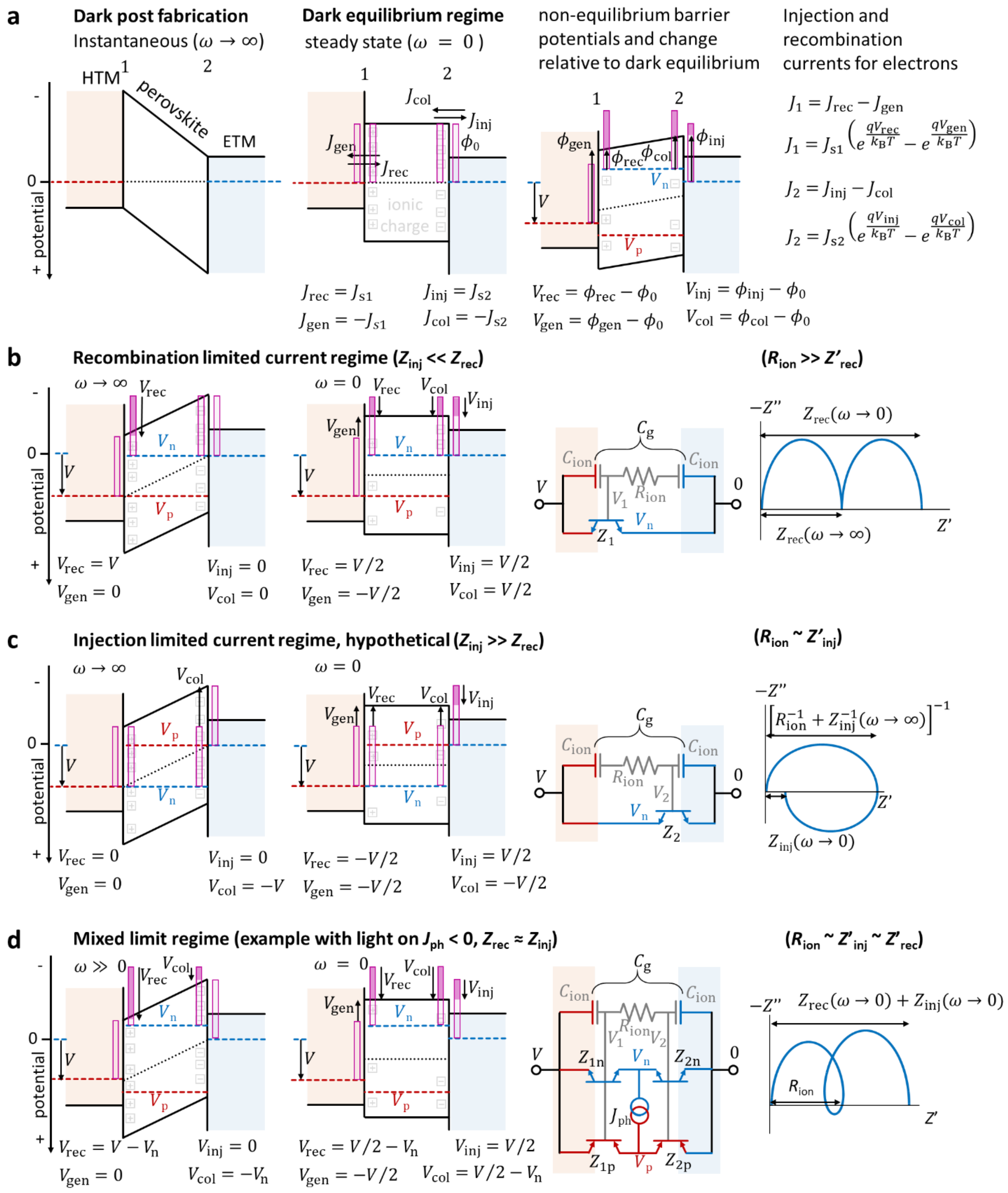

**Fig. S4 Simplified energy level diagrams and equivalent circuit models.** The conduction and valence bands of the perovskite layer are sandwiched by the hole transporting material (HTM, pink) and the electron transporting material (ETM, light blue), the vertical axis represents electrochemical potential energy which points down. The ionic accumulation layers are assumed negligibly thin. The equilibrium height of the energy barrier for electron injection/collection and recombination/generation in the dark is given by $\phi_0$ and ionic charge is represented by the light grey squares. The electron and hole quasi Fermi levels are indicated by the dashed blue and red lines, the other symbols are defined in the main text. The equivalent circuit diagrams are colour coded blue, red and grey to indicate the paths for electrons, holes and ions. **a**, The energy levels of the conduction and valence bands in the dark before and after ionic equilibration. The ideal Schottky-Mott limit electronic energy barriers are indicated, these change with applied potential and ionic redistribution. Energy levels after application of a voltage (*V*) shown instantaneously ($\omega\rightarrow\infty$) and at steady state ($\omega\rightarrow0$) and corresponding circuit models for devices in the: (**b**) recombination limited regime where $J_{s1} << J_{s2}$, (**c**) the injection limited regime where $J_{s1} >> J_{s2}$, and (**d**) the mixed limit regime. Example model Nyquist plots are also shown for each regime, the mixed limit plot corresponds to a special case where $R_{ion}$ is comparable to the real parts of $Z_{rec}$ and $Z_{inj}$.

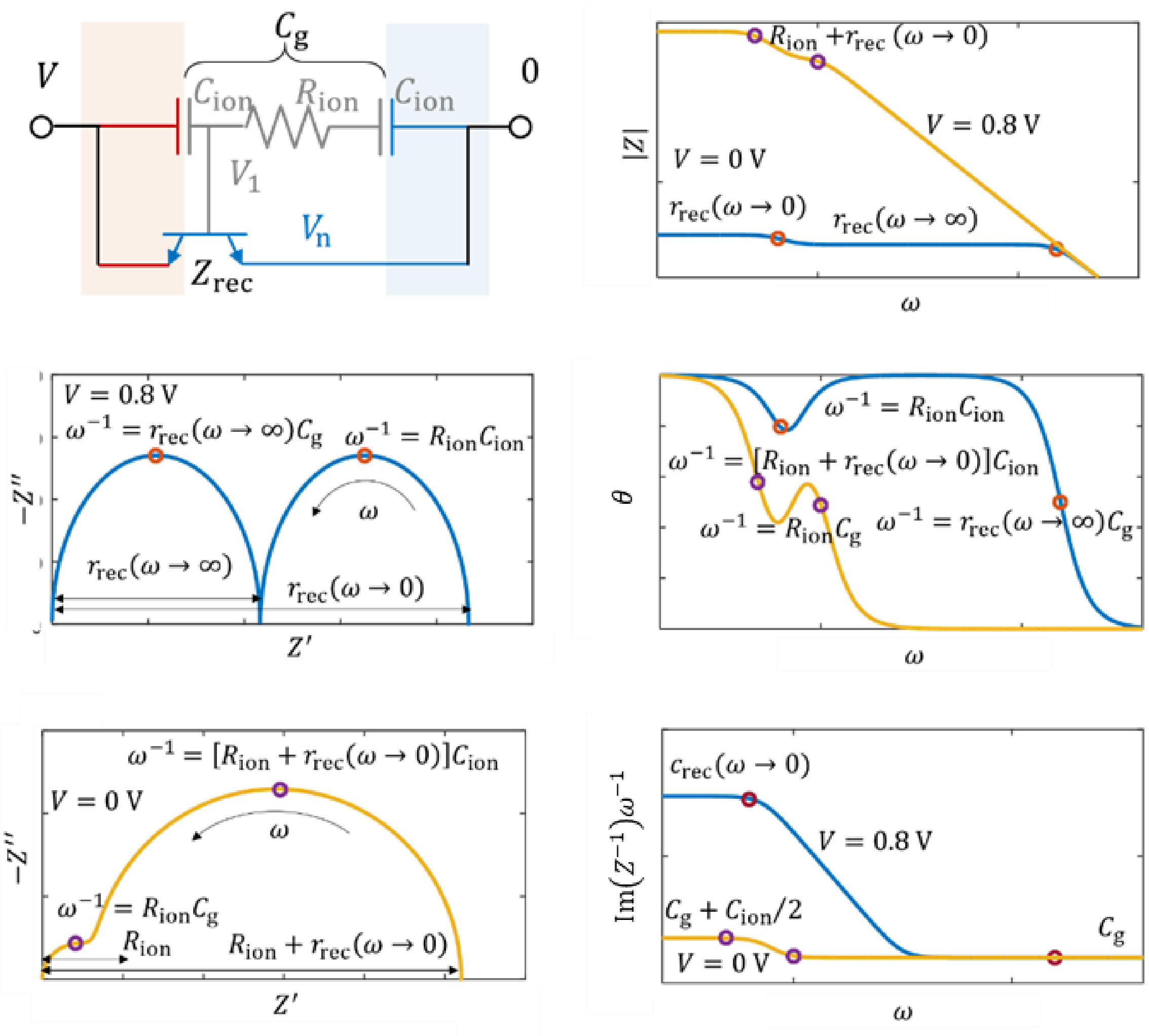

**Fig. S5 Interpretation of recombination limited impedance spectra.** Example of equivalent circuit model Nyquist plots and impedance spectra (magnitude $|Z(\omega)|$, phase $\theta$, and apparent capacitance $\mathrm{Im}(Z^{-1})\omega^{-1}$ for a recombination limited circuit showing the characteristic time constants at 0 V and 0.8 V. The time constants ($\omega^{-1}$) of various spectral features are indicated.

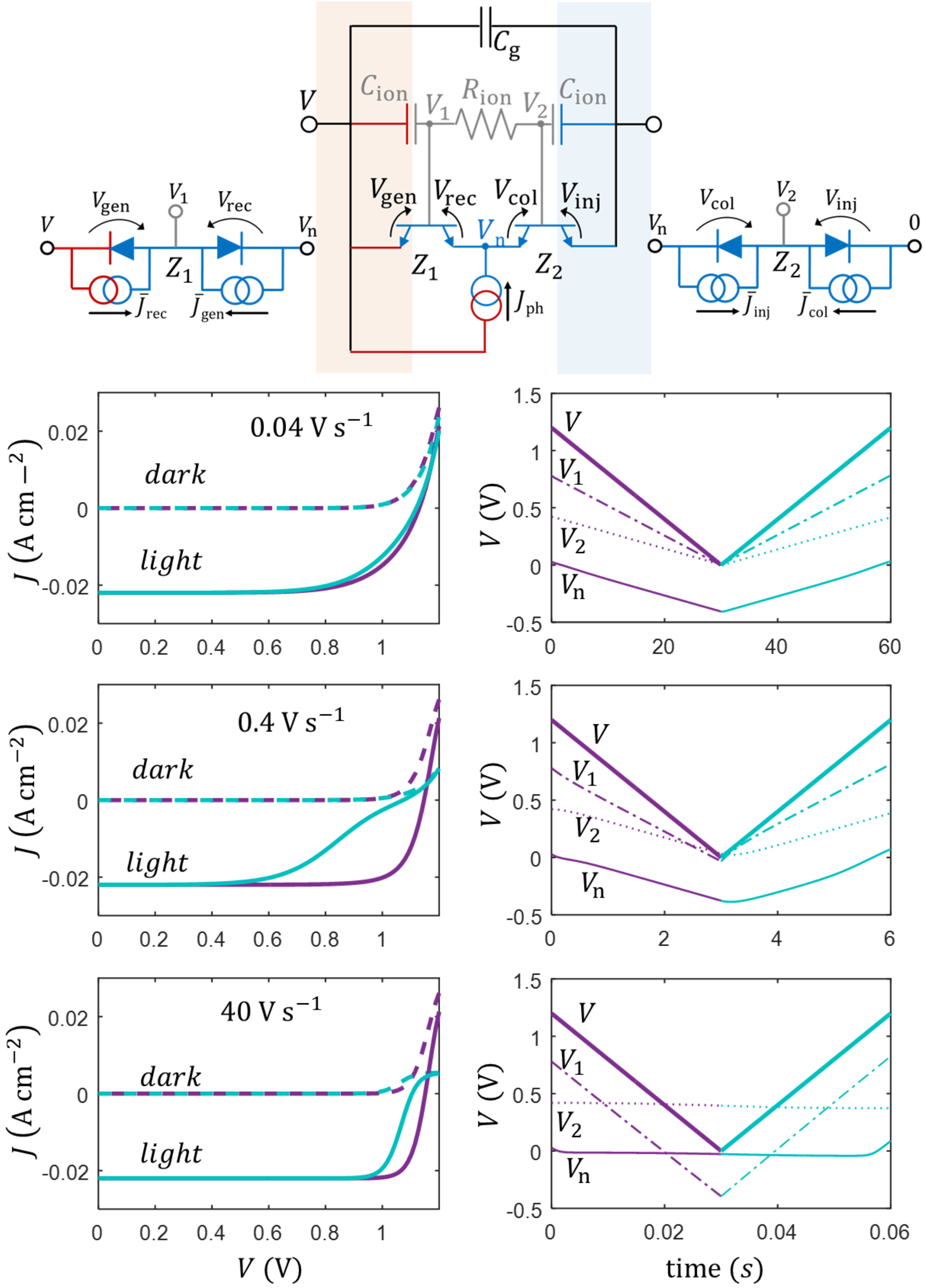

**Fig. S6 Circuit model cyclic voltammograms based on parameters from fit to experimental impedance data in Fig. 2.** The circuit model is shown above the plots, corresponding to the central column in Fig. S1 (Ebbers-Moll representations of interfacial transistors also shown) with impedance spectroscopy fitting parameters in Table S1. $J_{ph}$ = 22 mA cm$^{-2}$ (solid lines) and $J_{ph}$ = 0 (dashed lines) with a scan rates of $s$ = 0.04, 0.4, and 40 V s$^{-1}$ from 1.2 to 0 V for the reverse scan (purple) followed by a forward scan from 0 to 1.2 V (light blue). Applied voltage $V$, ionic interface potentials $V_1$ and $V_2$ and electron potential $V_n$ *vs* time are also shown for the illuminated $J_{ph}$ = 22 mA cm$^{-2}$ cases. The $s$ = 0.4 V s$^{-1}$ case is close to the measured current voltage curve seen in Fig. S8.

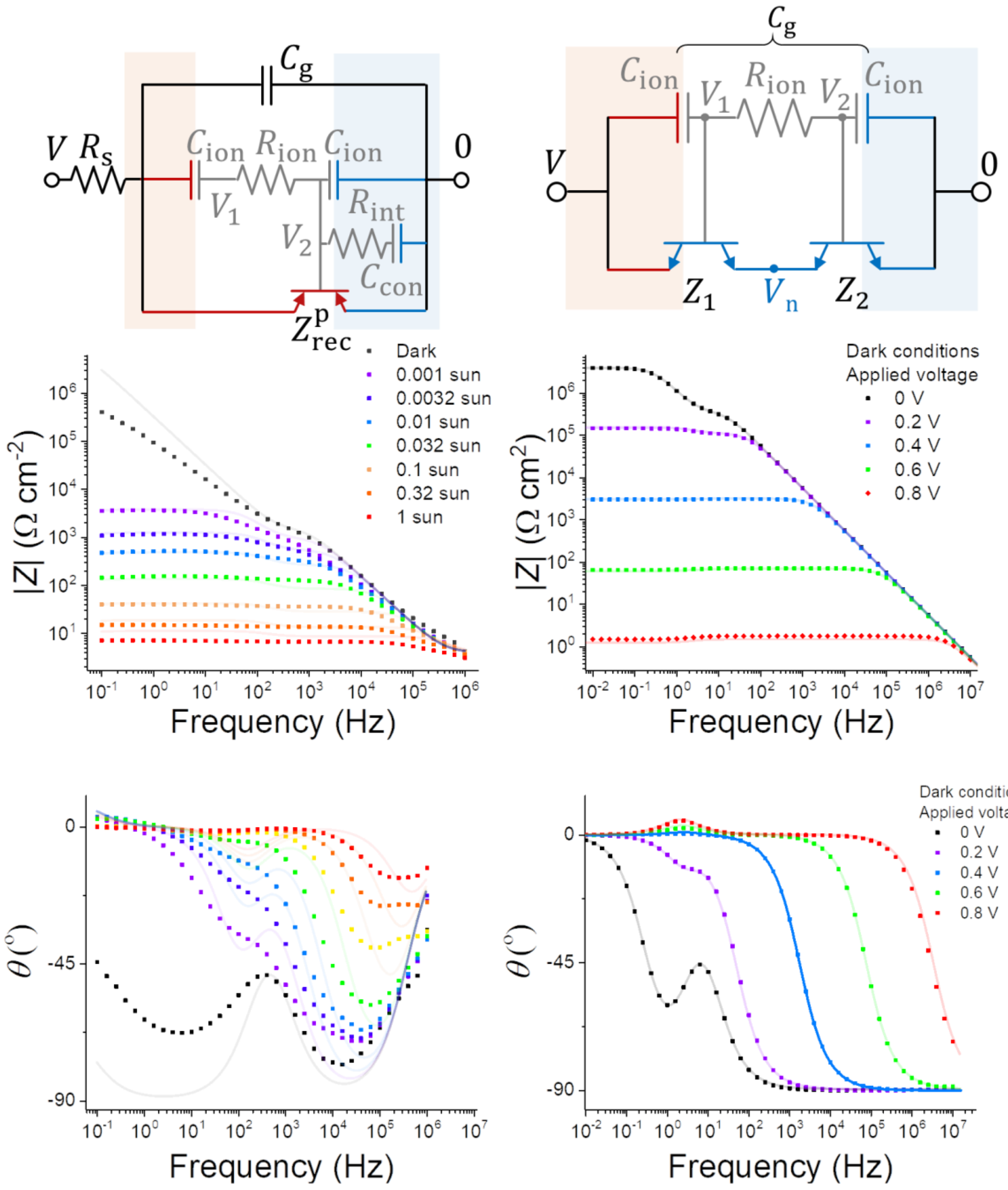

**Fig. S7 Circuit models and complete impedance spectra corresponding to Fig. 5a (left-hand column) and Fig. 5b (right-hand column).** The solid lines show the global fit to all the data using the parameters listed in Table S1. The left-hand column shows a global fit to the impedance measurements of the spiro-OMeTAD/ $FA_{0.85}MA_{0.15}PbI_3/SnO_x$ in Fig. 5a (measured around open circuit with different bias light intensities, see Methods and Table S1 caption for $V_{OC}$ values) assuming a model in which ions may penetrate or reversibly react at the recombination interface. The drift diffusion model parameters used to create the simulated impedance measurements in the right-hand column (Fig. 5b) are identical to those listed in Table S2 except that the recombination lifetimes of the contacts were reduced by 10,000 times so that $\tau_n = \tau_p = 5 \times 10^{-14}$ s, and the mobility of the majority carrier species in the contacts were reduced by 100 times so that $\mu_h = 0.2$ $cm^2 V^{-1} s^{-1}$ in the p-type contact and $\mu_e = 0.2$ $cm^2 V^{-1} s^{-1}$ in the n-type contact.

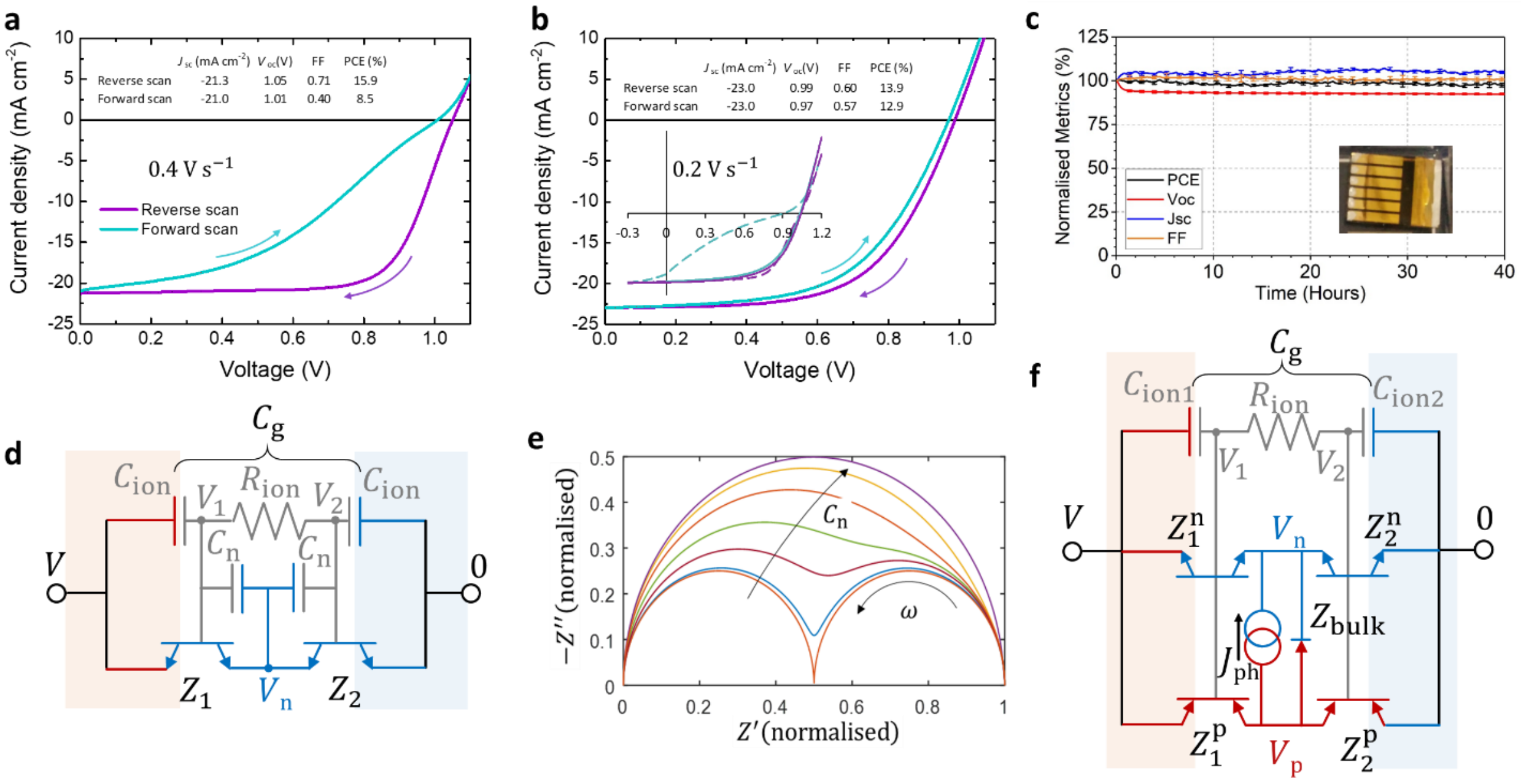

**Fig. S8 Solar cell data and circuit models described in the Methods and Note S3. a**, Current-voltage sweeps of the spiro-OMeTAD/ $Cs_{0.05}FA_{0.81}MA_{0.14}PbI_{2.55}Br_{0.45}/TiO_2$ solar cell in Fig. 2a and b measured under AM1.5 illumination with a sweep rate of 0.4 V s$^{-1}$. **b**, Current-voltage sweeps of the spiro-OMeTAD/ $FA_{0.85}MA_{0.15}PbI_3/SnO_2$ solar cell in Fig. 5a measured under AM1.5 illumination with a sweep rate of 0.2 V s$^{-1}$. For comparison the inset shows cyclic voltamograms for both cells (**a** -dashed lines, **b** - solid lines) measured both measured with a continuous 0.2 V s$^{-1}$ sweep cycle under the white LEDs (one sun equivalent, normalised photocurrent) used for the impedance measurements. **c**, Normalised power conversion efficiency (PCE), $V_{OC}$, short circuit current ($J_{SC}$), and fill factor (FF) as a function of illumination time of a device prepared using the same procedure as that measured in **a**. **d**, Equivalent circuit model including the effects of screening by electrons in the perovskite on the interfacial capacitances ($C_n$). **e**, Normalised Nyquist plot, calculated from the circuit model shown in **d**, indicating the effect of increasing $C_n$ on the shape of the spectrum. The example is calculated with the same parameters as those shown in Fig. S5 where $J_{s1} << J_{s2}$ with an applied voltage of 0.5 V, and varying $C_n$ from $10^{-12}$ – $10^{-7}$ F cm$^{-2}$. **f**, A general solar cell circuit model including, free electrons and holes, photogeneration, and the effects of bulk recombination.

## Supplementary Tables

**Table S1. Global fit parameters for the measured and simulated impedance data presented in the study.** The applied voltages used as inputs for the circuit model (Fig. 2e) of the experimental data in Fig. 2a and b (also Fig. S1b and c) were: 1.107 V (1 sun), 1.066 V (0.32 sun), 1.016 V (0.1 sun), 0.955 V (0.032 sun), 0.894 V (0.01 sun), 0.846 V (0.0032 sun), 0.795 V (0.001 sun) with a steady state ideality factor of $m_{ss}$ = 1.79. The applied voltages for the simulated measurements in Fig. 2c and d were: 0.931 V (1 sun), 0.876 V (0.32 sun), 0.822 V (0.1 sun), 0.766 V (0.032 sun), 0.711 V (0.01 sun), 0.656 V (0.0032 sun), 0.600 V (0.001 sun) with a steady state ideality factor of $m_{ss}$ = 1.84, $V_{bi}$ = 1.3 V. * The value inferred from the simulation input parameters is $R_{\text{ion}} = d_{\text{intrinsic}}/(q\mu_{\text{a}}N_{\text{ion}})$ = 3.1 ×10$^5$ Ω cm$^2$, close to the value extracted from the fit to the simulated impedance measurements using the expression $R_{\text{ion}} = c_{\text{rec}}(\bar{V})4m_{\text{ss}}(1-f_{\text{c}}/2)k_{\text{B}}T/(qf_{\text{c}}C_{\text{ion}}J_{\text{rec}}(\bar{V})$ = 3.8 ×10$^5$ Ω cm$^2$ (see main text, and, for the experimental data the inset in Fig. 2b). The deviation between the simulation input value and the fit value of $R_{ion}$ in the table below arises due to factors not accounted for by the circuit model which the fit attempts to compensate for, particularly the capacitive screening of interfaces by the electronic charge at the higher light intensities (see Note S3). The applied voltages used as inputs to the circuit model (Fig. S7, left column) for the experimental data in Fig. 5a were: 1.061 V (1 sun), 1.012 V (0.32 sun), 0.948 V (0.1 sun), 0.865 V (0.032 sun), 0.777 V (0.01 sun), 0.713 V (0.0032 sun), 0.638 V (0.001 sun), with a steady state ideality factor of $m_{ss}$ = 2.43. The applied voltages used as inputs for the circuit model (Fig. 4g, Fig. S7, right column) of the simulated impedance measurements in Fig. 5b were: 0, 0.2, 0.4, 0.6, and 0.8 V, with a steady state ideality factor of $m_{ss}$ = 1.93. The ideality factor for charge injection/collection was assumed to be unity. Fit uncertainties approximately correspond to the number of decimal places shown.

| Parameter | Experimental data (Fig. 2a,b) | Simulated data (Fig. 2c,d) | Experimental data (Fig. S1d-f) | Experimental data (Fig. 5a) | Simulated data (Fig. 5b) |
|---|---|---|---|---|---|
| $R_s$ (Ω cm$^2$) | - | - | - | 3.2 | - |
| $C_g$ (F cm$^{-2}$) | 4.4 × 10$^{-8}$ | 2.8 × 10$^{-8}$ | 4.4 × 10$^{-8}$ | 1.0 × 10$^{-7}$ | 2.8 × 10$^{-8}$ |
| $R_{ion}$ (Ω cm$^2$) | 6.7 × 10$^4$ | 3.8 ×10$^5$ * | 6.7 × 10$^4$ | 2.2 × 10$^3$ | 3.8 ×10$^5$ |
| $R_{int}$ (Ω cm$^2$) | - | - | - | 4.1 × 10$^6$ | - |
| $C_{ion}$ (F cm$^{-2}$) | 7.2 × 10$^{-6}$ | 2.6 × 10$^{-7}$ | 7.2 × 10$^{-6}$ | 8.6 × 10$^{-7}$ | 2.6 × 10$^{-7}$ |
| $C_{con}$ (F cm$^{-2}$) | - | - | - | 7.8 × 10$^{-7}$ | - |
| $J_{s1}$ (A cm$^{-2}$) | 6.1 × 10$^{-13}$ | 7.1 × 10$^{-11}$ | 7.0 × 10$^{-13}$ | - | 1.19 × 10$^{-8}$ |
| $J_{s2}$ (A cm$^{-2}$) | - | - | 3.1 × 10$^{-9}$ | 6.0 × 10$^{-9}$ | 1.50 × 10$^{-8}$ |
| $f_c$ | 0.70 | 0.77 | 0.70 | 0.65 | 0.996 |

**Table S2 Drift-diffusion simulation parameters.** These parameters were used for all the simulated data (simulated as described in reference [25]), except where explicitly stated. The 1 sun equivalent $V_{OC}$ resulting from this parameters set is 0.931 V, the resulting $J_{SC}$ is 20.3 mA/cm$^2$.

| **Parameter name** | **Symbol** | **p-type** | **Intrinsic** | **n-type** | **Unit** |
|---|---|---|---|---|---|
| Layer thickness | $d$ | 200 | 500 | 200 | nm |
| Band gap | $E_g$ | 1.6 | 1.6 | 1.6 | eV |
| Built in voltage | $V_{bi}$ | 1.3 | 1.3 | 1.3 | V |
| Relative dielectric constant | $\varepsilon_s$ | 20 | 20 | 20 | |
| Mobile ionic defect density | $N_{ion}$ | 0 | $10^{19}$ | 0 | cm$^{-3}$ |
| Ion mobility | $\mu_a$ | - | $10^{-10}$ | - | cm$^2$ V$^{-1}$ s$^{-1}$ |
| Electron mobility | $\mu_e$ | 0.02 | 20 | 20 | cm$^2$ V$^{-1}$ s$^{-1}$ |
| Hole mobility | $\mu_h$ | 20 | 20 | 0.02 | cm$^2$ V$^{-1}$ s$^{-1}$ |
| Donor doping density | $N_A$ | $3.0 \times 10^{17}$ | - | - | cm$^{-3}$ |
| Acceptor doping density | $N_D$ | - | - | $3.0 \times 10^{17}$ | cm$^{-3}$ |
| Effective density of states | $N_0$ | $10^{20}$ | $10^{20}$ | $10^{20}$ | cm$^{-3}$ |
| Band-to-band recombination rate coefficient | $k_{btb}$ | $10^{-12}$ | $10^{-12}$ | $10^{-12}$ | cm$^{-3}$ s$^{-1}$ |
| SRH trap energy | $E_t$ | $E_{CB}$-0.8 | - | $E_{CB}$-0.8 | eV |
| SRH time constants | $\tau_n$, $\tau_p$ | $5 \times 10^{-10}$ | - | $5 \times 10^{-10}$ | s |
| Generation rate | $G$ | - | $2.5 \times 10^{21}$ | - | cm$^{-3}$ s$^{-1}$ |

**Table S3. Changes in interfacial barrier potentials and small perturbation impedances due to ionic redistribution considering only free electrons.** The terms in the equations are described in the main text, Note S2 and illustrated in Fig. 4g and Fig. S4. In the small perturbation regime an oscillating voltage $v$ is superimposed on the steady state cell bias potential $\bar{V}$. Complete expressions considering holes are given in the in Table S4 and Note S6, considering asymmetric interfacial capacitances, and screening within the perovskite. The symbols covered by a bar (e.g. $\bar{V}$) indicate the steady state value of the at quantity when $\omega \rightarrow 0$. *Assumes that mobile ionic charge does not penetrate or react at interfaces and the $C_{ion}$ is the same at each interface.

| Change in barrier potential for: | (V) | | (V) | | response to small voltage perturbation, $v$ (V) |
|---|---|---|---|---|---|
| Electron generation | $V_{gen}$ | = | $V_1 - V$ | = | $-\bar{V}\bar{A} - vA$ |
| Electron recombination | $V_{rec}$ | = | $V_1 - V_n$ | = | $\bar{V}(1 - \bar{A} - \bar{B}_n) + v(1 - A - B_n)$ |
| Electron collection | $V_{col}$ | = | $V_2 - V_n$ | = | $\bar{V}(\bar{A} - \bar{B}_n) + v(A - B_n)$ |
| Electron injection | $V_{inj}$ | = | $V_2$ | = | $\bar{V}\bar{A} + vA$ |

| Small voltage perturbation parameters | | | | | |
|---|---|---|---|---|---|
| Fraction of ionic screening potential within contact layer | $f_c$ | = | $1 - \frac{C_{ion}}{C_{per}}$ | = | $1 - \frac{\text{total interface capacitance}}{\text{perov. space charge capacitance}}$ |
| Fraction of voltage change at interface due to ionic redistribution* | $A$ | = | $\frac{v_C}{v}$ | = | $\frac{f_c}{2 + i\omega R_{ion} C_{ion}}$ |
| Potential due to ions at interface 1 (V) | | | $v_1$ | = | $v(1 - A)$ |
| Potential due to ions at interface 2 (V) | | | $v_2$ | = | $vA$ |
| Fractional change in voltage of electron quasi Fermi level | $B_n$ | = | $\frac{v_n}{v}$ | = | $\frac{\bar{J}_{rec} + A(\bar{J}_{gen} - \bar{J}_{rec} + \bar{J}_{col} - \bar{J}_{inj})}{\bar{J}_{rec} + \bar{J}_{col}}$ |

| Interfacial currents | | | (A cm$^{-2}$) |
|---|---|---|---|
| Electron generation | $J_{gen}$ | = | $J_{s1} e^{\frac{qV_{gen}}{m_1 k_B T}}$ |
| Electron recombination | $J_{rec}$ | = | $J_{s1} e^{\frac{qV_{rec}}{m_1 k_B T}}$ |
| Electron collection | $J_{col}$ | = | $J_{s2} e^{\frac{qV_{col}}{m_2 k_B T}}$ |
| Electron injection | $J_{inj}$ | = | $J_{s2} e^{\frac{qV_{inj}}{m_2 k_B T}}$ |

| Interfacial impedances | | | ($\Omega$ cm$^2$) |
|---|---|---|---|
| Electron generation impedance | $Z_{gen}$ | = | $\frac{(1 - B_n)}{A} \frac{m_1 k_B T}{q\bar{J}_{gen}}$ |
| Electron recombination impedance | $Z_{rec}$ | = | $\frac{(1 - B_n)}{(1 - A - B_n)} \frac{m_1 k_B T}{q\bar{J}_{rec}}$ |
| Interface 1 electron impedance | $Z_1$ | $=$ | $\left(\frac{1}{Z_{\mathrm{rec}}}+\frac{1}{Z_{\mathrm{gen}}}\right)^{-1}$ |
| Electron collection impedance | $Z_{\mathrm{col}}$ | $=$ | $\frac{B_{\mathrm{n}}}{(B_{\mathrm{n}}-A)}\frac{m_2 k_{\mathrm{B}} T}{q\bar{J}_{\mathrm{col}}}$ |
| Electron injection impedance | $Z_{\mathrm{inj}}$ | $=$ | $\frac{B_{\mathrm{n}}}{A}\frac{m_2 k_{\mathrm{B}} T}{q\bar{J}_{\mathrm{inj}}}$ |
| Interface 2 electron impedance | $Z_2$ | $=$ | $\left(\frac{1}{Z_{\mathrm{inj}}}+\frac{1}{Z_{\mathrm{col}}}\right)^{-1}$ |

**Table S4. Changes in interfacial barrier potentials and small perturbation impedances due to ionic redistribution considering both free electrons and holes, and including bulk recombination.** The superscripts n and p are used to processes involving free electrons or holes respectively, *they are not exponents*. The terms in the equations are described in the main text and illustrated in Fig. 4. In the small perturbation regime an oscillating voltage $v$ is superimposed on the cell potential $V$. The electron and hole quasi Fermi levels, $V_n$ and $V_p$ have corresponding small perturbation oscillations $v_n$ and $v_p$. The ideality factors of interface 1 and 2 are given by $m_1$ and $m_2$ respectively. $A_1$ and $A_2$ arise because the capacitances of each interface are different, $C_{ion1}$ and $C_{ion2}$. The symbols covered by a bar (e.g. $\bar{V}$) indicate the steady state value of the quantity when $\omega \rightarrow 0$. *Assumes that mobile ionic charge does not penetrate or chemically react at interfaces.

| Change in barrier potential for: | | (V) | | Response to small voltage perturbation, *v* (V) |
|---|---|---|---|---|
| Electron generation | $V_{gen}^{n}$ | $= V_1 - V$ | = | $-\bar{V}\bar{A}_1 - vA_1$ |
| Electron recombination | $V_{rec}^{n}$ | $= V_1 - V_n$ | = | $\bar{V}(1-\bar{A}_1-\bar{B}_n)+v(1-A_1-B_n)$ |
| Electron collection | $V_{col}^{n}$ | $= V_2 - V_n$ | = | $\bar{V}(\bar{A}_2-\bar{B}_n)+v(A_2-B_n)$ |
| Electron injection | $V_{inj}^{n}$ | $= V_2$ | = | $\bar{V}\bar{A}_2 + vA_2$ |
| Hole generation | $V_{gen}^{p}$ | $= -V_2$ | = | $-\bar{V}\bar{A}_2 - vA_2$ |
| Hole recombination | $V_{rec}^{p}$ | $= V_p - V_2$ | = | $\bar{V}(\bar{B}_p-\bar{A}_2)+v(B_p-A_2)$ |
| Hole collection | $V_{col}^{p}$ | $= V_p - V_1$ | = | $\bar{V}(\bar{B}_p+\bar{A}_1-1)+v(B_p+A_1-1)$ |
| Hole injection | $V_{inj}^{p}$ | $= V - V_1$ | = | $\bar{V}\bar{A}_1 + vA_1$ |
| Bulk recombination | $V_{bulk}$ | $= V_p - V_n$ | = | $\bar{V}(\bar{B}_p-\bar{B}_n)+v(B_p-B_n)$ |

| Small voltage perturbation parameters | | | | |
|---|---|---|---|---|
| Fraction of ionic screening potential within contact layers | $f_c$ | $= 1 - \frac{C_{ion}}{C_{per}}$ | = | $1 - \frac{\text{total interface capacitance}}{\text{perov. space charge capacitan}}$ |
| Fraction voltage change at interface 1 due ion redistribution* | $A_1$ | $= \frac{v_{C_{ion1}}}{v}$ | = | $\frac{f_c}{1 + C_{ion1}/C_{ion2} + i\omega R_{ion} C_{ion1}}$ |
| Fraction voltage change at interface 2 due ion redistribution* | $A_2$ | $= \frac{v_{C_{ion2}}}{v}$ | = | $\frac{f_c}{1 + C_{ion2}/C_{ion1} + i\omega R_{ion} C_{ion2}}$ |
| Potential due to ions at interface 1 (V) | | $v_1$ | = | $v(1-A_1)$ |
| Potential due to ions at interface 2 (V) | | $v_2$ | = | $vA_2$ |
| Fractional change in voltage of electron quasi Fermi level | $B_n$ | $= \frac{v_n}{v}$ | = | Lengthy analytical expression, solved using Kirchhoff's laws |
| Fractional change in voltage of hole quasi Fermi level | $B_p$ | $= \frac{v_p}{v}$ | = | Lengthy analytical expression, solved using Kirchhoff's laws |

| Interfacial currents | (A cm$^{-2}$) |
|---|---|

| | | | |
|---|---|---|---|
| Ideality factor of interface 1 | $m_1$ | | |
| Ideality factor of interface 2 | $m_2$ | | |
| Electron generation | $J_{\mathrm{gen}}^{\mathrm{n}}$ | = | $J_{\mathrm{s1}} e^{\frac{qV_{\mathrm{gen}}^{\mathrm{n}}}{m_1 k_{\mathrm{B}} T}}$ |
| Electron recombination | $J_{\mathrm{rec}}^{\mathrm{n}}$ | = | $J_{\mathrm{s1}} e^{\frac{qV_{\mathrm{rec}}^{\mathrm{n}}}{m_1 k_{\mathrm{B}} T}}$ |
| Electron collection | $J_{\mathrm{col}}^{\mathrm{n}}$ | = | $J_{\mathrm{s2}} e^{\frac{qV_{\mathrm{col}}^{\mathrm{n}}}{m_2 k_{\mathrm{B}} T}}$ |
| Electron injection | $J_{\mathrm{inj}}^{\mathrm{n}}$ | = | $J_{\mathrm{s2}} e^{\frac{qV_{\mathrm{inj}}^{\mathrm{n}}}{m_2 k_{\mathrm{B}} T}}$ |
| Hole generation | $J_{\mathrm{gen}}^{\mathrm{p}}$ | = | $J_{\mathrm{s1}} e^{\frac{qV_{\mathrm{gen}}^{\mathrm{p}}}{m_2 k_{\mathrm{B}} T}}$ |
| Hole recombination | $J_{\mathrm{rec}}^{\mathrm{p}}$ | = | $J_{\mathrm{s1}} e^{\frac{qV_{\mathrm{rec}}^{\mathrm{p}}}{m_2 k_{\mathrm{B}} T}}$ |
| Hole collection | $J_{\mathrm{col}}^{\mathrm{p}}$ | = | $J_{\mathrm{s2}} e^{\frac{qV_{\mathrm{col}}^{\mathrm{p}}}{m_1 k_{\mathrm{B}} T}}$ |
| Hole injection | $J_{\mathrm{inj}}^{\mathrm{p}}$ | = | $J_{\mathrm{s2}} e^{\frac{qV_{\mathrm{inj}}^{\mathrm{p}}}{m_1 k_{\mathrm{B}} T}}$ |
| Bulk recombination | $J_{\mathrm{bulk}}$ | = | $J = J_0 \left( e^{\frac{qV_{\mathrm{bulk}}}{k_{\mathrm{B}} T}} - 1 \right)$ |

| **Interfacial impedances** | | | **(Ω cm²)** |
|---|---|---|---|
| Electron generation impedance | $Z_{\mathrm{gen}}^{\mathrm{n}}$ | = | $\frac{(1 - B_{\mathrm{n}})}{A_1} \frac{m_1 k_{\mathrm{B}} T}{q\bar{J}_{\mathrm{gen}}^{\mathrm{n}}}$ |
| Electron recombination impedance | $Z_{\mathrm{rec}}^{\mathrm{n}}$ | = | $\frac{(1 - B_{\mathrm{n}})}{(1 - A_1 - B_{\mathrm{n}})} \frac{m_1 k_{\mathrm{B}} T}{q\bar{J}_{\mathrm{rec}}^{\mathrm{n}}}$ |
| Interface 1 electron impedance | $Z_1^{\mathrm{n}}$ | = | $\left( \frac{1}{Z_{\mathrm{rec}}^{\mathrm{n}}} + \frac{1}{Z_{\mathrm{gen}}^{\mathrm{n}}} \right)^{-1}$ |
| Electron collection impedance | $Z_{\mathrm{col}}^{\mathrm{n}}$ | = | $\frac{B_{\mathrm{n}}}{(B_{\mathrm{n}} - A_2)} \frac{m_2 k_{\mathrm{B}} T}{q\bar{J}_{\mathrm{col}}^{\mathrm{n}}}$ |
| Electron injection impedance | $Z_{\mathrm{inj}}^{\mathrm{n}}$ | = | $\frac{B_{\mathrm{n}}}{A_2} \frac{m_2 k_{\mathrm{B}} T}{q\bar{J}_{\mathrm{inj}}^{\mathrm{n}}}$ |
| Interface 2 electron impedance | $Z_2^{\mathrm{n}}$ | = | $\left( \frac{1}{Z_{\mathrm{inj}}^{\mathrm{n}}} + \frac{1}{Z_{\mathrm{col}}^{\mathrm{n}}} \right)^{-1}$ |
| Hole generation impedance | $Z_{\mathrm{gen}}^{\mathrm{p}}$ | = | $\frac{B_{\mathrm{p}}}{A_2} \frac{m_2 k_{\mathrm{B}} T}{q\bar{J}_{\mathrm{gen}}^{\mathrm{p}}}$ |
| Hole recombination impedance | $Z_{\mathrm{rec}}^{\mathrm{p}}$ | = | $\frac{B_{\mathrm{p}}}{(B_{\mathrm{p}} - A_2)} \frac{m_2 k_{\mathrm{B}} T}{q\bar{J}_{\mathrm{rec}}^{\mathrm{p}}}$ |
| Interface 2 hole impedance | $Z_2^{\mathrm{p}}$ | = | $\left( \frac{1}{Z_{\mathrm{rec}}^{\mathrm{p}}} + \frac{1}{Z_{\mathrm{gen}}^{\mathrm{p}}} \right)^{-1}$ |
| Hole collection impedance | $Z_{\mathrm{col}}^{\mathrm{p}}$ | = | $\frac{(1 - B_{\mathrm{p}})}{(1 - A_1 - \mathrm{B}_p)} \frac{m_1 k_{\mathrm{B}} T}{q\bar{J}_{\mathrm{col}}^{\mathrm{p}}}$ |
| Hole injection impedance | $Z_{\mathrm{inj}}^{\mathrm{p}}$ | = | $\frac{(1-B_{\mathrm{p}})}{A_1}\frac{m_1 k_{\mathrm{B}}T}{q\bar{J}_{\mathrm{inj}}^{\mathrm{p}}}$ |
| Interface 1 hole impedance | $Z_1^{\mathrm{p}}$ | = | $\left(\frac{1}{Z_{\mathrm{inj}}^{\mathrm{p}}}+\frac{1}{Z_{\mathrm{col}}^{\mathrm{p}}}\right)^{-1}$ |
| Bulk recombination impedance | $Z_{\mathrm{bulk}}$ | = | $\frac{k_{\mathrm{B}}T}{q\bar{J}_{\mathrm{bulk}}}$ |
| Impedance of hole circuit branch | $Z_{\mathrm{p}}$ | = | $Z_1^{\mathrm{p}}+Z_2^{\mathrm{p}}$ |
| Impedance of electron circuit branch | $Z_{\mathrm{n}}$ | = | $Z_1^{\mathrm{n}}+Z_2^{\mathrm{n}}$ |
| Total impedance of active layer interfaces | $Z_{\mathrm{np}}$ | = | $\left(\frac{1}{Z_{\mathrm{n}}}+\frac{1}{Z_{\mathrm{p}}}\right)^{-1}$ |